\documentclass[12pt]{iopart}

\usepackage{graphicx}
\usepackage{bm}
\begin{document}

\title[Gravitational-wave Extraction using Independent Component Analysis]{Gravitational-wave Extraction using Independent Component Analysis}

\author{Rika Shimomura, Yuuichi Tabe \& Hisaaki Shinkai}

\address{Faculty of Information Science and Technology, Osaka Institute of Technology,\\
  Kitayama 1-79-1, Hirakata City, Osaka 573-0196, Japan
}
\ead{hisaaki.shinkai@oit.ac.jp}
\vspace{10pt}
\begin{indented}
\item[]May 4, 2025
\end{indented}

\begin{abstract}
Independent component analysis (ICA) is a method to extract a set of time-series data using ``statistical independency" of each component. 
We applied ICA to extract gravitational wave (GW) signals directly from the detector data.  
Our idea is to extract a coherent signal that is included in multiple detectors and find it by shifting the data set around its arrival time.
In this article, we report several tests using injected signals, and show that this method works for inspiral-wave events with a signal-to-noise ratio of $> 15$.  We then applied the method to actual LIGO-Virgo-KAGRA O1-O3 events, and showed that the identification of the arrival time can be estimated more precisely than previously reported. 
This approach does not require templates of waveform, therefore it is attractive for testing theories of gravity, and for finding unknown GW. 
\end{abstract}

%
%
%
%
%

\section{Introduction}

Almost a decade has passed since the LIGO-Virgo collaboration directly detected the first gravitational wave (GW), GW150914 \cite{GW150914}. Until now,  the LIGO-Virgo-KAGRA (LVK) collaboration has reported 90 events as the GWTC-3 catalog, which is the result of the O3 observing run that ended in March 2020\cite{GWTC3}.  The catalog consists of 90 events from compact binary coalescence (CBC), of which 85 are from binary black holes (BHs). The LVK collaboration plans to announce the next catalog in the coming months, in which we expect the number of detections to double. In short, we will enter an era of GW astronomy. 

As the number of events increases, various aspects of astronomy and physics can be discussed.  For example, it is unclear how super-massive BHs are formed in the center of galaxies. This will require BH formation processes together with its formation and merging rates, which will be explained by GW statistics (see references in \cite{GWTC3population, shinkai2017}).  We are not sure how general relativity (GR) is consistent with quantum theory.  This viewpoint requires the validity of GR, for which the best test can be performed in strong-gravity regimes, such as the merger of BHs (see references in \cite{GWTC3TGR, mockdata}). 

GW signals are quite weak and are often buried in the noisy outputs of detectors.  The main method for detecting GWs from CBCs is matched-filtering analysis, which uses a template bank of waveforms and measures the correlation between a template and the observed signal.  However, this method is only applicable to the prepared set of templates and is not effective for others. 
For example, GW searches of burst events from supernova explosions and/or stochastic events from phase transitions in the early universe have not yet been systematically performed because of the lack of templates.  We are testing GR using templates based on GR, but are not able to test other gravity theories. 
Therefore, it is desirable to develop a new method that can identify GW signals without using prepared templates.

Several new ideas have been proposed in this regard. One idea is to process time series data with mathematical and/or statistical techniques, or another is to use machine learning approaches (see a comparison of them for extracting ringdown mock data \cite{mockdata} and references therein).  This idea can be applied both for removing noise and  extracting GW signals and is expected to be useful for supporting current standard data analysis methods. 

In this article, we propose a new method for extracting GW signals directly from the outputs of detector data, using independent component analysis (ICA).
ICA is a method for separating a set of time series data into a new set, as each component has ``statistical independence" (see a review by Hyv\"arinen {\it et al.} \cite{ICAarticle, ICAbook}).  A well-known example is ``blind signal separation," a method to identify the voices of $n$ persons in a cocktail party from their mixed sound files of $n$ microphones. See \cite{BellandSejnowski} for $n=10$ case.  
ICA provides a linear representation of non-Gaussian data such that the components are statistically independent or as independent as possible.  The fundamental measure is the separation from the Gaussian.  Thus, in GW data analysis, we expect that GW extraction using ICA is effective, at least when the background noise is Gaussian.  

We note the differences between ICA and principal component analysis (PCA) for which applications to GW analysis have been reported recently  \cite{Saleem2021, arXiv:2208.07757, Rui2024, Watarai2024}.  Both have the common goal of extracting data as in the lower-dimensional space from the original data, but their criteria are different.  PCA is based on the principle of uncorrelation, which maximizes the variance of the data. On the other hand, ICA is based on statistical independence; in other words, no information can be obtained from other components, which is a stronger requirement than uncorrelation.  PCA is suitable for dimensional reduction to capture the main direction of large variance, data visualization, noise removal, etc, whereas ICA is effective for separating mixed signals and extracting hidden independent components.

The use of ICA for GW data analysis was first reported by De Rosa {\it et al.} \cite{Rosa2012}. 
The application of non-Gaussian noise subtraction was proposed by Morisaki {\it et al.} \cite{UT_ICA2016}. 
De Rosa {\it et al.} demonstrated ICA for injected signals in mimicking two interferometer data and reported that preprocessing ICA before the matched filtering technique lowers the level of noise (increasing the signal-to-noise ratio (SNR)).  The KAGRA collaboration reported an application to their real data (iKAGRA data in 2016) to obtain an  enhanced SNR using physical environmental channels (seismic channels) as known signals\cite{KAGRA_ICA2020}.  

In this article, we report our trials to extract GW signals directly using ICA.  We demonstrate how the injected data can be extracted from Gaussian noise or from real detector output, and also show our trials of actual GW event data.  GW events are fundamentally identified using the coherence of multiple interferometers, taking into account the difference in arrival time (up to 10~ms between Livingston and Hanford and 30~ms between Hanford and Virgo).
We developed a tool for extracting real GW signals that are available by shifting the data-set of multiple-detector data around its arrival time. Consequently,  the arrival time differences between the detectors can be estimated more precisely than previously reported. 

The remain of this report is structured as follows. 
In Section II, we explain the fundamental concept of ICA and our basic procedures. 
In Section III, we present some tests using injected data for Gaussian noise and actual detector noise. 
In Section IV, we present the results of the GW extractions of binary black-hole mergers in O1-O3.  Our conclusions and outlook are presented in Section V.

\section{Independent Component Analysis}

\subsection{The fundamental procedures of ICA}

Suppose we receive the time-series signal ${\bm x}(t)\equiv(x_1(t), \cdots, x_n(t))^T$ with $n$ detectors from $n$ source signals ${\bm s}(t) \equiv(s_1(t), \cdots, s_n(t))^T$, and they are mixed linearly by 
\begin{equation}
{\bm x}(t) = A {\bm s}(t), \label{eqICA1} 
\end{equation}
where $A$ is a time-independent matrix that represents the superposition of source signals. 
ICA is a method for extracting a (candidate) signal $\tilde{\bm s}(t)$ from ${\bm x}(t)$. 

We write the problem as 
\begin{equation}
\tilde{\bm s}(t) = W {\bm x}(t), \label{eqICA2} 
\end{equation}
and set the goal to find a time-independent matrix $W$, so as $\tilde{\bm s}(t)$ contains meaningful signal information. 
We, heareafter, call ${\bm x}(t)$ and $\tilde{\bm s}(t)$ input signals (to ICA) and output signals (from ICA), respectively. 

In the GW case, we expect that a real GW signal is commonly included in all components of ${\bm x}(t)$.  Our goal is to find a GW signal in a single component of the output signal $\tilde{\bm s}(t)$, whereas the other components are noise. 

ICA uses the idea of ``statistical independence" of each source signal component $\tilde{\bm s}$.  
Roughly speaking, ``statistical independence" can be evaluated as how far from Gaussianity. 
(Hence, ICA is not appropriate for extracting Gaussian signals since their superposition is Gaussian.)   

A standard method of ICA starts from whitening the input data, which we write
 \begin{equation}
 {\bm z}(t) = V {\bm x}(t), \label{eq:whiten}
\end{equation}
where $V$ makes ${\bm x}(t)$ no correlation and its variance unity. 
We then write $i$-th component of $\tilde{\bm s}(t)$ as 
\begin{equation}
s_i(t) = {\bm w}_i^T  {\bm z}(t),  
\end{equation}
where $ {\bm w}_i$ is $i$-th row of $W$, and searching $W$ as it gives $s_i(t)$ with statistical independence.

One possible measure of non-Gaussianity is to use the kurtosis of ${\bm w}_i^T {\bm z}$, and a well-known algorithm (FastICA) was first developed using kurtosis to find  $ {\bm w}_i$ using an iterative method  (see Appendix A for further explanation).  However, this method is known to be sensitive to outliers.  We also met this fact, and decided to use alternative FastICA known as the $g$-function method, which uses a mimic function to the kurtosis. The target procedure is to find  $ {\bm w}_i$ that satisfies 
\begin{equation}
{\bm w}_i = E[{\bm z}g({\bm w}_i ^{T} {\bm z})] - E[g^{\prime}({\bm w}_i ^T {\bm z})] {\bm w}_i  \label{eq.gfunc}
\end{equation}
where $g(y) = {\rm tanh}~y$.

Our procedure can be summarized as follows. Let $p$ be the number of iterations.
\begin{enumerate}
\item Whiten the data $x(t)$ using the power spectral density of each detector and apply filtering if necessary. 
\item Normalize $x(t)$ to $z(t)$ (mean zero, variance one). [eq. (\ref{eq:whiten})]
\item 
Set index $i=1$.
\item Randomly choose an initial weight matrix ${\bm w}_{i}^{(p)}$.
\item Obtain ${\bm w}_{i}^{(p+1)}$ from (\ref{eq.gfunc}) as
\begin{equation}
{\bm w}_{i}^{(p+1)} = E[{\bm z}g({\bm w}^{T(p)}_{i} {\bm z})] - E[g^{\prime}({\bm w}^{T(p)}_{i}  {\bm z})] {\bm w}^{(p)}_i. 
\end{equation}
\item Orthogonalize ${\bm w}_{i}^{(p+1)} $ from the other components, 
\begin{equation}
    \tilde{\bm w}_{i}^{(p+1)}  = {\bm w}_{i}^{(p+1)}  - \sum_{j=1}^{i-1}({\bm w}_{i}^{T(p+1)}{\bm w}_j){\bm w}_j.
\end{equation}
\item 
Normalize $ \tilde{\bm w}_{i}^{(p+1)}$ as  ${\bm w}_{i}^{(p+1)} = \tilde{\bm w}_{i}^{(p+1)}/\| \tilde{\bm w}_{i}^{(p+1)} \|$.
\item 
If ${\bm w}_{i}^{(p+1)} $ does not converge to the previous ${\bm w}_{i}^{(p)}$, then return to (v).
\item 
Set $i$ to $ i + 1$. If $i$ is smaller than the number of components of ${\bm x}$, then return to (iv).
\end{enumerate}

To confirm that our converged ${\bm w}_i$ was unique, we repeated this sequence at least five times by changing the initial trial weight matrix (procedure iv).
In most cases, a converged solution is directly obtained.

Note that the output signals from the ICA do not have amplitude information because we normalize the data first. The phase of the output signals can be reversed owning the signature of $\mbox{det}(W)$. 
Further studies are required to identify them. 

The order of $\tilde{s}_1, \tilde{s}_2, \cdots $ represents the order of independent signal that the code found, and
we found that the order sometimes changes due to the initial ${\bm w}_{i}^{(p=0)}$. 
This order does not depend on the amplitude of the input signal. 
We also found that the inspiral GW signal in ICA output has the largest amplitude whatever it is in $\tilde{s}_1, \tilde{s}_2, \cdots $. 
We think we need not care about the order, but in order to avoid confusion, we will show the figures in which GW signals are in $\tilde{s}_1$.

\subsection{Measure for identification of GW}
For the injection tests (Sec. III), we evaluated the output signal by comparing it with the injected signal, using its waveform and spectrum.  When we test with an injection of an inspiral-wave signal, the waveform 
$h_{\rm insp}$ is 
\begin{equation}
h_{\rm insp}(t; t_c, M_c)=A_{\rm insp} \left(\frac{5}{c (t_c - t)}\right)^{1/4}  \sin \left\{-2 \left(\frac{5 G M_c}{c^3}\right)^{-5/8} (t_c - t)^{5/8}  \right\}, 
    \label{eq:inspiral}
\end{equation}
where $c$, $G$, $M_c$, $t_c$ are the speed of light, gravitational constant, chirp mass, and merger time, respectively. The factor of the amplitude $A_{\rm insp}$ can be written as $A_{\rm insp}=({1}/{r}) \left({G M_c}/{c^2}\right)^{5/4}$, where $r$ is the distance to the source, but we just adjust $A_{\rm insp}$ as a parameter. 

For the real GW extraction (Sec. IV), we compared the components of 
the output signals  $\tilde{s}_i(t)$, by defining the ``signal strength" ${\cal A}$ [Eq. (\ref{SES})] using the area of the signal in the graph 
 \begin{equation}
A_i = {\sum_{t=t_s}^{t_c} \tilde{s}_i(t) \cdot \Delta t}, \label{eq:A}
\end{equation}
where $t_c$ is the time of coalescence and $t_s=t_c - 15$~ms for binary black-hole data. 

Let $\tilde{s}_1(t)$ be the candidate of the GW signal and the others ($\tilde{s}_2(t)$ for two detector events; $\tilde{s}_2(t)$ and $\tilde{s}_3(t)$ for three detector events) be detector noises.  We evaluate the ratio 
\begin{eqnarray}
{\cal A}= \left\{ 
\begin{array}{ll}
\displaystyle \frac{A_1}{A_2} &\mbox{~for~2~detector event} \\
\displaystyle \frac{2 A_1}{A_2+A_3} & \mbox{~for~3~detector event, } 
\end{array}
\right.
\label{SES}
\end{eqnarray}
as ``signal strength" to the other remained signals. 
A larger ${\cal A}$ suggests that $\tilde{s}_1(t)$ has a larger amplitude than the others. 

We also compare the candidate signal $\tilde{s}_1(t)$ and the estimated inspiral waveform using the chirp mass $M_c$ reported in GWOSC (Gravitational Wave Open Science Center) website \cite{LVKopendata}, $h_{\rm insp}(t; t_c, M_c)$, by taking the residuals
\begin{equation}
{\cal R} = \sum_{t=t_s}^{t_c} |\tilde{s}_1(t) - h_{\rm insp}(t; t_c, M_c)|  \cdot \label{eq:residual}
\end{equation}
Smaller ${\cal R}$ suggests better extraction than the others. 

\subsection{Finding the arrival time}
One of the key procedures of our proposal is to determine the arrival time of the GW signal to each detector by continually shifting the input data.  For normal applications of the ICA method, we have not observed such an implementation, but it is necessary for GW signals.  

In this article, we only demonstrate the applications for known events up to O3, so the merger time, $t_c$, is taken as a given suggested value.  We then search for the GW signal around $t_c$ by shifting the data from each detector up to $\pm 30$~ms ($\sim$ distance between Hanford and Virgo), and search for the combination that shows the largest ${\cal A}$. 
As we show  in Table \ref{table:modelGWreal}, 
the result of the shifted time, for example, 
$\Delta t_{\rm HL} (=t_{\rm L}-t_{\rm H})$ 
between Hanford and Livingston, can be identified with the difference in the arrival time of GW, and its comparison with those in published articles will also support how our method works. 

\section{Demonstrations with injected GW signals}
In this section, we discuss how ICA extracts the GW with test problems. 
The target discussion is at what level we can apply ICA for signal extraction, that is,  the applicable characteristics of the signals.

\subsection{Injections of inspiral signal to Gaussian Noise}

The first test is the injection of an inspiral-wave signal in Gaussian noise. 

We prepared two different Gaussian-type noises, $n_{1 \rm G}$ and $n_{2 \rm G}$, and injected $h_{\rm insp}$ into them as  
\begin{equation}
\mbox{Model~1~:~} 
\left\{ \begin{array}{l}
x_1(t) =n_{1 \rm G}(t)+ h_{\rm insp}(t; t_c, M_c), \\
x_2(t) =n_{2 \rm G}(t) + h_{\rm insp}(t; t_c, M_c).
\end{array}
\right.
\label{eq:model1}
\end{equation}
If we apply the same Gaussian noise for both $x_1(t)$ and $x_2(t)$, then the test is the same as the well-known blind signal separation problem.  Model 1, however, is  somewhat challenging for weakly injected signals.

\begin{table}
\begin{center}
\caption{ \label{table:model1}
Results of Model 1 [injection of inspiral wave to the Gaussian noise]:  the signal strength ${\cal A}$, the fitting parameter $\alpha$ and $\beta$ of the Fourier spectrum of the extracted signal are listed. Note that $\alpha=15.3$ and  $\beta=0.63$ for the injected signal. }
\footnotesize
\begin{tabular}{cccccc}
\br
$|h_{\mbox{\footnotesize insp}} (t=1)|$ & ${\cal A}$ & $\alpha$& $\beta$&  ref.
\\
\mr
5.0 & {2.21} & 13.7 & 0.665 & Fig.\ref{fig:model1}(a4) \\
2.5& {1.67}& 7.53 & 0.706  & Fig.\ref{fig:model1}(b4)\\
1.0& {1.28} & $-57.5$ & 0.661  & Fig.\ref{fig:model1}(c4)\\
\br
\end{tabular}
\end{center}
\end{table}

\begin{figure}
\begin{tabular}{p{0.5\textwidth}p{0.5\textwidth}}
\includegraphics[width=.40\textwidth]{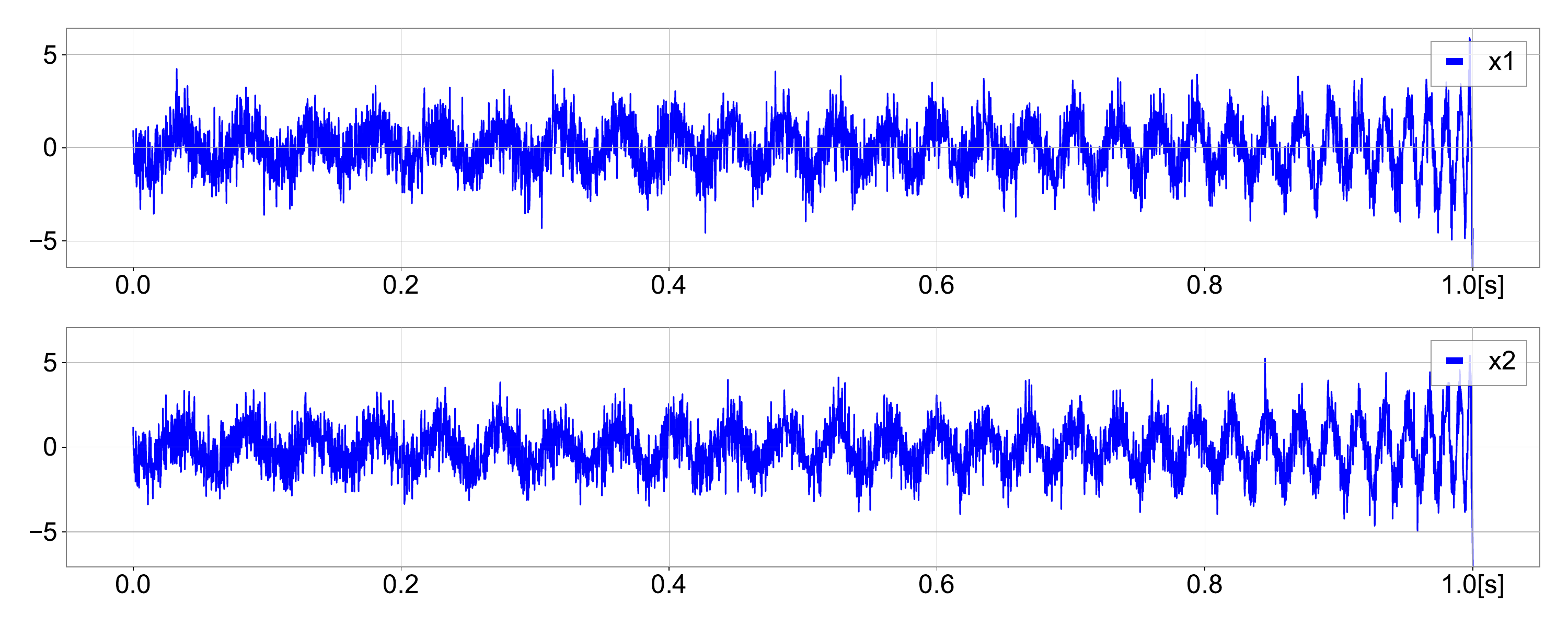}
&
\includegraphics[width=.40\textwidth]
{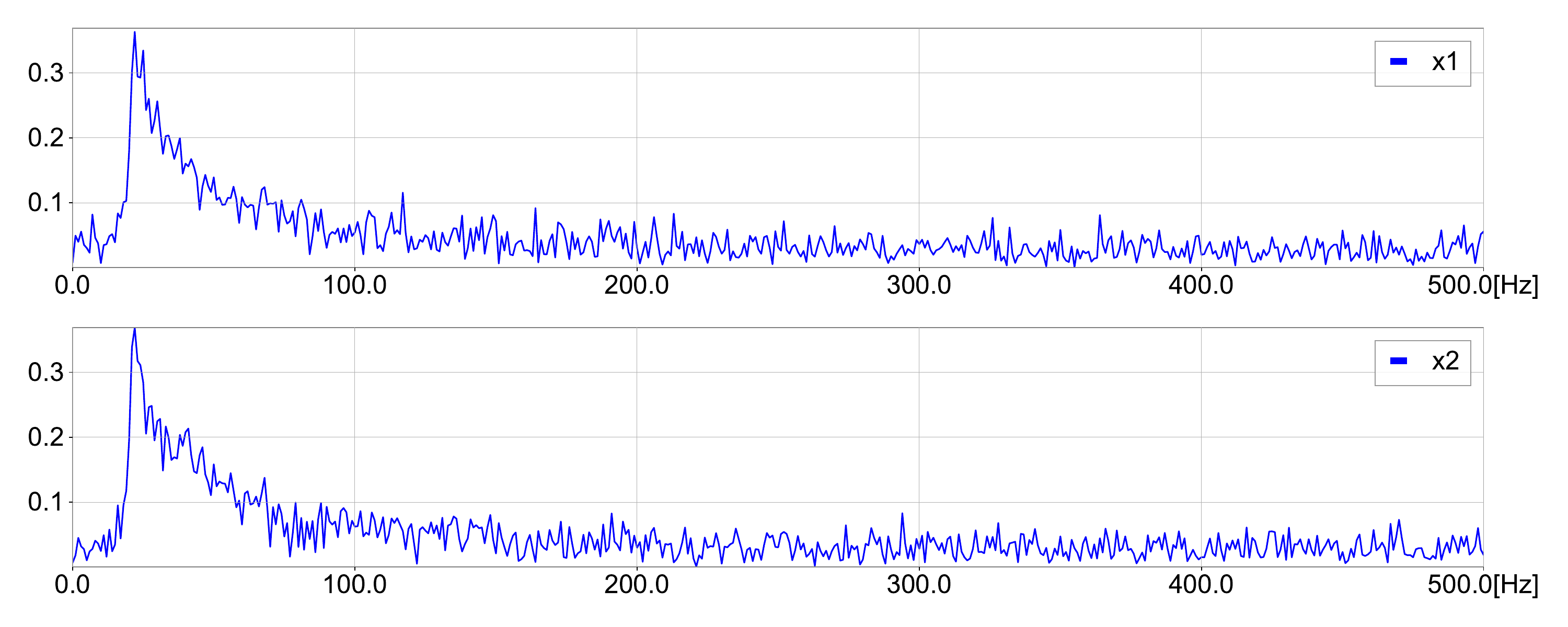}
\\
{\small (a1) Input signals with $|h_{\mbox{\footnotesize insp}}(t=1)|=5.0$.}
&
{\small (a2) Fourier spectrum of (a1).}
\\
\includegraphics[width=.40\textwidth]
{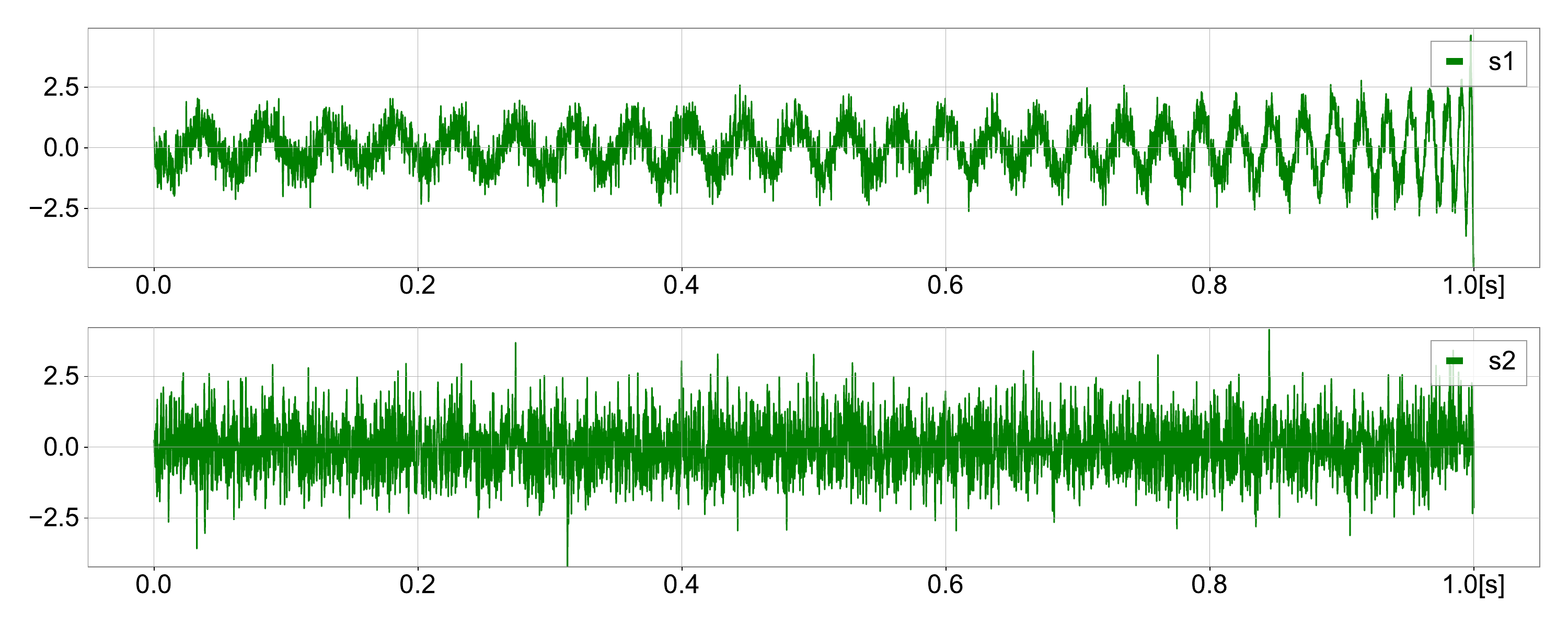}
&
\includegraphics[width=.40\textwidth]
{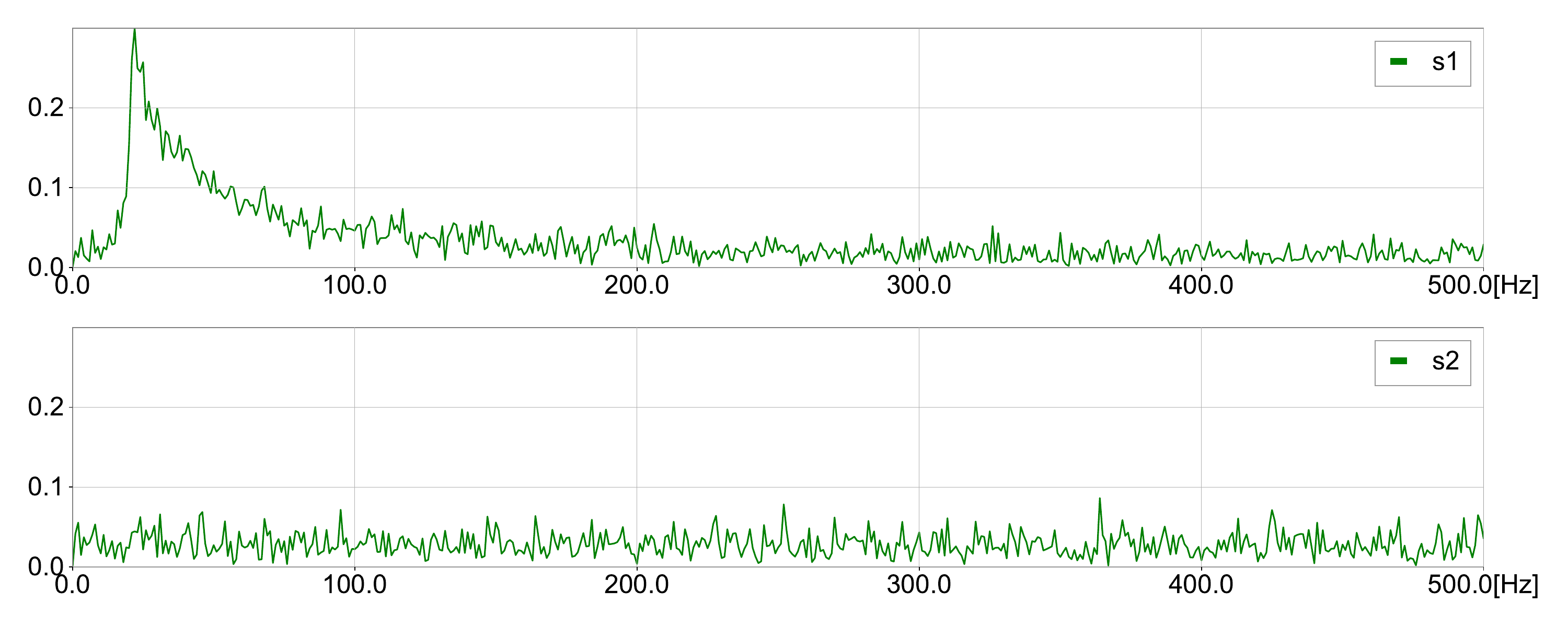}
\\
{\small (a3) Output of ICA for (a1). }&
{\small (a4) Fourier spectrum of (a3).}
\\[1em]
\includegraphics[width=.40\textwidth]
{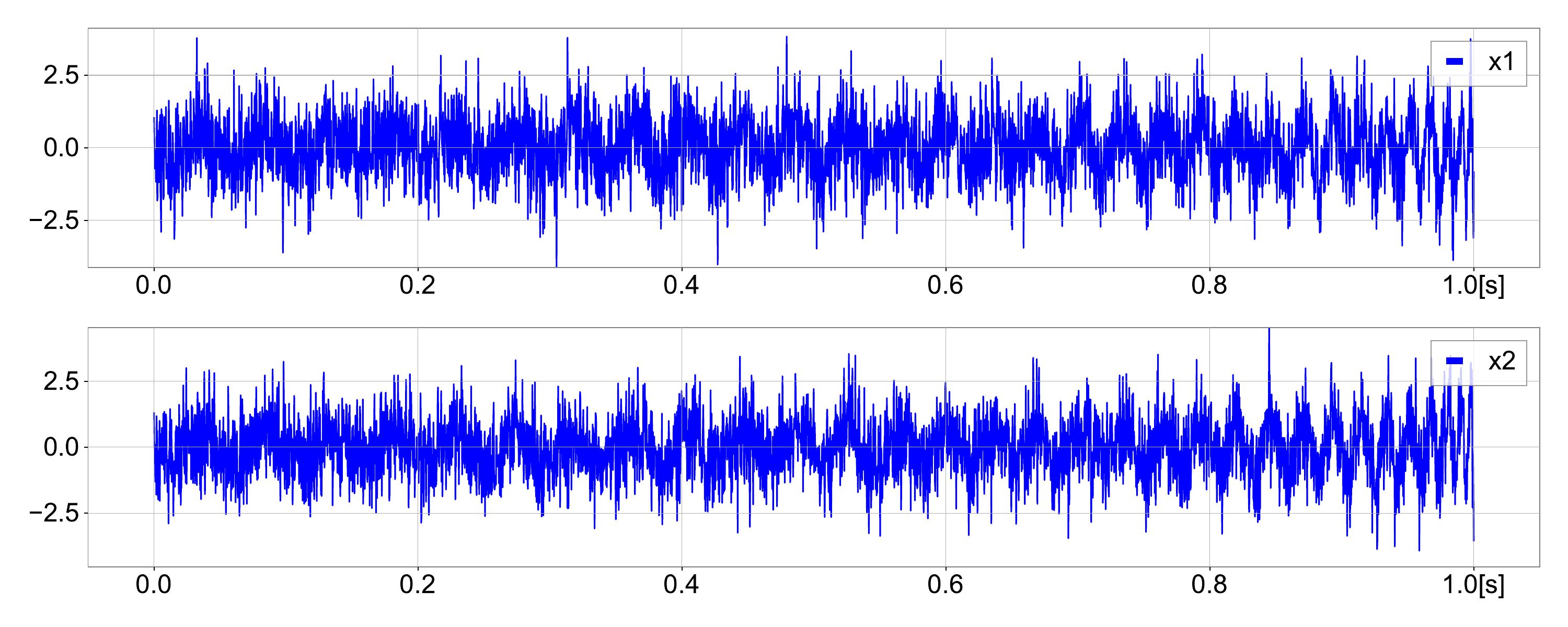}
&
\includegraphics[width=.40\textwidth]
{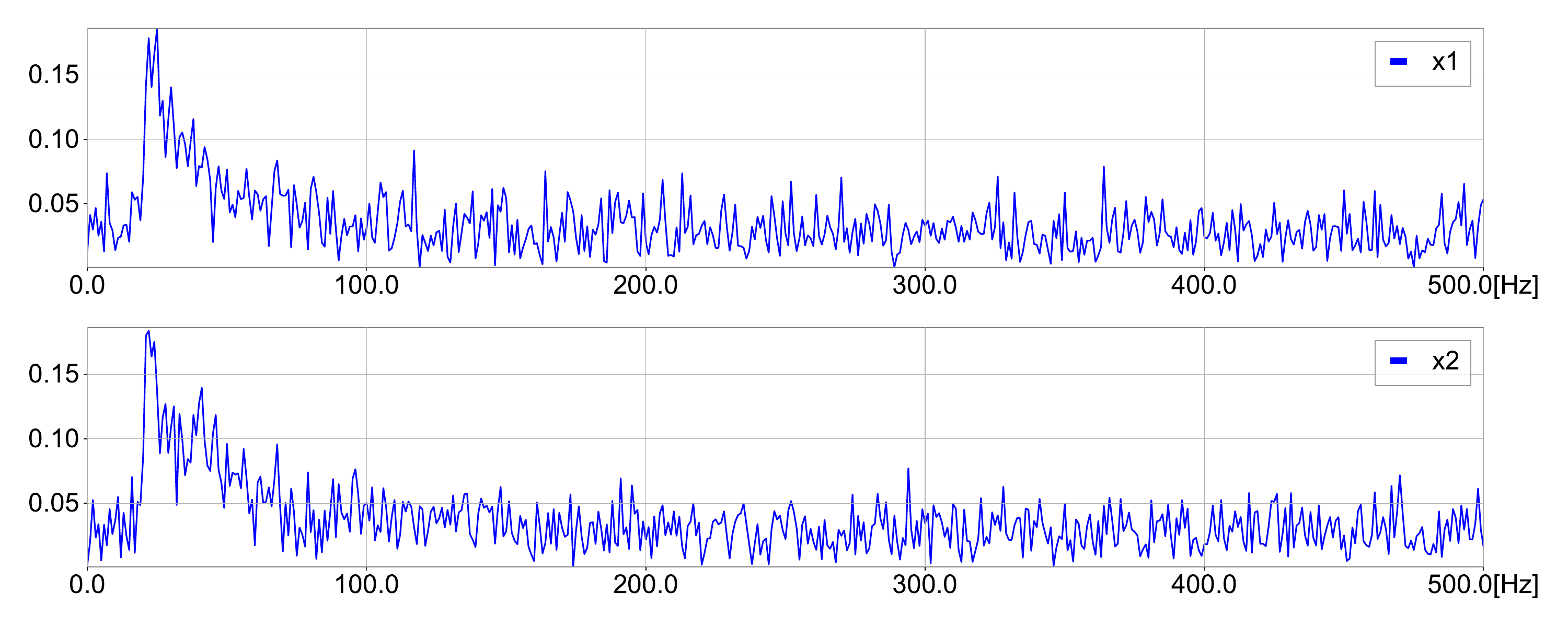}
\\
{\small (b1) Input signals with $|h_{\mbox{\footnotesize insp}}(t=1)|=2.5$.}
&
{\small (b2) Fourier spectrum of (b1).}
\\
\includegraphics[width=.40\textwidth]
{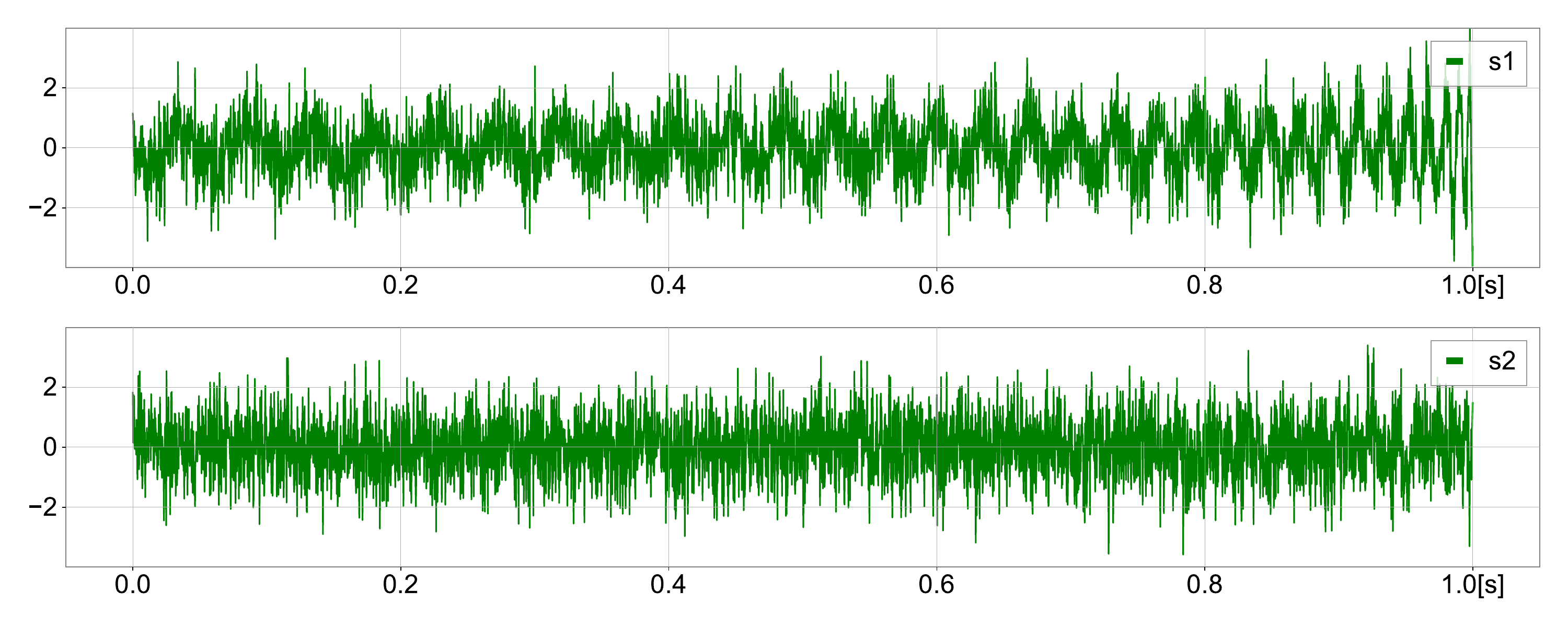}
&
\includegraphics[width=.40\textwidth]
{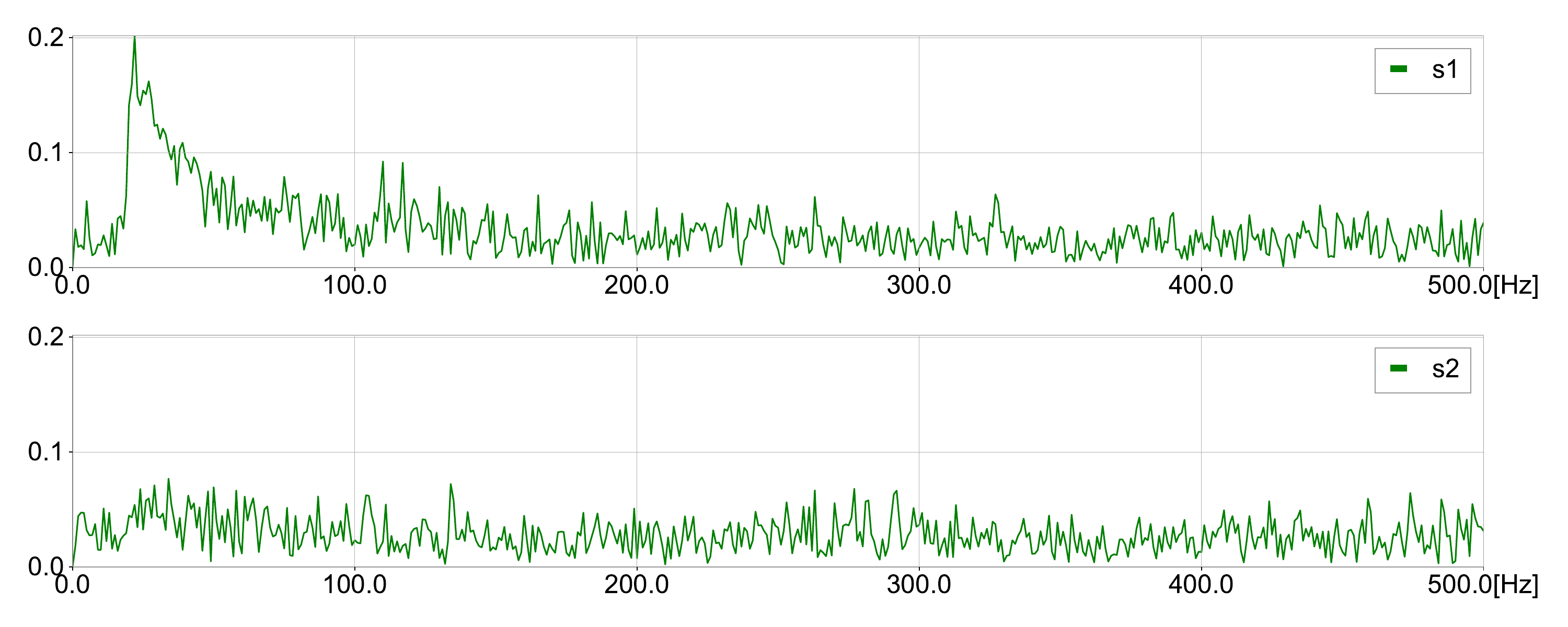}
\\
{\small (b3) Output of ICA for (b1). }&
{\small (b4) Fourier spectrum of (b3).}
\\[1em]
\includegraphics[width=.40\textwidth]
{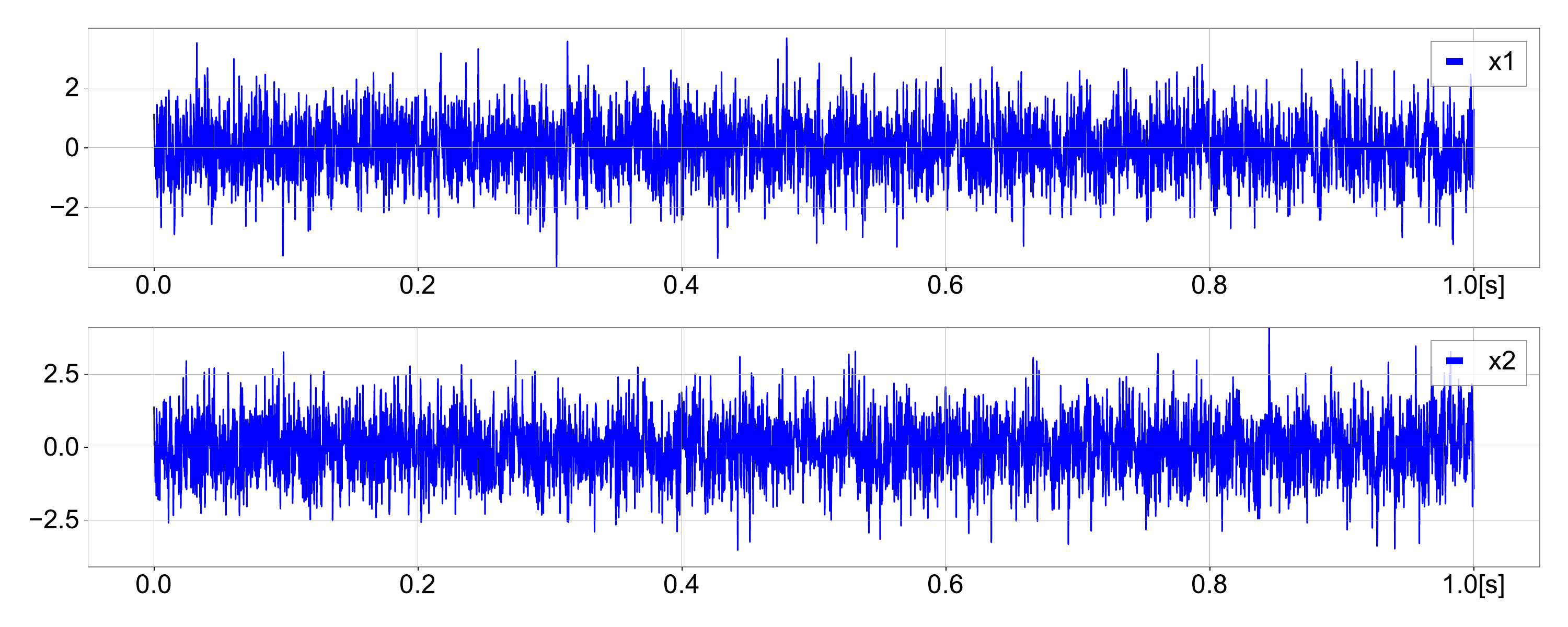}
&
\includegraphics[width=.40\textwidth]
{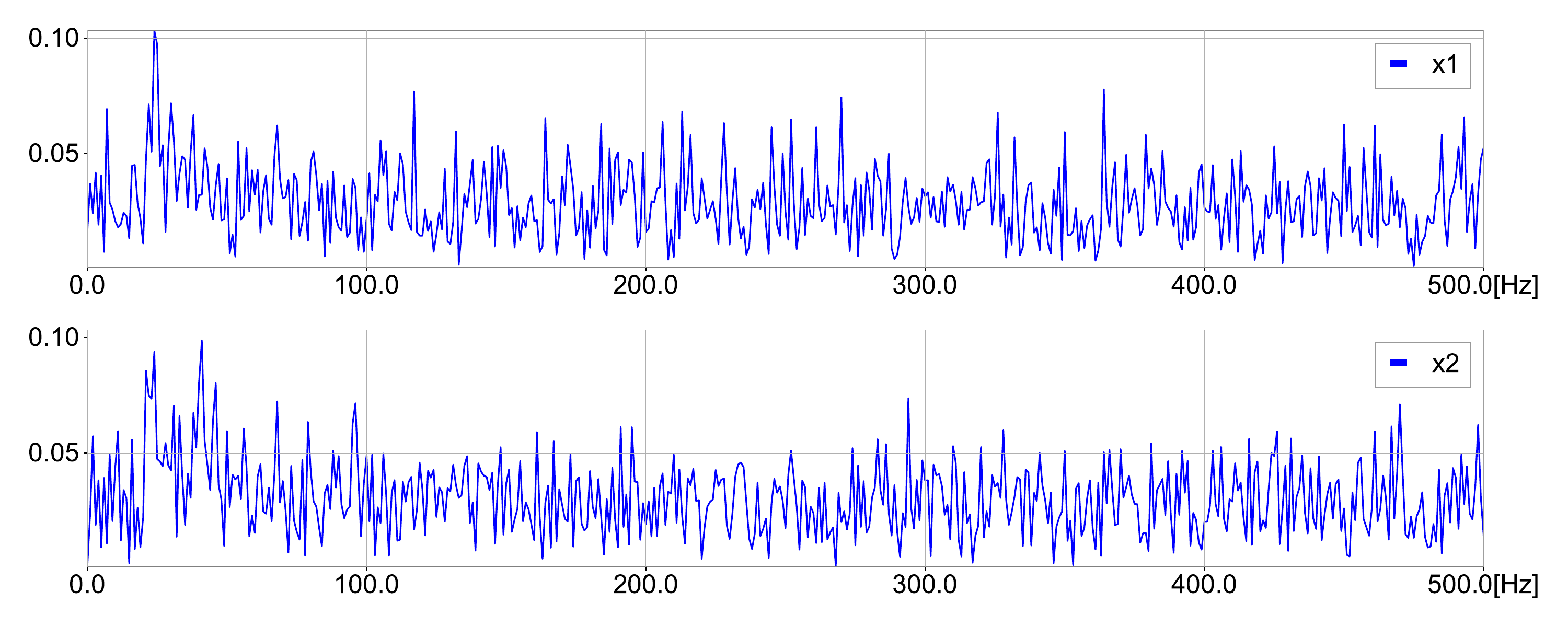}
\\
{\small (c1) Input signals with $|h_{\mbox{\footnotesize insp}}(t=1)|=1.0$.}
&
{\small (c2) Fourier spectrum of (c1).}
\\
\includegraphics[width=.40\textwidth]
{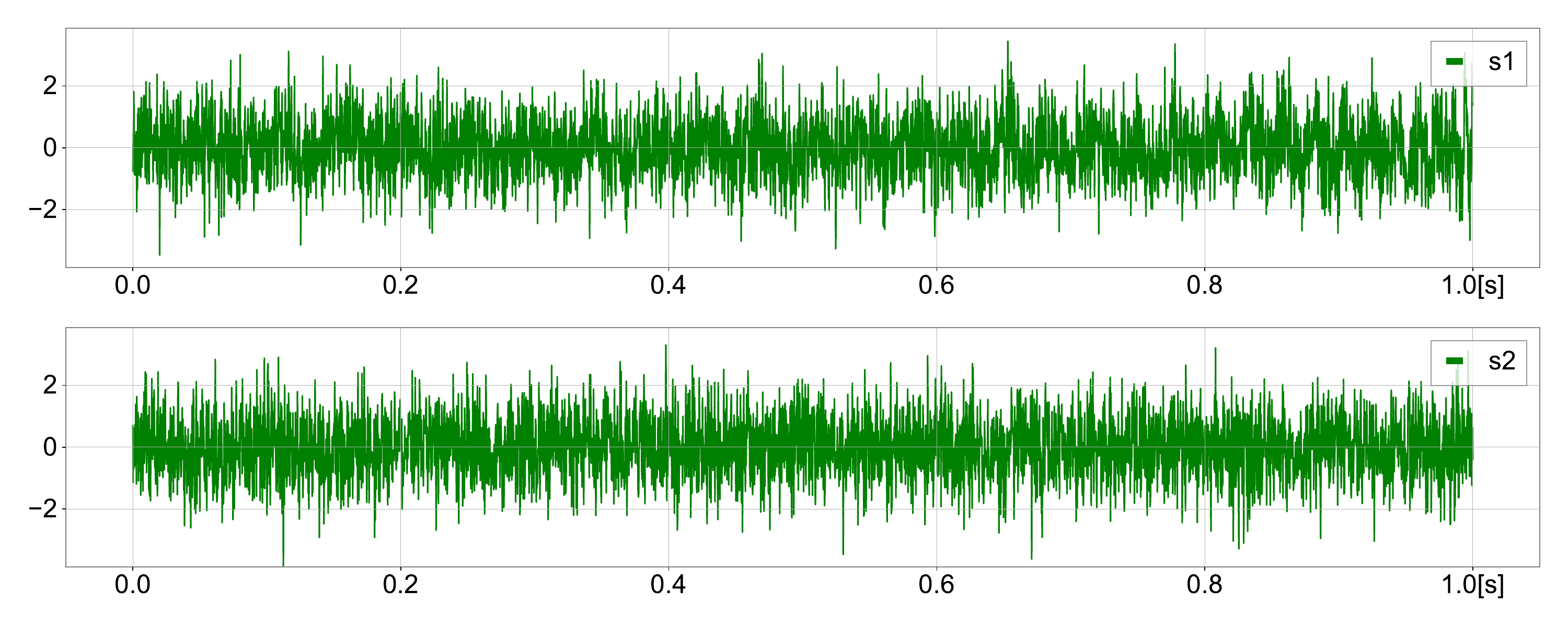}
&
\includegraphics[width=.40\textwidth]
{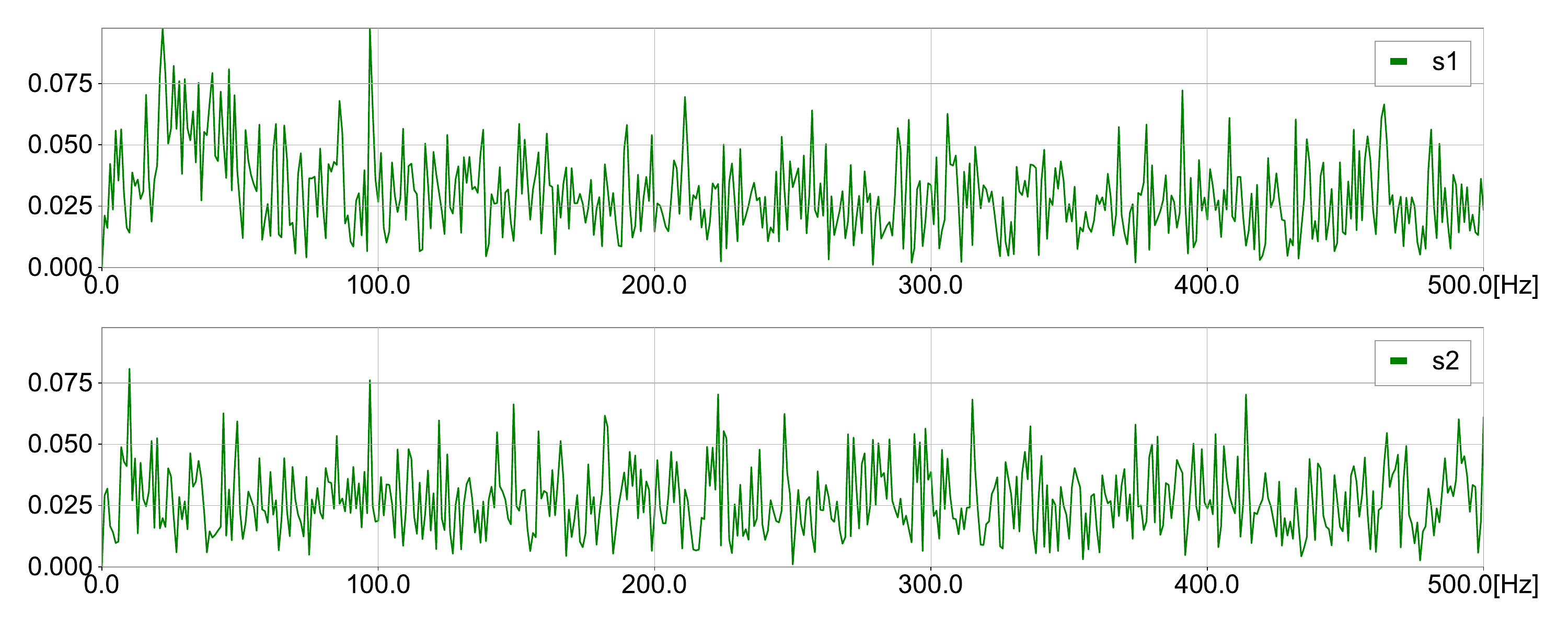}
\\
{\small (c3) Output of ICA for (c1). }&
{\small (c4) Fourier spectrum of (c3).}
\\
\end{tabular}
\caption{An example of GW extraction using ICA, the case of Model 1 [eq. (\ref{eq:model1})], injection of an inspiral wave to Gaussian background. Figures are of $|h_{\mbox{\footnotesize insp}}(t=1)|/ \overline{|n_{\mbox{\footnotesize G}}(t)|}=5.0$ (a1-a4), $2.5$ (b1-b4), and $1.0$ (c1-c4). 
We use one-second length data $t=[0,1]$, with $t_c=1+1/4096$. 
\label{fig:model1}}
\end{figure}

We prepare one-second data with a sampling rate of 4096, and two Gaussian noises with average amplitudes of  
$ \overline{|n_{1 \rm G}|^2} = \overline{|n_{2 \rm G}|^2}=1$. For inspiral signals, 
we set $M_c=26.12M_\odot$, which is of $30 M_\odot $-$30 M_\odot $ binary at the observed frame, 
$t_c=1.0+1/4096$ ({\it i.e.} the moment of the merger is out of the range), and compare three cases of its amplitude $|h_{\rm insp} (t=1)|=5.0,  2.5$, and 1.0.(Note that the signals are amplified five times during one second.)

The results for the largest amplitude case are shown in Figure  \ref{fig:model1}. 
Figure \ref{fig:model1} (a1) shows the input signals, which show that the inspiral behavior is observable by eye over 30~Hz. 
We also show the Fourier spectrum in (a2) and the output results of ICA in (a3) together with the Fourier spectrum in (a4).  Note that the ICA results do not include  information on the strength of each mode, that is, the amplitude of the output data does not indicate its strength in the input data.  

To see how the extracted signal matches  the injected signal
we approximate the power spectrum of the output signal [Figure \ref{fig:model1}(a4)] 
with a function $(f-\alpha)^{-\beta}$ for $f=[20, 300]$~Hz where $\alpha, \beta$ are constants, and compare them with those of the injected signal, $(f-15.3)^{-0.63}$. 
A comparison of the three cases is shown in Table \ref{table:model1}.  We performed the same extractions 10 times by changing the initial guess of ${\bm w}_p$ randomly, and show the average of the fitting parameters $\alpha$ and $\beta$ in the table. 
As expected, if the amplitude of the injected signals is large, the ICA clearly identifies the signal.  We can see that  identification is difficult when the amplitude of the injected signal is at the same level as the noise [Figure \ref{fig:model1}(b)(c)]. 

\subsection{Injections of GW signal to real detector data}
Next, we demonstrate the signal extraction of the injected waves from the real detector data. We used the detector data around GW150914 (GPS time of $t_c=1126259462.4$)
of LIGO-Hanford, $n_{\rm H}(t)$, and LIGO-Livingston, $n_{\rm L}(t)$, which we downloaded from GWOSC. 

We made two tests. One is the injection of a sinusoidal wave, 
\begin{equation}
\mbox{Model~2~:~} 
\left\{ \begin{array}{l}
x_1(t) = n_{\rm H}(t) + \sin(2\pi ft), \\
x_2(t) = n_{\rm L}(t) + \sin(2\pi ft)
\end{array}
\right.
\label{eq:model2}
\end{equation}
and the other is an inspiral wave [eq.(\ref{eq:inspiral})], 
\begin{equation}
\mbox{Model~3~:~} 
\left\{ \begin{array}{l}
s_1(t) = n_{\rm H}(t) + h_{\rm insp}(t; t_0, M_c),\\
s_2(t) = n_{\rm L}(t) + h_{\rm insp}(t; t_0, M_c). 
\end{array}
\right.
\label{eq:model3}
\end{equation}

\begin{figure}
\begin{tabular}{p{0.50\textwidth}p{0.50\textwidth}}
\includegraphics[width=.40\textwidth]
{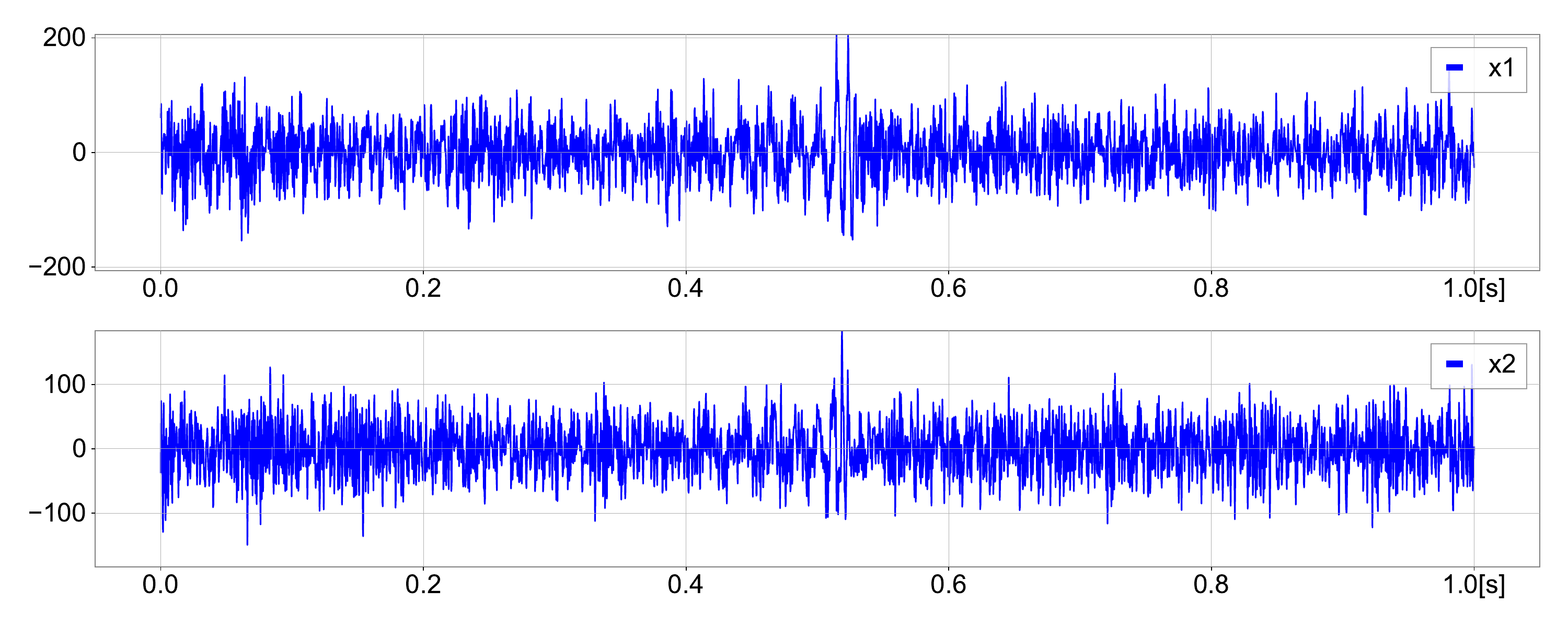}
&
\includegraphics[width=.40\textwidth]
{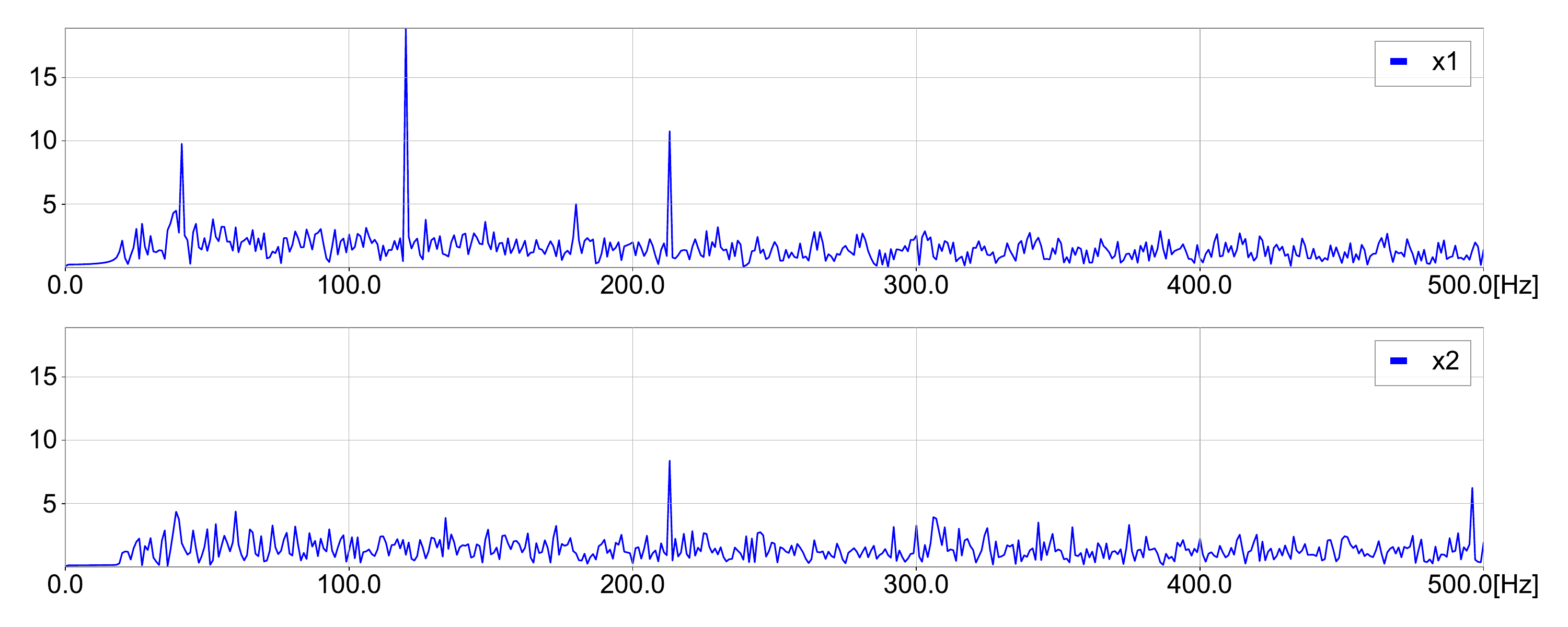}
\\
{\small (a1) Input signals of SNR 20. }&
{\small (a2) Fourier spectrum of (a1). }\\
\includegraphics[width=.40\textwidth]
{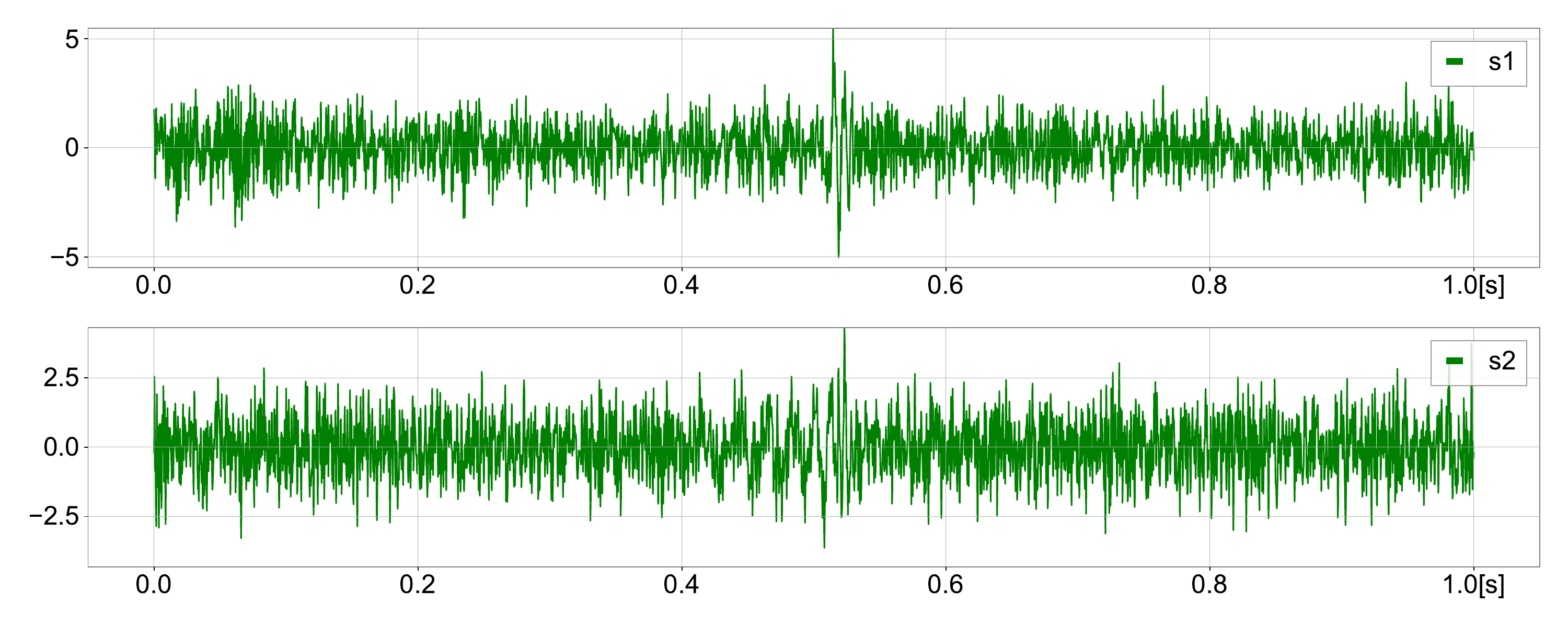}
&
\includegraphics[width=.40\textwidth]
{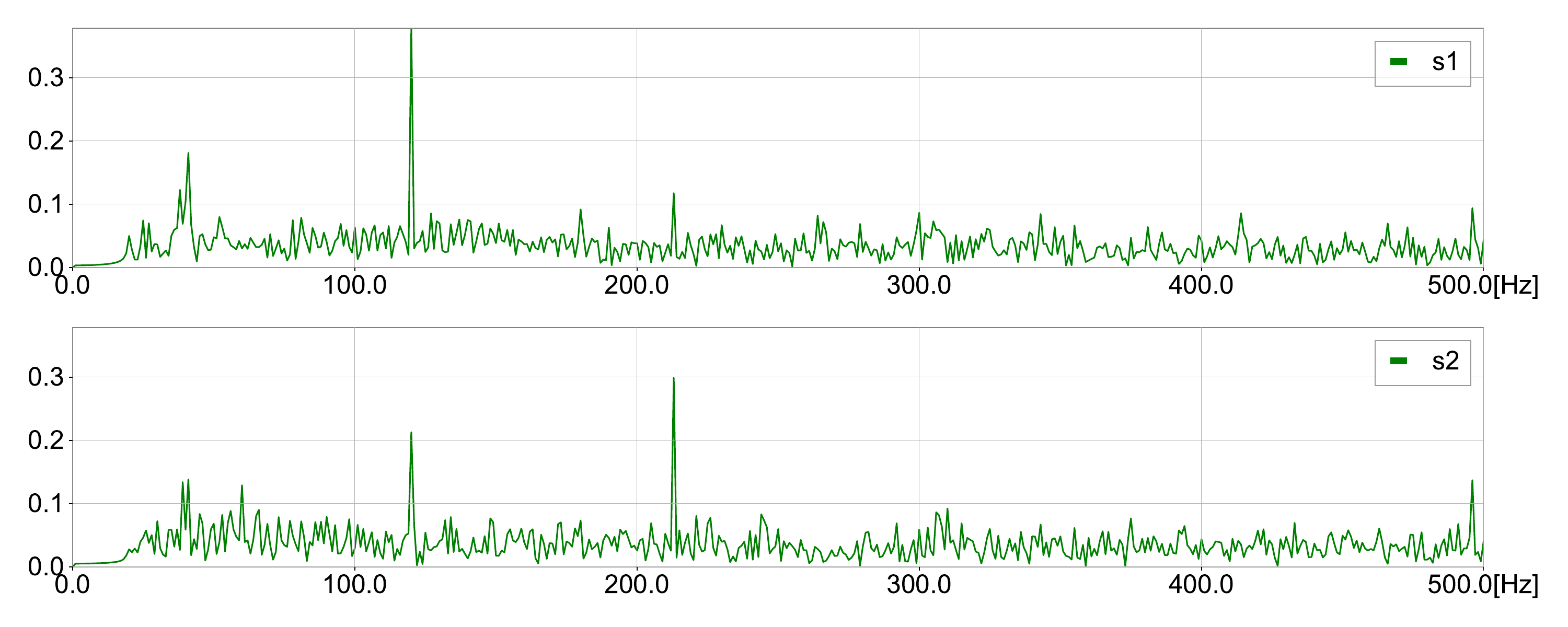}
\\
{\small (a3) Output of ICA for (a1).} &
{\small (a4) Fourier spectrum of (a3).}
\\[1em]
\includegraphics[width=.40\textwidth]
{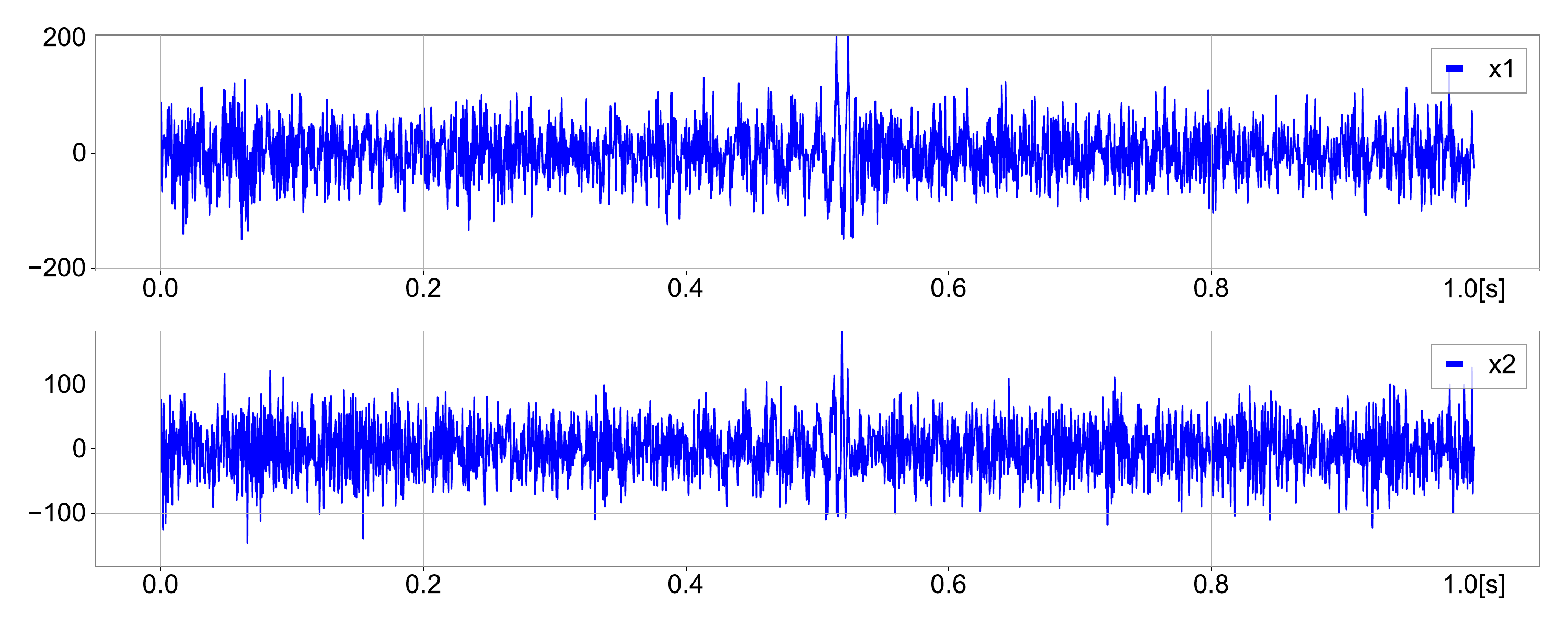}
&
\includegraphics[width=.40\textwidth]
{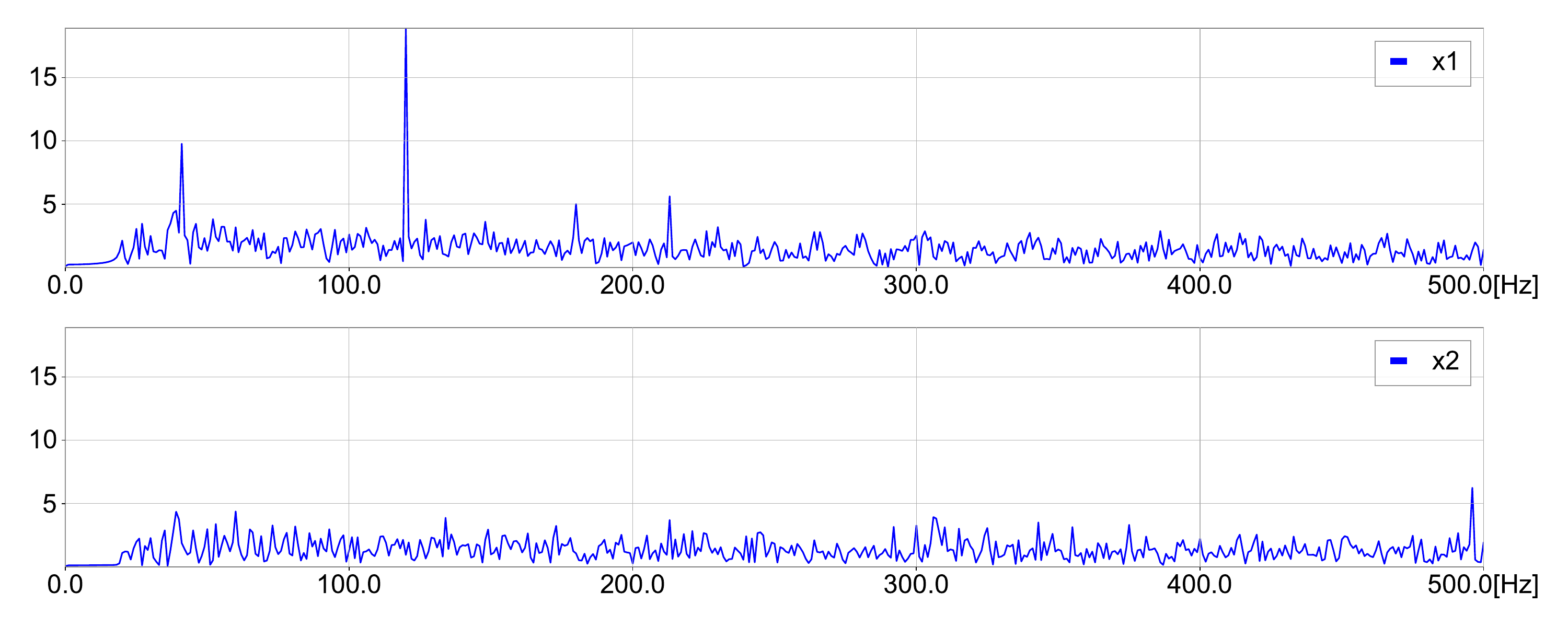}
\\
{\small (b1) Input signals of SNR 10. }&
{\small (b2) Fourier spectrum of (b1). }\\
\includegraphics[width=.40\textwidth]
{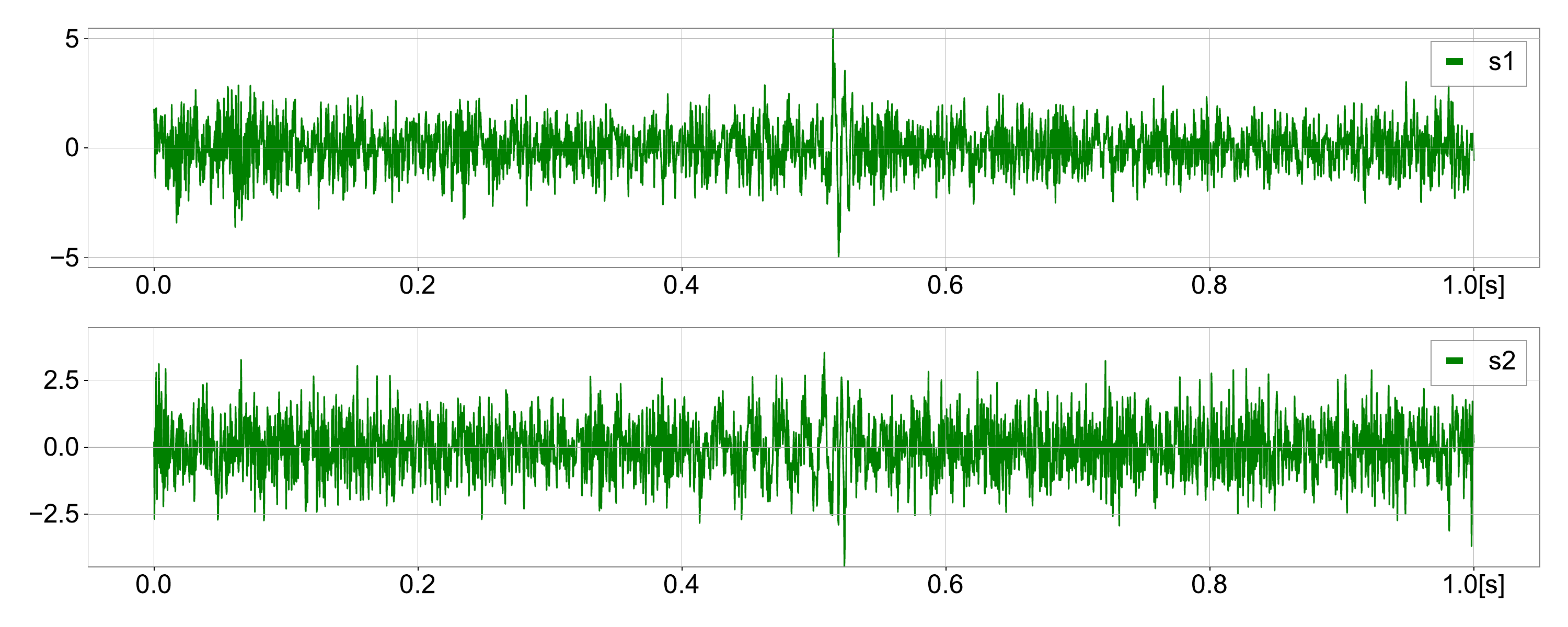}
&
\includegraphics[width=.40\textwidth]
{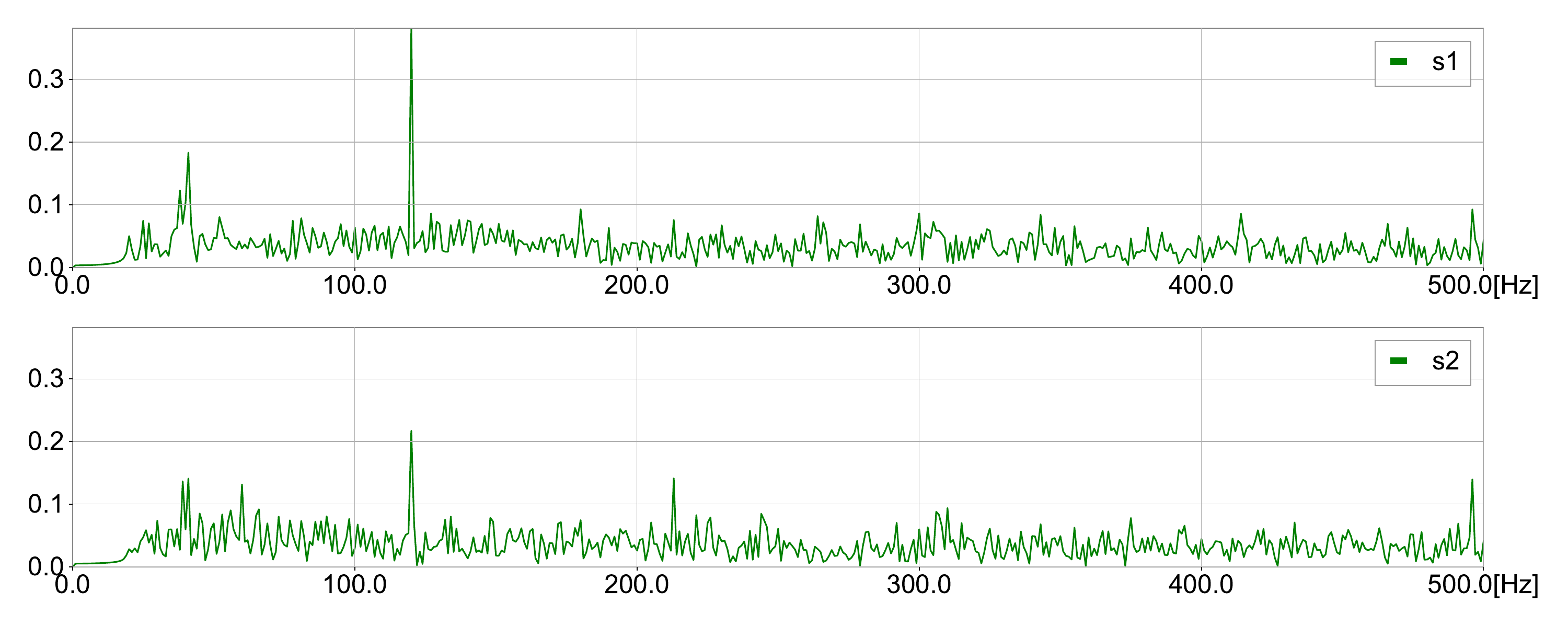}
\\
{\small (b3) Output of ICA for (b1). }&
{\small (b4) Fourier spectrum of (b3).}
\\[1em]
\includegraphics[width=.40\textwidth]
{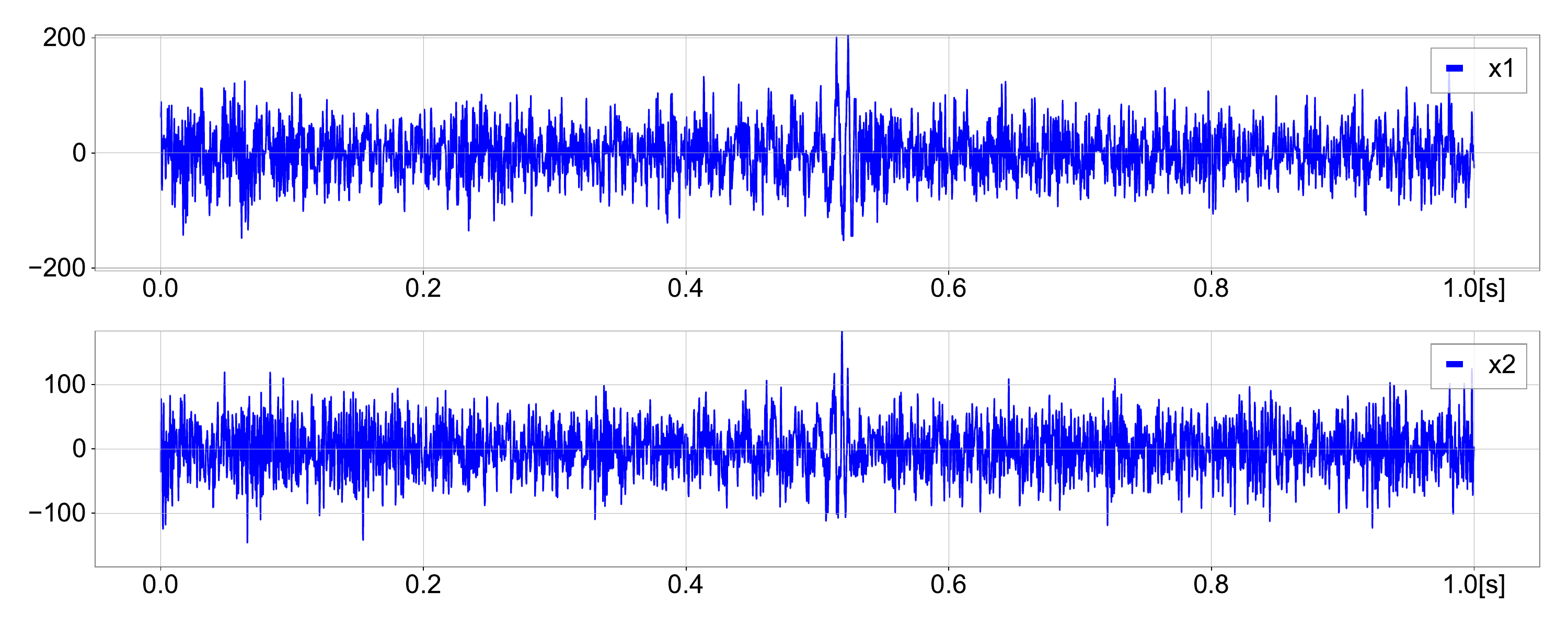}
&
\includegraphics[width=.40\textwidth]
{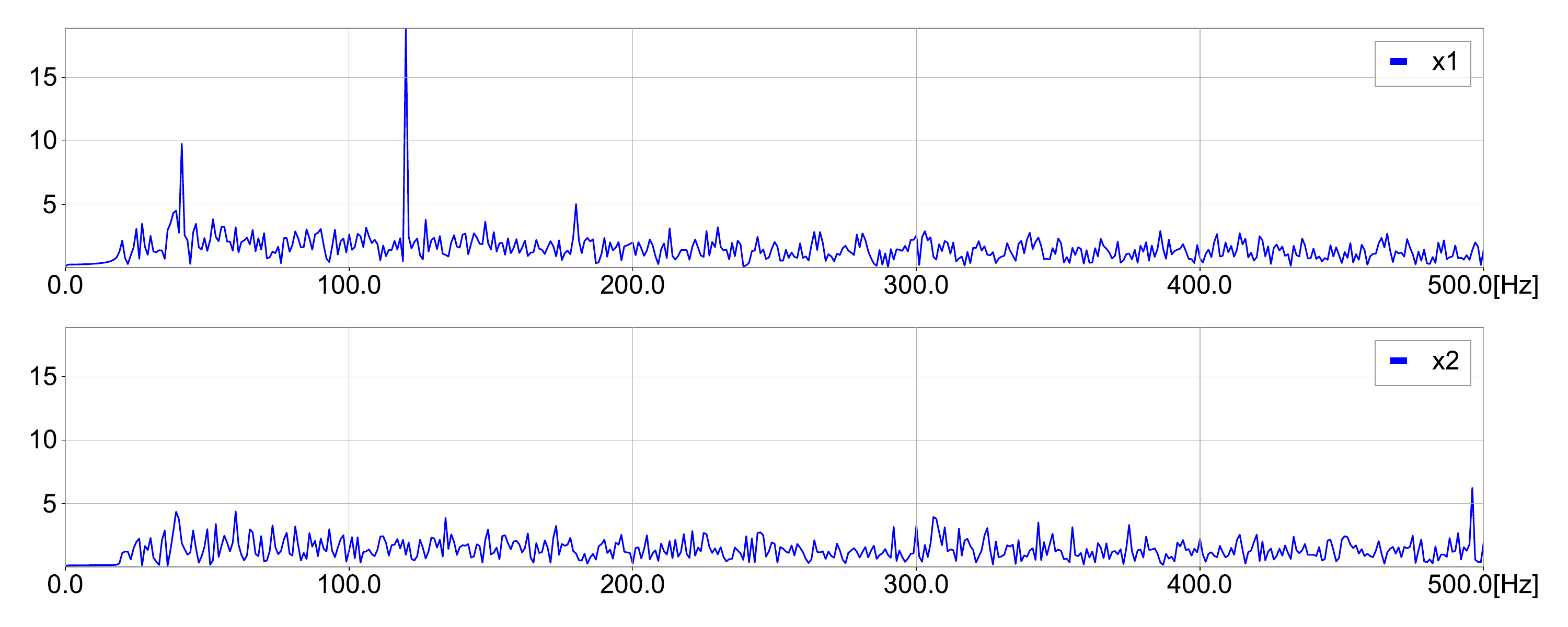}
\\
{\small (c1) Input signals of SNR 5.} &
{\small (c2) Fourier spectrum of (c1). }\\
\includegraphics[width=.40\textwidth]
{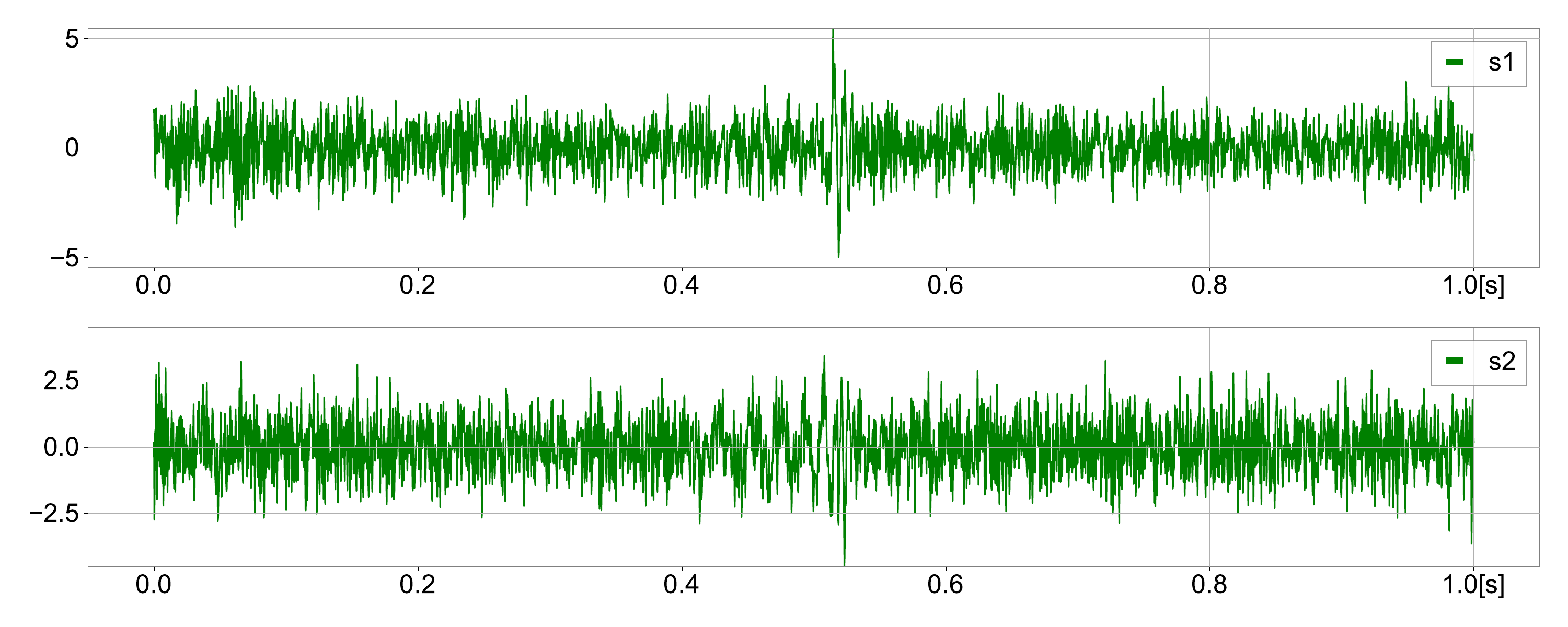}
&
\includegraphics[width=.40\textwidth]
{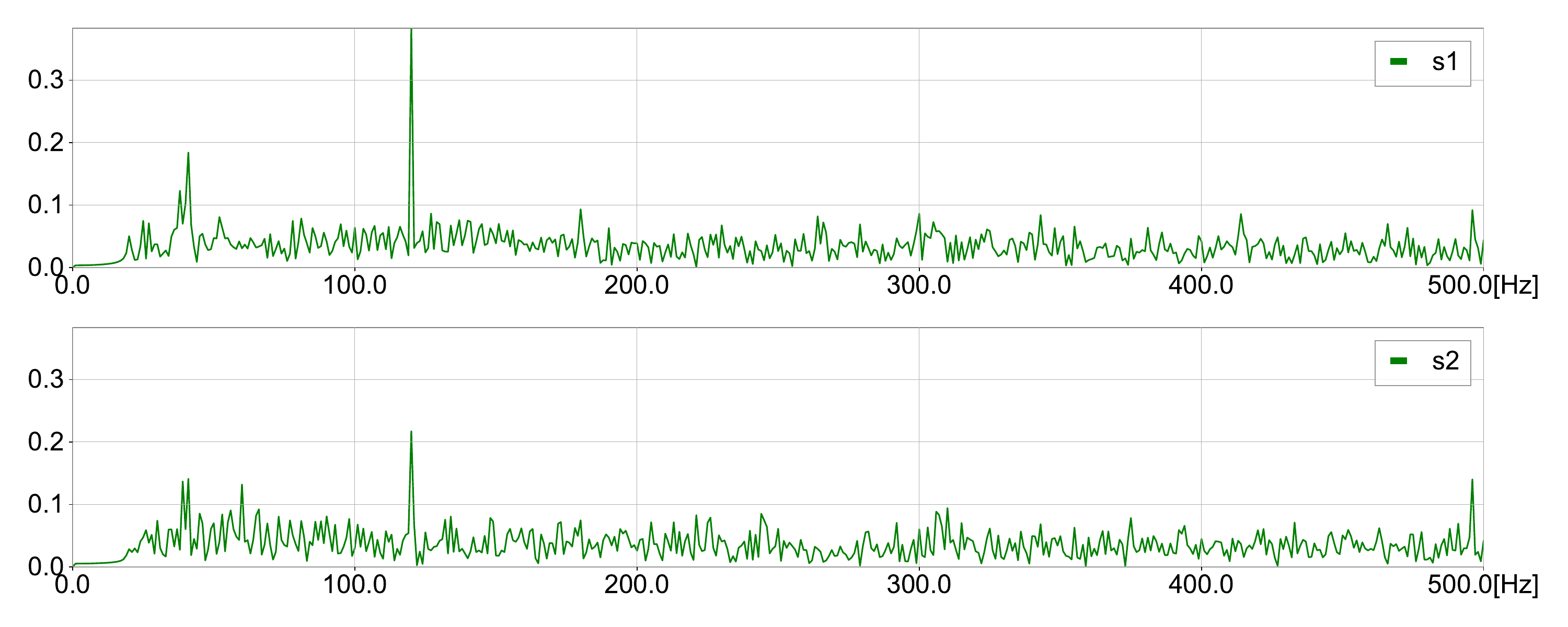}
\\
{\small (c3) Output of ICA for (c1). }&
{\small (c4) Fourier spectrum of (c3).}
\end{tabular}
\caption{
Test of GW extraction using ICA: the case of Model 2 [eq. (\ref{eq:model2})], injections of a sinusoidal wave of 213~Hz to the Hanford and Livingston data around the event GW190914. 
(Note that the original data include large noises at 60~Hz and 120~Hz.)
The signal-to-noise ratio (SNR) is 20 (a1-a4), 10 (b1-b4), and 5 (c1-c4).  
We see the injected wave is clearly extracted for (a) and (b), but not for (c).
\label{fig:model2}}
\end{figure}

\begin{figure}
\begin{tabular}{p{0.50\textwidth}p{0.50\textwidth}}
\includegraphics[width=.40\textwidth]
{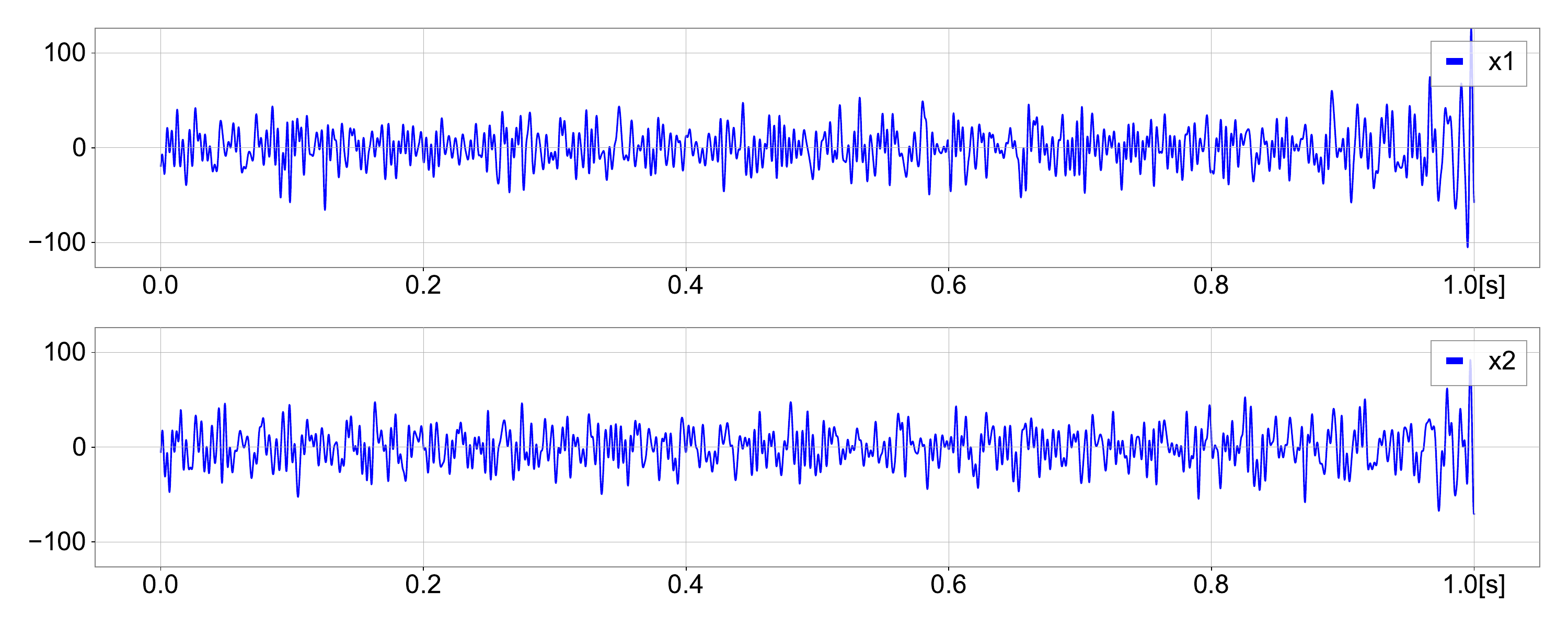}
&
\includegraphics[width=.40\textwidth]
{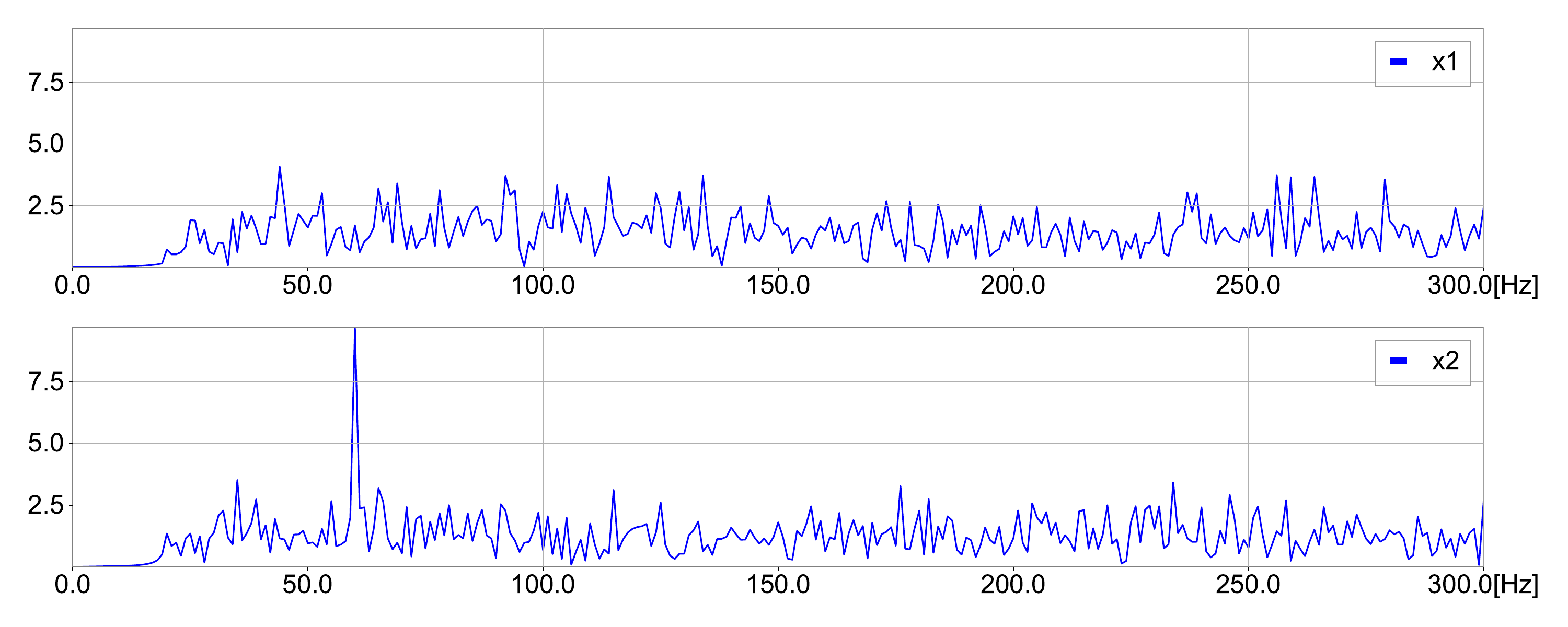}
\\
{\small (a1) Input signals of SNR 20.9. }&
{\small (a2) Fourier spectrum of (a1).} \\
\includegraphics[width=.40\textwidth]
{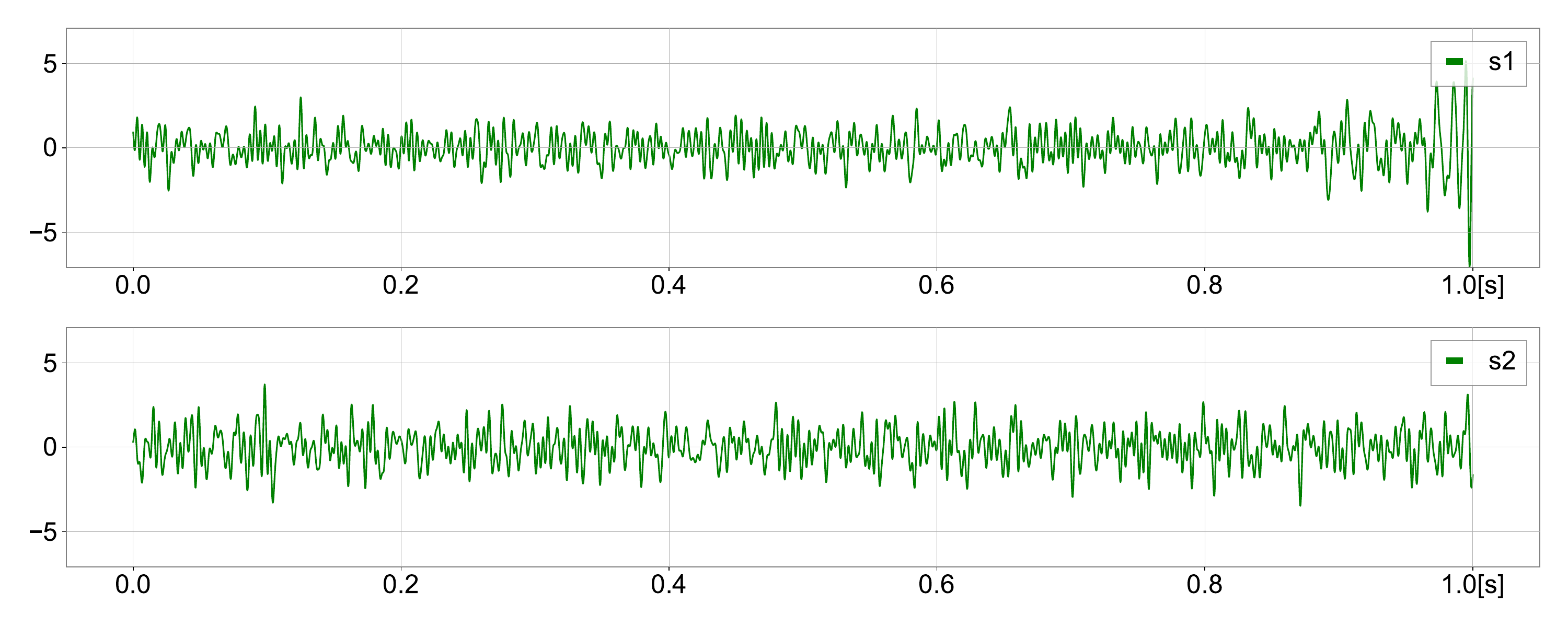}
&
\includegraphics[width=.40\textwidth]
{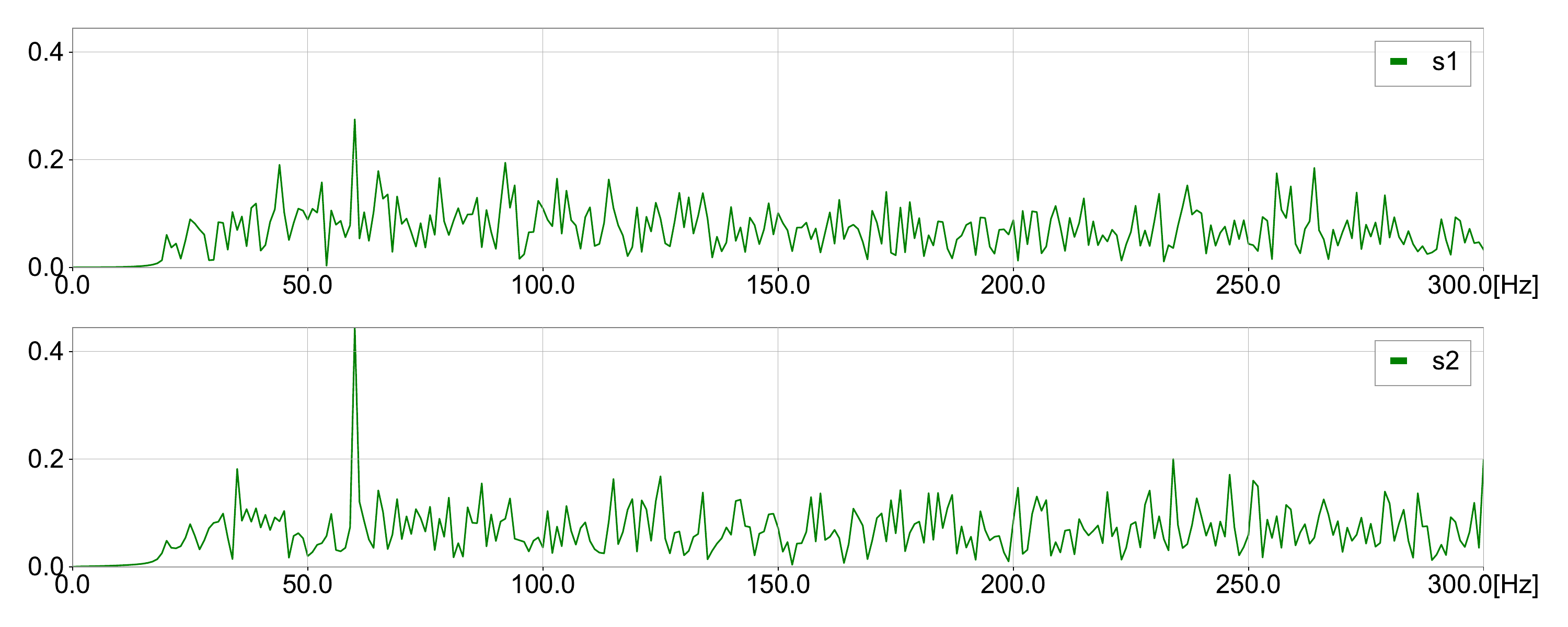}
\\
{\small (a3) Output of ICA for (a1).} &
{\small (a4) Fourier spectrum of (a3).}
\\[1em]
\includegraphics[width=.40\textwidth]
{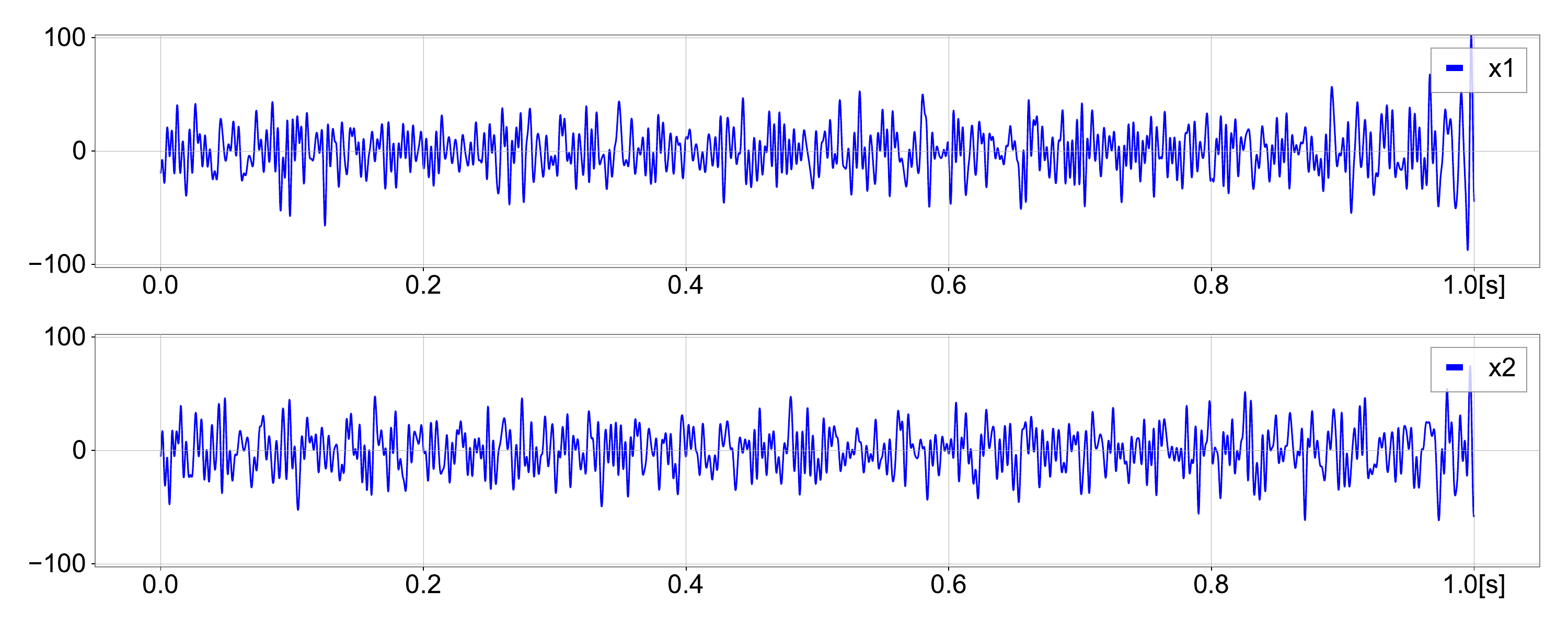}
&
\includegraphics[width=.40\textwidth]
{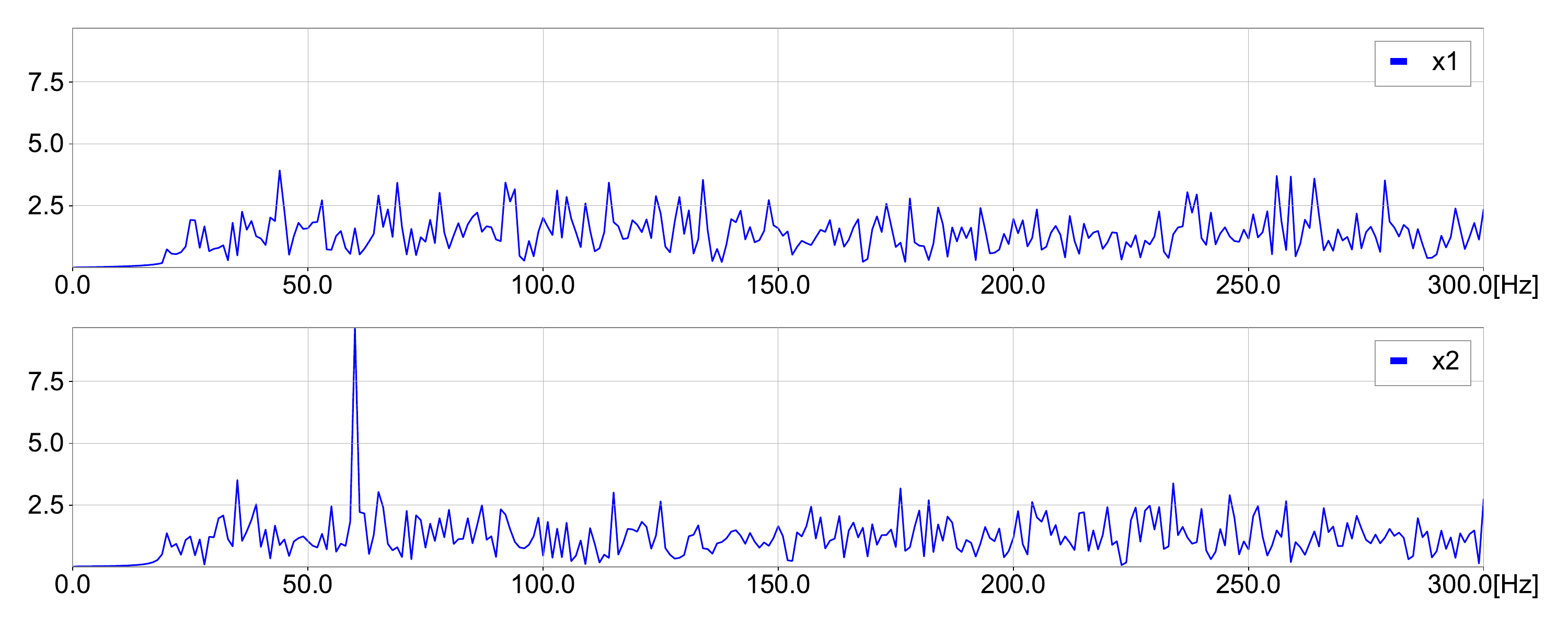}
\\
{\small (b1) Input signals of SNR 16.8. }&
{\small (b2) Fourier spectrum of (b1). }\\
\includegraphics[width=.40\textwidth]
{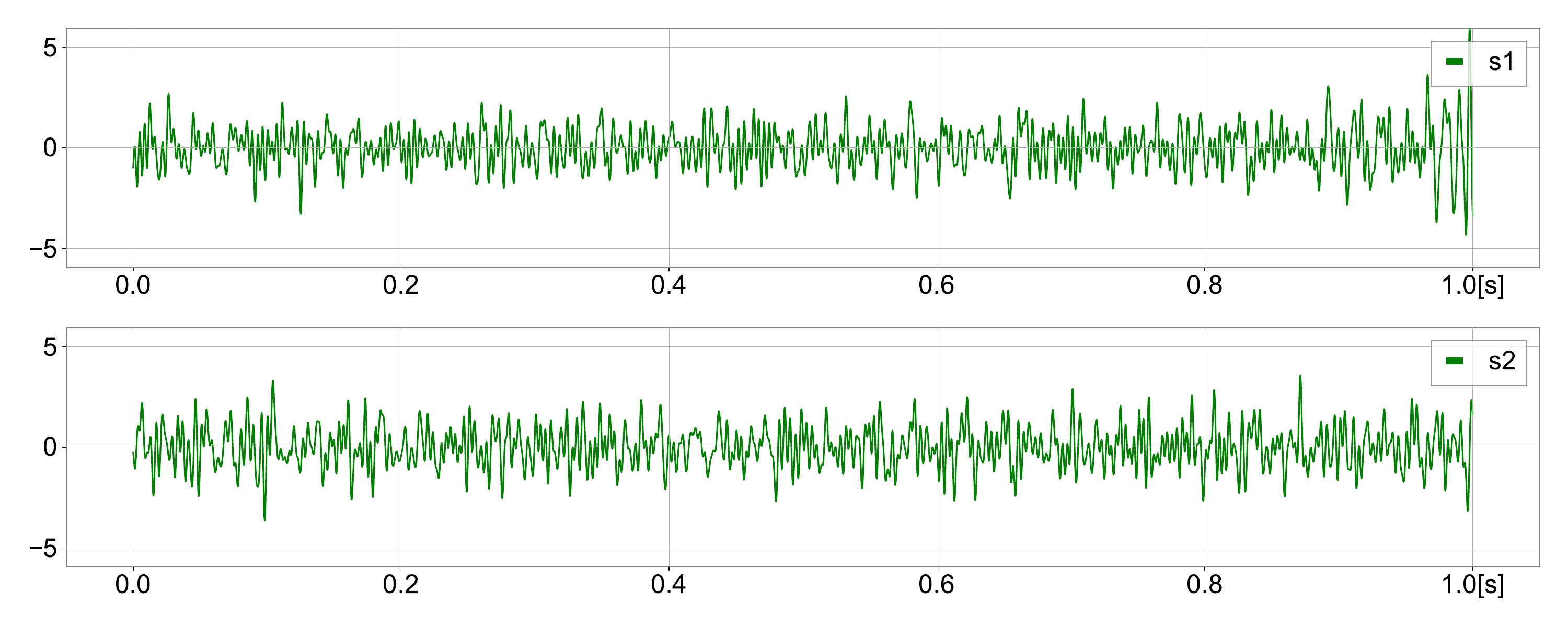}
&
\includegraphics[width=.40\textwidth]
{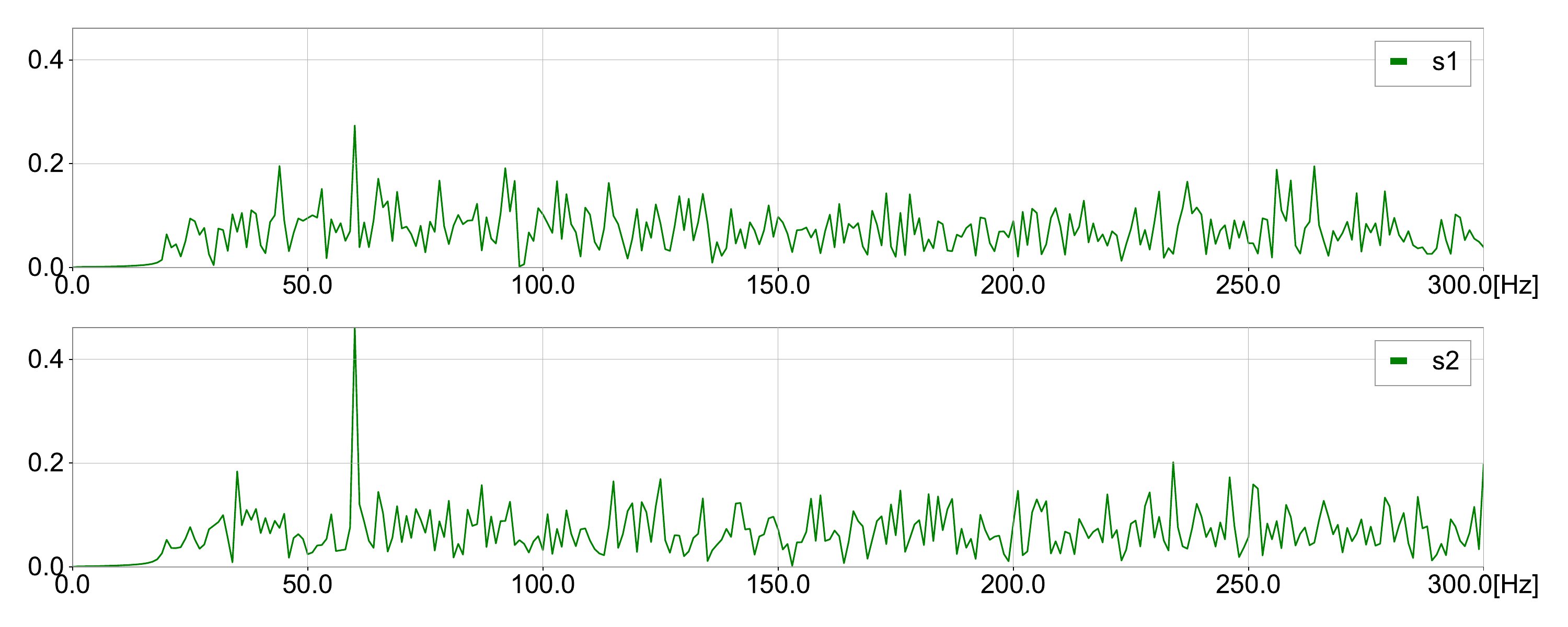}
\\
{\small (b3) Output of ICA for (b1). }&
{\small (b4) Fourier spectrum of (b3).}
\\[1em]
\includegraphics[width=.40\textwidth]
{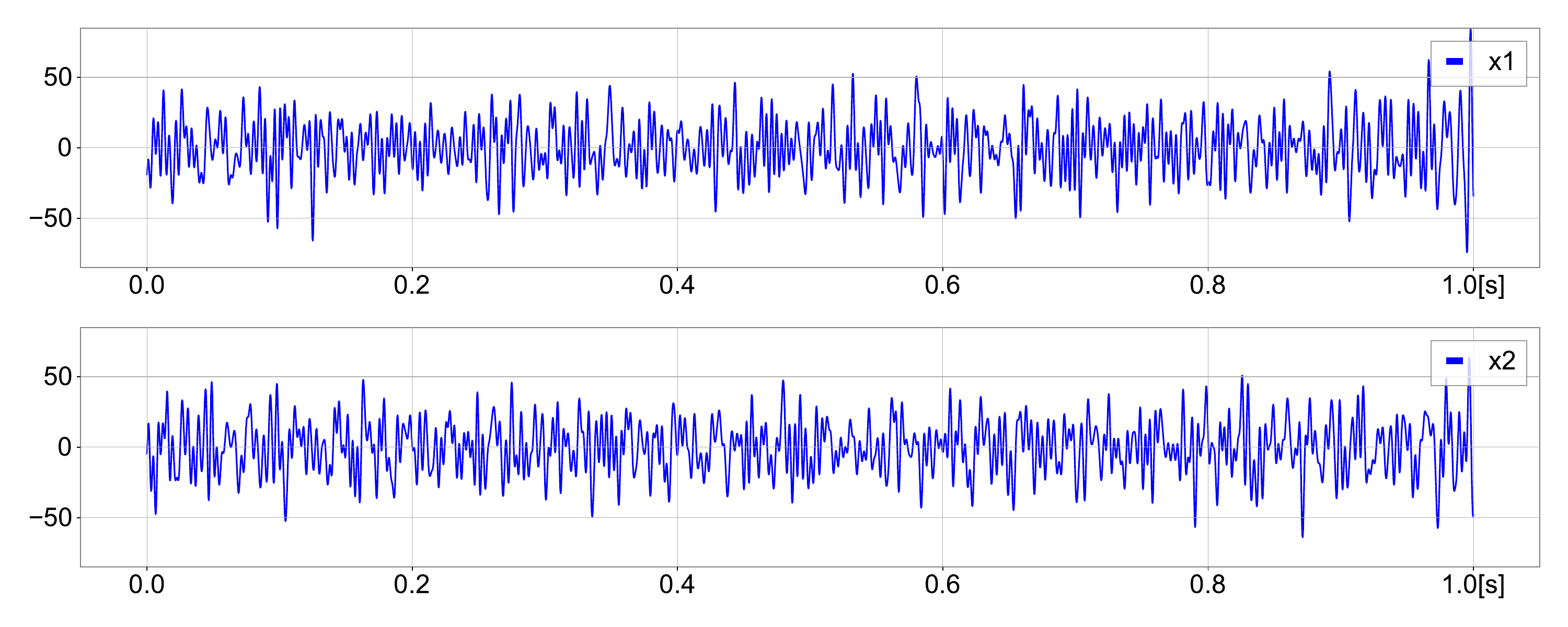}
&
\includegraphics[width=.40\textwidth]
{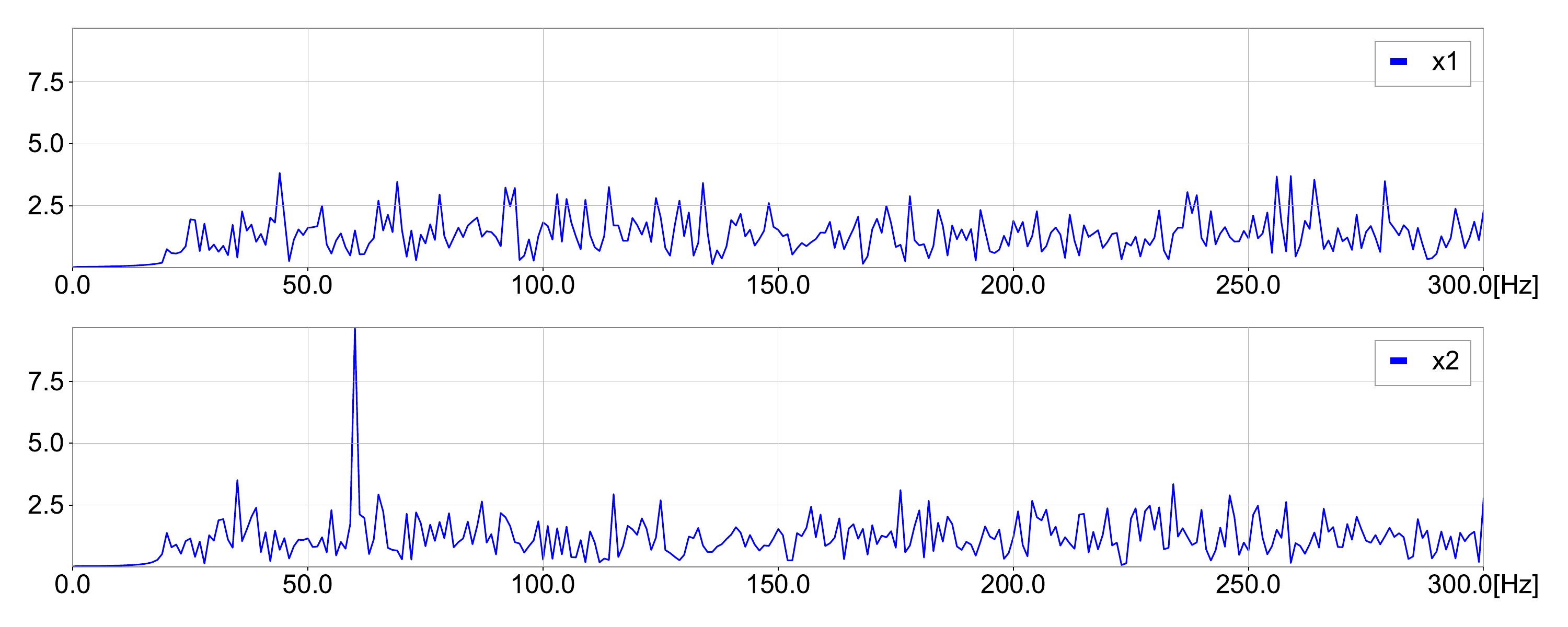}
\\
{\small (c1) Input signals of SNR 10.5. }&
{\small (c2) Fourier spectrum of (c1) }\\
\includegraphics[width=.40\textwidth]
{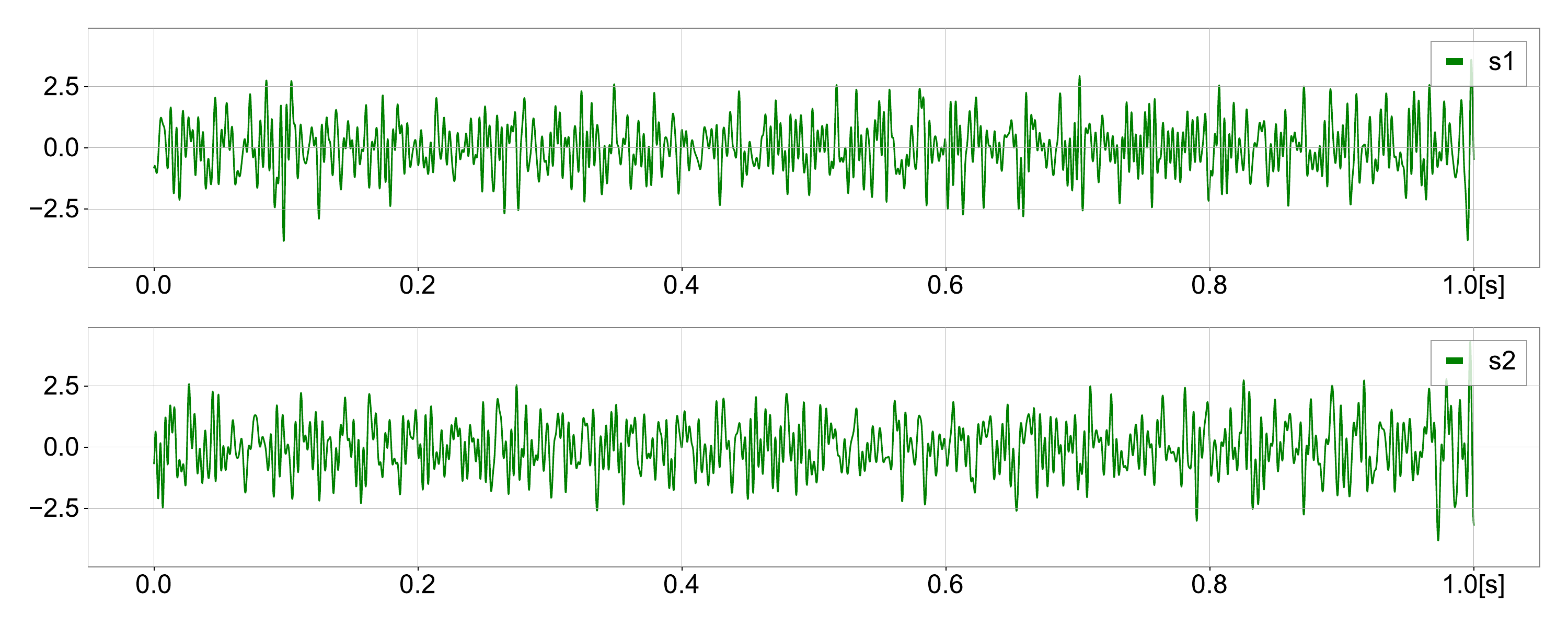}
&
\includegraphics[width=.40\textwidth]
{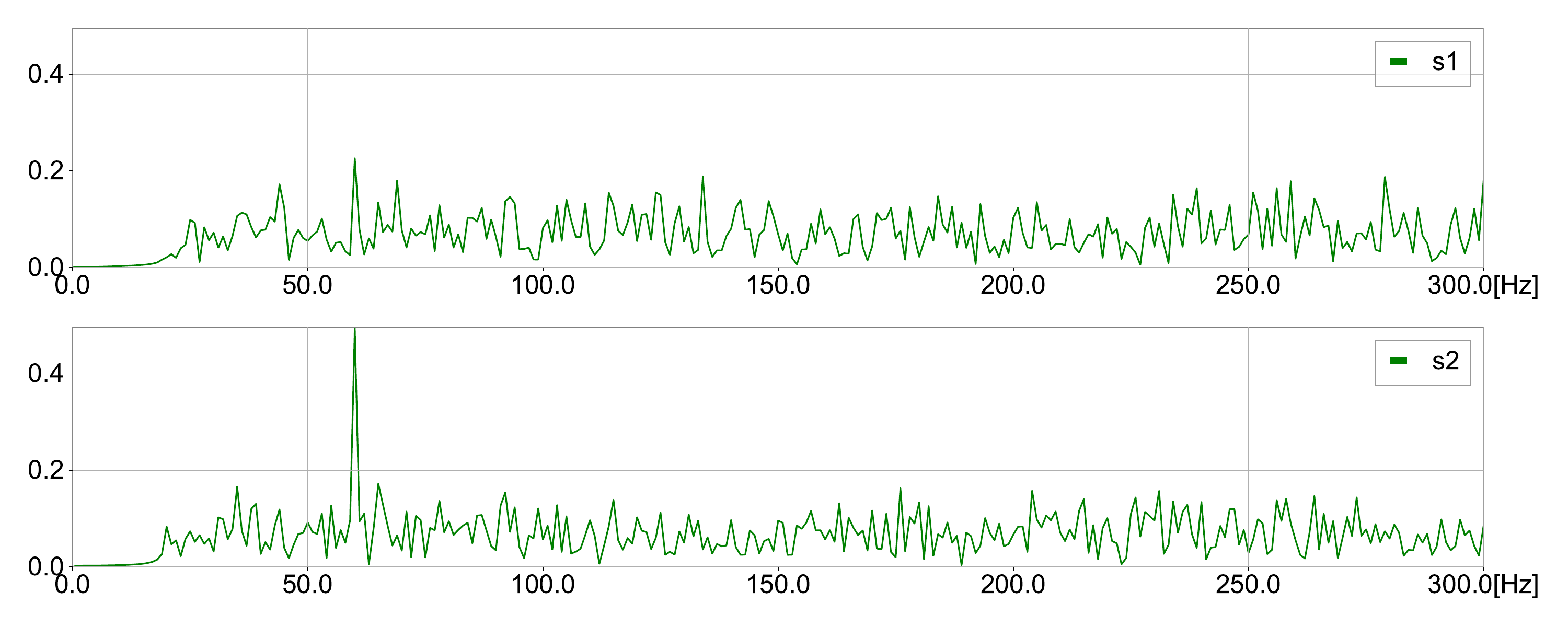}
\\
{\small (c3) Output of ICA for (c1). }&
{\small (c4) Fourier spectrum of (c3).}
\end{tabular}
\caption{
Test of GW extraction using ICA: the case of Model 3 [eq. (\ref{eq:model3})], injections of a inspiral wave signal to the Hanford and Livingston data one second before the event GW150914. 
The SNR of the injected signal is 20.9 (a1-a4), 16.8 (b1-b4), and 10.5 (c1-c4). We see the inspiral feature in Fourier spectrum of the output data for SNR $>$ 15. Note that we filtered for $f>300$~Hz and also that the input data has strong line noise at 60~Hz. 
\label{fig:model3}}
\end{figure}

\begin{table}
\begin{center}
\caption{\label{table:model2} 
Results of Model 2 [injection of sinusoidal wave to the real detector data]: SNR and excess power spectrum of 213~Hz in the extracted data over the average of 150-2050 Hz. }
\footnotesize
\begin{tabular}{cccc}
\br
SNR to $h_H$~ & SNR to $h_L$~ & excess power  & ref.
\\
\mr
~~20.8&~~20.0&~~7.78& Fig.\ref{fig:model2}(a4)\\
~~10.4&~~10.0&~~3.76&Fig.\ref{fig:model2}(b4)\\
~~5.2&~~5.0&~~1.93&Fig.\ref{fig:model2}(c4)\\
\br
\end{tabular}
\end{center}
\end{table}

For Model 2, we used two second data around the merger time of GW150914, and we set $f=213$~Hz which is independent of the known line noises.  Figure \ref{fig:model2} and Table \ref{table:model2} show the results of Model 2. 
By changing the amplitude, we show the cases with signal-to-noise ratio (SNR) of 20 (Fig.(a)), (b)10  and (c) 5. 
Figures (a1), (b1), and (c1) are the input data of $x_1(t)$ and $x_2(t)$, where we show only the center one-second length. 
Figures (a2), (b2), and (c2) show the Fourier spectra of (a1), (b1), and (c1), respectively. 
The model uses actual detector data, which have different noise levels and include many noises. Therefore, if we inject the same signal into both Hanford (x1) and Livingston (x2) data, their appearances differ. 
Figures (a3), (b3), and (c3) are the outputs of ICA, and (a4), (b4), and (c4) are their Fourier spectra. 
To judge whether ICA extracted the injected wave, we show in Table  \ref{table:model2} the ratio of the power spectrum of 213~Hz over those of the average of 150-250~Hz.  
The reason of the imperfect separation is, we think, due to the original strong line noises at 60 Hz and 120 Hz. 
We think we are safe to say the extraction was effective for the cases SNR $\geq 10$. 

For Model 3, we used one second data before one second of merger time $t_c$ of GW150914. The injected inspiral wave is $h(t; t_c-1~\mbox{s}, 26 M_\odot)$. 
The results are shown in Figure \ref{fig:model3}, with the same notation as in the previous figures. 
Note that we filtered {over} $f>300$~Hz and  that the input data had strong line noise at 60~Hz. 
The SNR of the injected signal is 20.9 (a1-a4), 16.8 (b1-b4), and 10.5 (c1-c4). The inspiral feature in the Fourier spectrum of the output data for an SNR $>$ 15.

\section{Applications to real GW extractions}
In this section, we demonstrate GW extractions of the real events.

\subsection{GW150914}
Here, we present the case of GW150914\cite{GW150914} in detail.  This is the first detection of GW and has been well tested as the standard.  The event was observed by the Hanford (H) and Livingston (L) observatories, and the announced network-SNR in GWOSC was 26.0. 

We used one-second data around $t_c$ with sampling rate of 4096.  As a preprocessing step, we whitened the data using each detector's power spectral density and filtered the data to [20, 300]~Hz.  
By shifting Livingston's data every 1/4096~s, we applied 
ICA with five different initial weight matrices {${\bm w}_i^{(p=0)}$} and searched the data-set that showed the maximum {signal strength}  ${\cal A}$  [eq. (\ref{SES})].  The calculations required only 90~s by a laptop to obtain the best  ${\cal A}$ result.

\begin{figure}
\begin{center}
\includegraphics[width=.6\textwidth]
{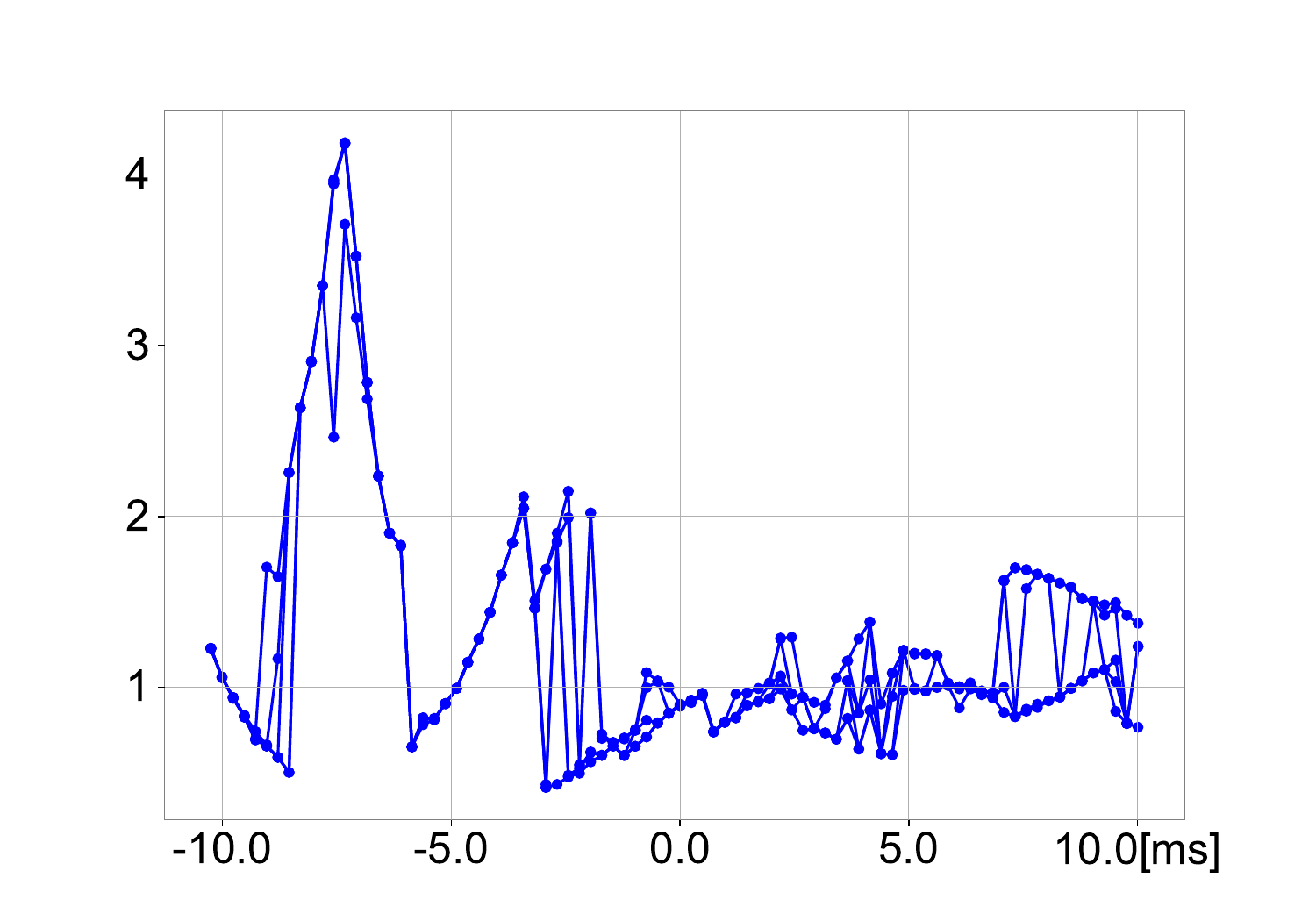}
\caption{
The strength of the extracted signal, ${\cal A}$ [eq. (\ref{SES})] as a function of $\Delta t_{\mbox{\footnotesize HL}}$ for the case of GW150914.  Five trials of different initial weight matrix are plotted at each $\Delta t_{\mbox{\footnotesize HL}}$. We see the maximum is at  $\Delta t_{\mbox{\footnotesize HL}}=-7.32^{+0.15}_{-0.15}$~ms.  Note that LIGO-Virgo paper \cite{GW150914} shows  $\Delta t_{\mbox{\footnotesize HL}}=-6.9^{+0.5}_{-0.4}$~ms. 
\label{fig:GW150914SESgraph}}
\end{center}
\end{figure}

We determined the maximum ${\cal A}$ when $\Delta t_{\rm HL}=-7.32$~ms.  
Figure \ref{fig:GW150914SESgraph} shows ${\cal A}$ as a function of $\Delta t_{\rm HL}$. 
The plot includes the results of our five different initial weight matrices ${\bm w}_i^{(0)}$, 
and we see that the values of ${\cal A}$ are slightly different from the initial value ${\bm w}_i^{(0)}$, but the $\Delta t$ of the maximum of ${\cal A}$ is independent of  ${\bm w}_i^{(0)}$, and is uniquely determined.  From Figure  \ref{fig:GW150914SESgraph}, we estimate $\Delta t_{\rm HL}$ with the error-bar $\pm 0.15$~ms.  Note that the LIGO-Virgo paper \cite{GW150914} denotes  $\Delta t_{\rm HL}=-6.9^{+0.5}_{-0.4}$~ms.  Our number is consistent with \cite{GW150914} and with a small error-bar.

\begin{figure}
\begin{tabular}{p{0.5\textwidth}p{0.5\textwidth}}
\includegraphics[width=.45\textwidth]
{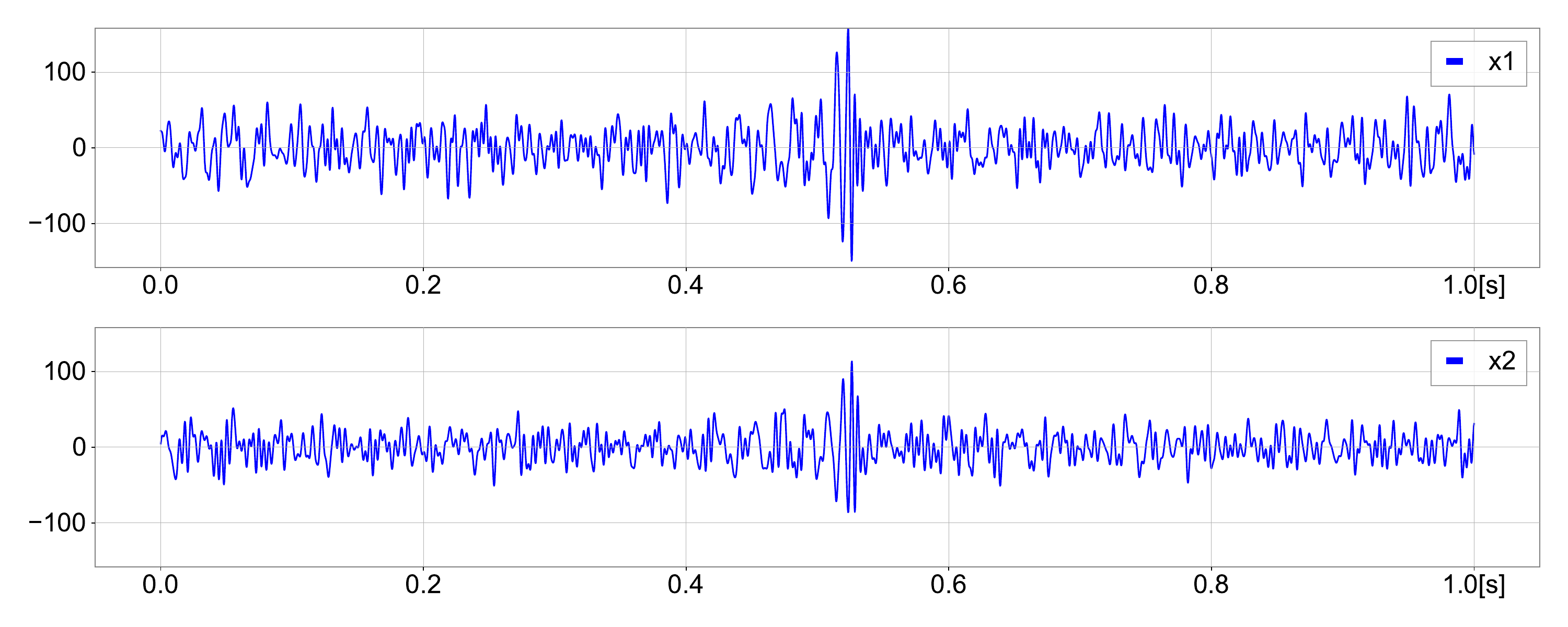}
&
\includegraphics[width=.45\textwidth]
{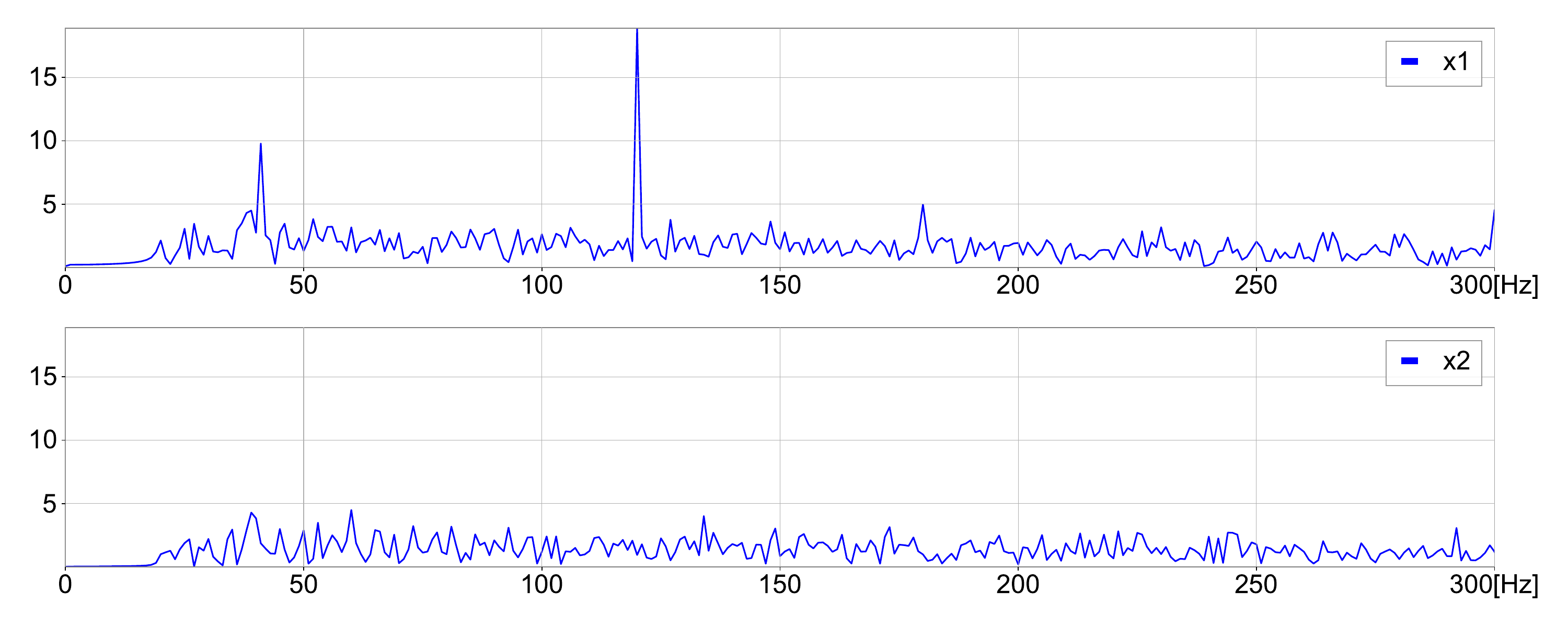}
\\
{\small (a) Input signals with $\Delta t_{\mbox{\footnotesize HL}}=-7.32$~ms. The data x1 and x2 are of Hanford and Livingston data, respectively.}&
{\small (b) Fourier spectrum of (a).}\\
\includegraphics[width=.45\textwidth]
{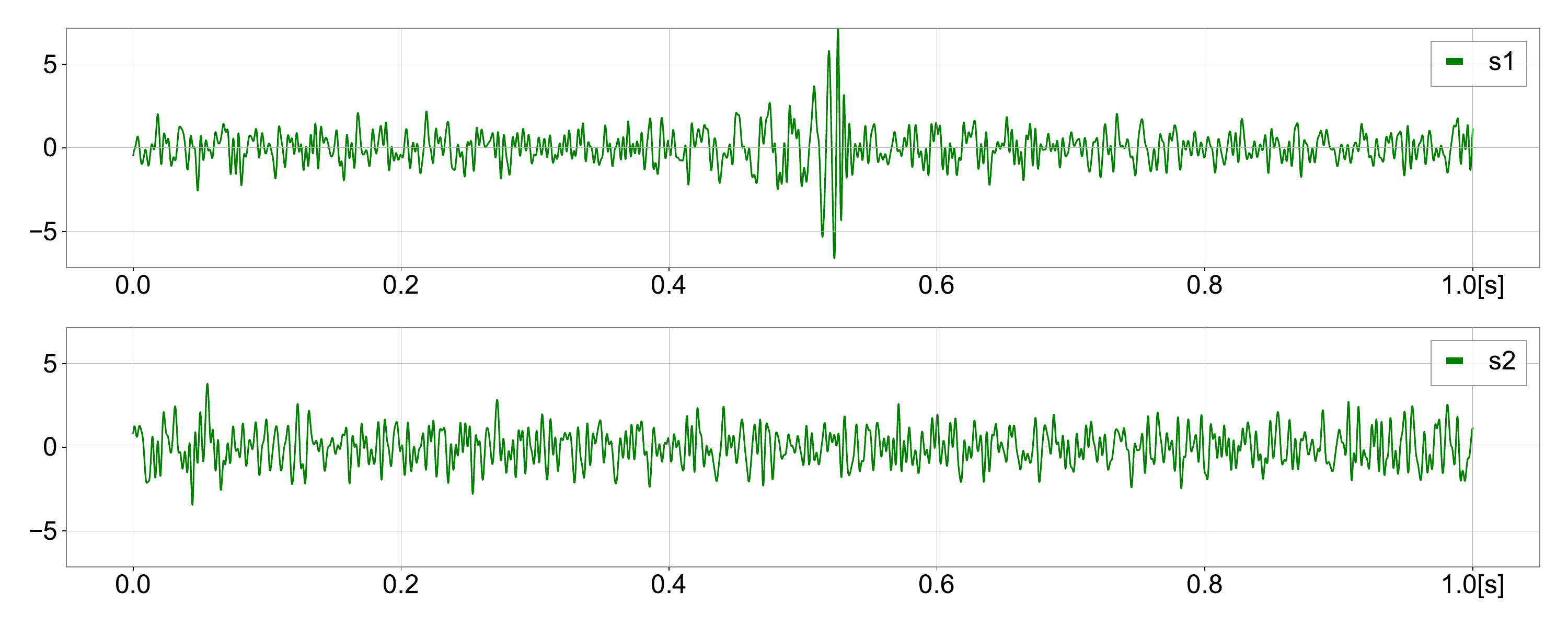}
&
\includegraphics[width=.45\textwidth]
{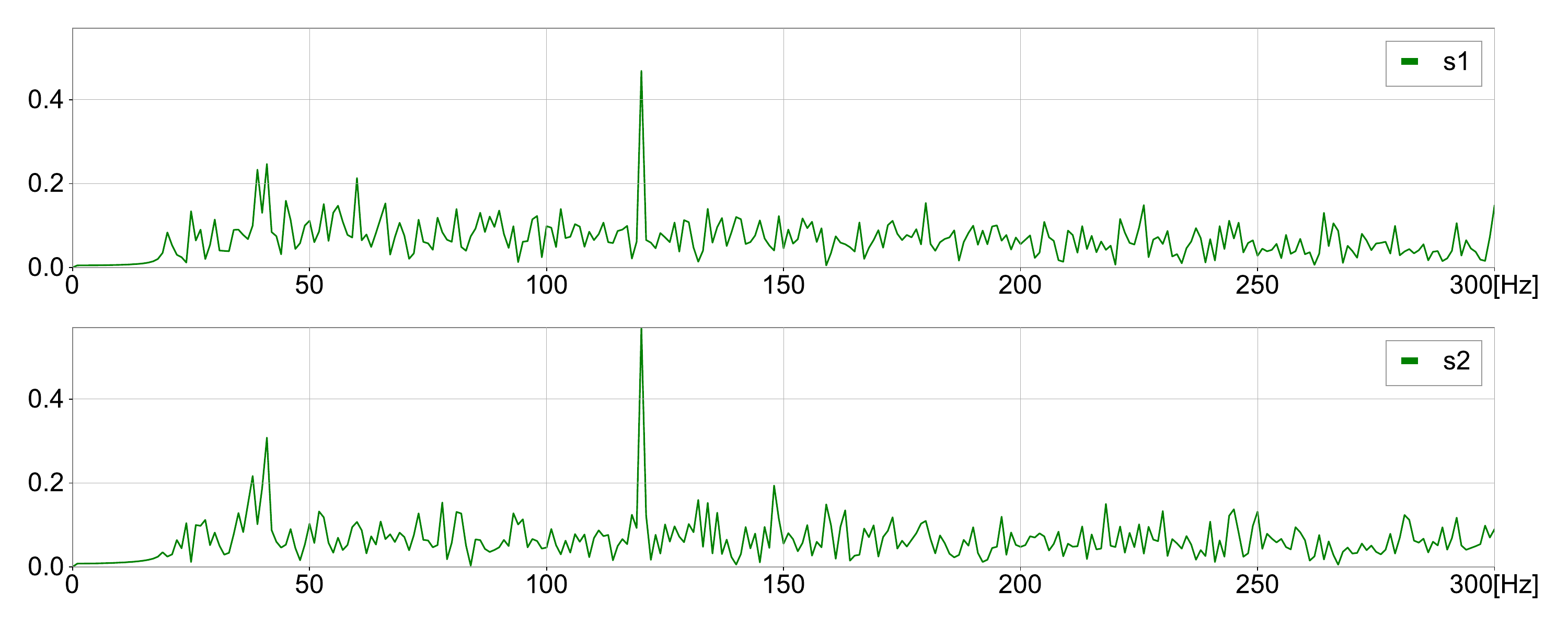}
\\
{\small (c) Output of ICA.} &
{\small (d) Fourier spectrum of (c).}\\
\end{tabular}
\caption{\label{fig:GW150914}
Application to GW150914. (a/b) The set of input data and its spectrum with $\Delta t_{\mbox{\footnotesize HL}}=-7.32$~ms which shows the largest ${\cal A}$ in the output of ICA.  The x1 and x2 indicates the data of Hanford and Livingston, respectively.   (c/d) The output of ICA, and its spectrum. We clearly see the signal (s1) is separted from the other (s2). Note that the input data have a line noise at 120~Hz.
}
\end{figure}

\begin{figure}
\begin{center}
\includegraphics[width=.8\textwidth]
{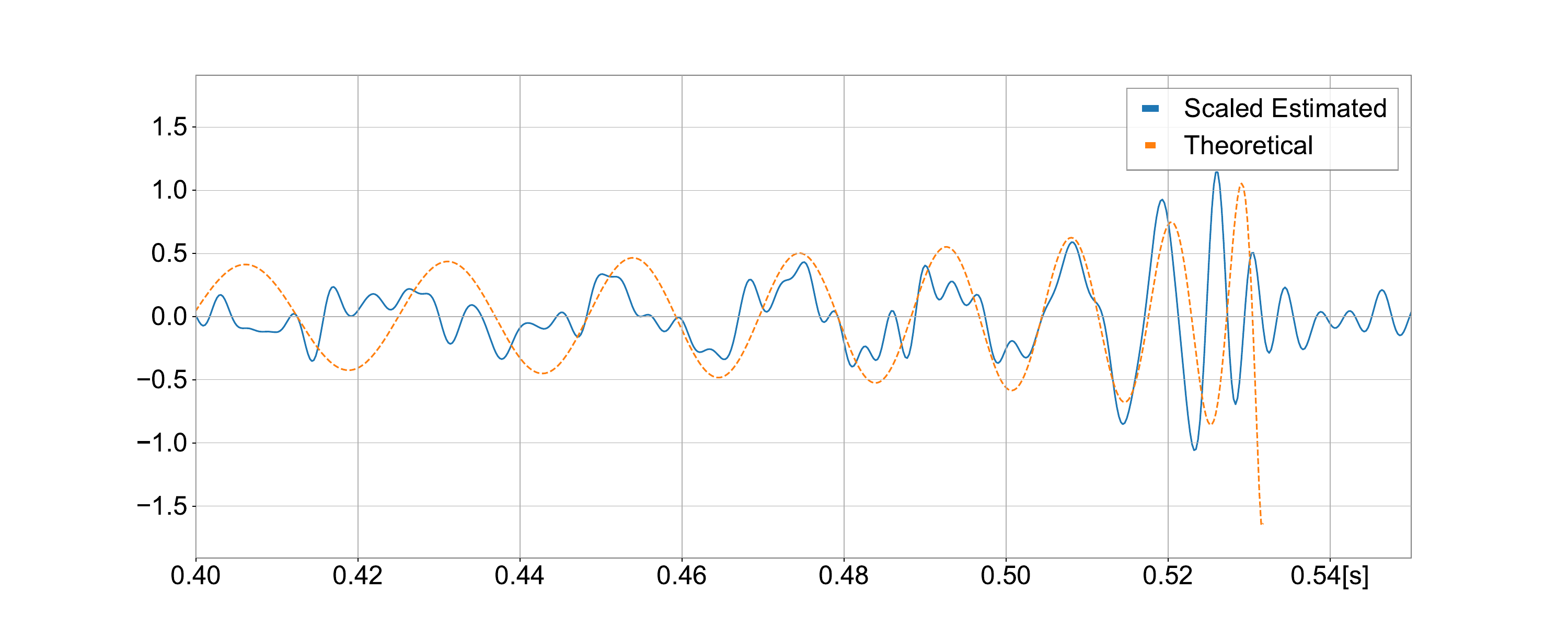}
\\
\includegraphics[width=.8\textwidth, height=2.0cm]
{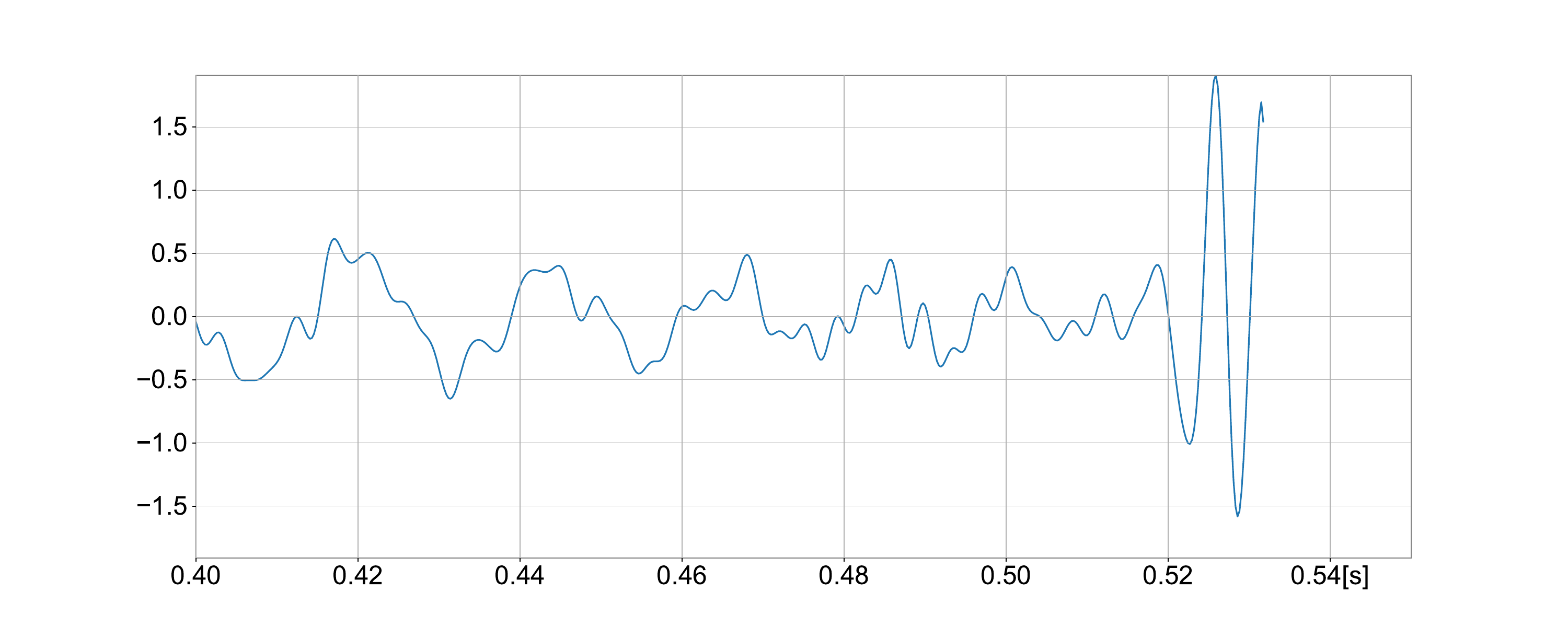}
\\
\caption{Extraction of GW150914. 
(Upper) The output data (Fig. \ref{fig:GW150914} (c) s1) is overlapped with $h_{\rm \footnotesize{insp}}(t; t_c, M_c=30.8M_\odot)$. (Lower) The residuals. 
\label{fig:GW150914overlap}}
\end{center}
\end{figure}

We plot the input data (x1, x2) and output data (s1, s2) for the largest ${\cal A}$ case in Figure \ref{fig:GW150914}. 
For the output data s1 [in Figure \ref{fig:GW150914}(c)], we next searched the 
best-matched inspiral waveform, $h_{\rm  insp}(t; t_c, M_c)$, by changing $M_c$ and amplitude, measuring its residuals, ${\cal R}$.  
We found $M_c=30.8 M_\odot$ shows the lowest residuals. 
This case is shown in Figure \ref{fig:GW150914overlap}, and we see how output s1 is similar to the expected inspiral signal. 

GWOSC webpage shows $M_c=28.6^{+1.7}_{-1.5}M_\odot$ as an estimated value in the source frame. 
Our $M_c$ is in the observed frame, which differs from the 
\begin{equation}
M_c^{\rm obs}=M_c^{\rm source} (1+z)
\end{equation}
where $z$ is the redshift parameter of the source.  From error-bars in $M_c^{\rm source}$ in GWOSC data, we calculate $z$ (hereafter we denote $z_{\rm ICA}$) as $z_{\rm ICA}=0.077\pm 0.06$.  The redshift $z$ of GW150914 in GWOSC webpage is $z=0.09\pm 0.03$.  Therefore our result is again consistent with them.

\subsection{Other events}
\begin{figure}
\begin{tabular}{p{0.5\textwidth}p{0.5\textwidth}}
\includegraphics[width=.45\textwidth]
{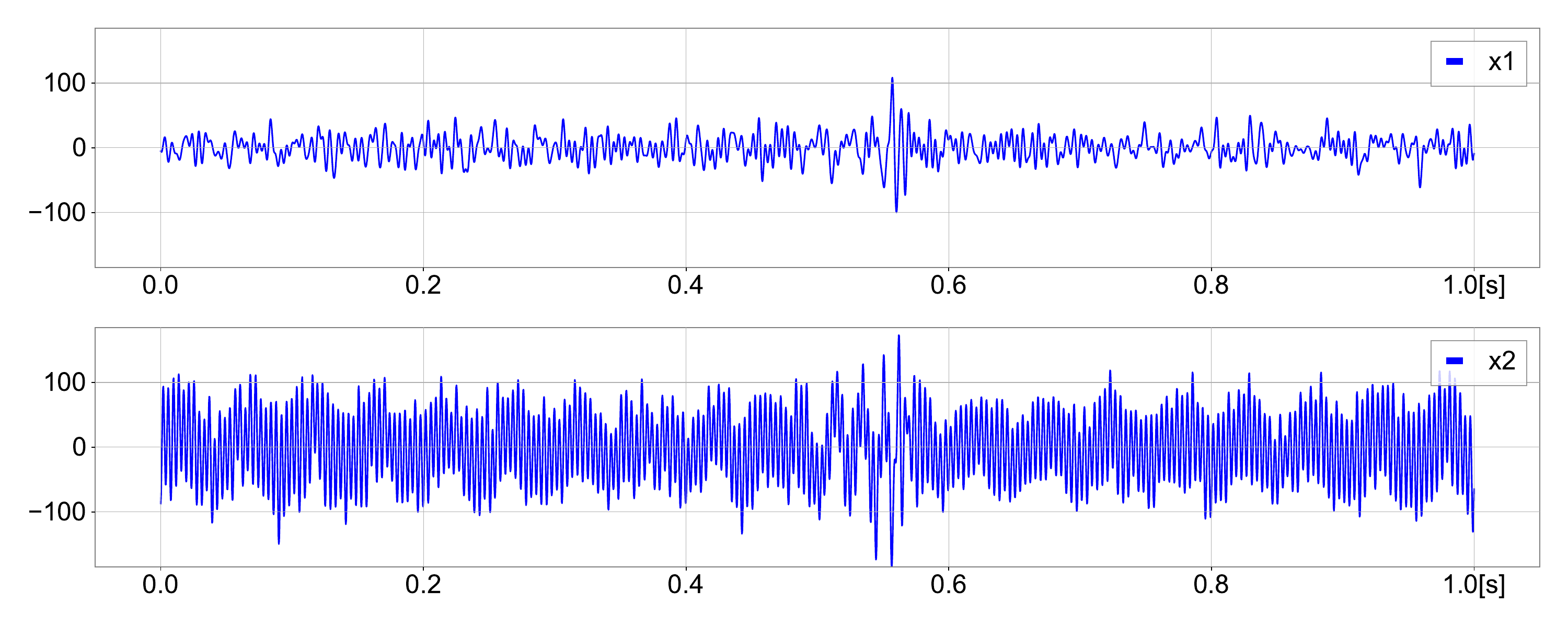}
&
\includegraphics[width=.45\textwidth]
{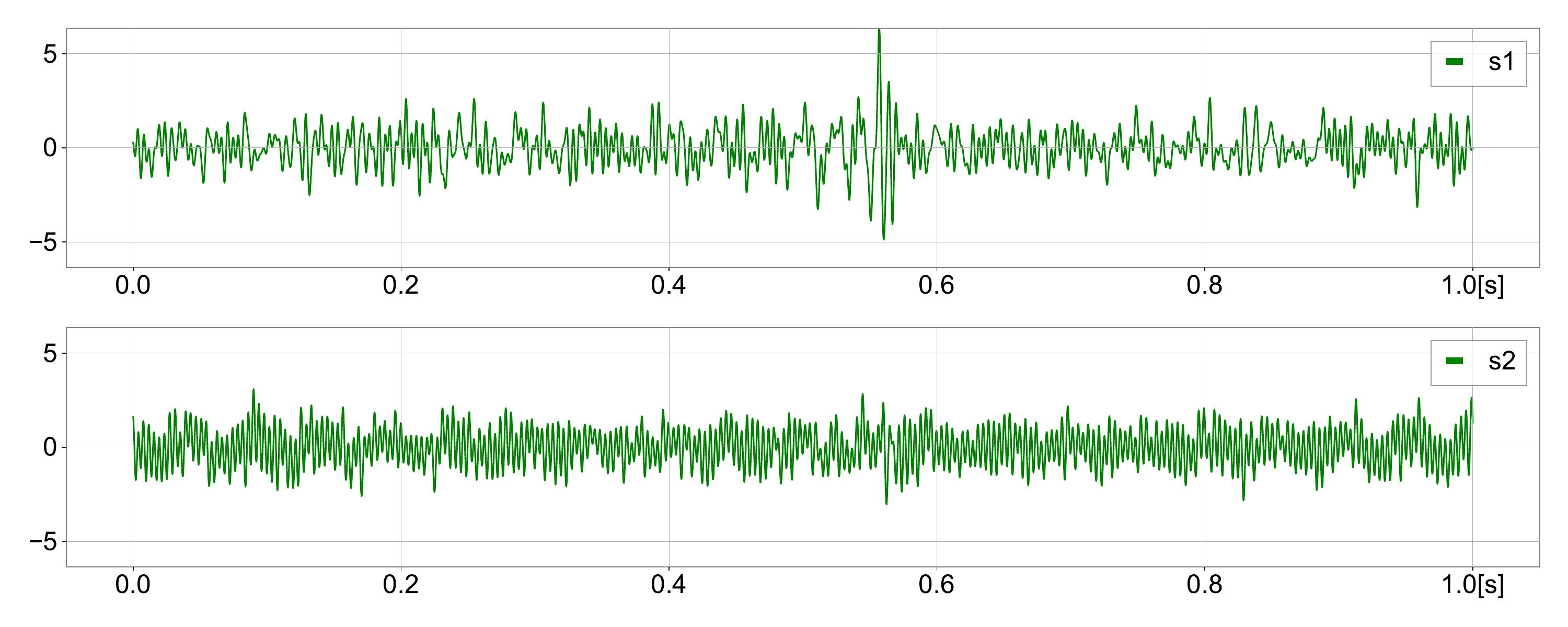}
\\
{\small (a1) Input signals of GW190521\_074359 with $\Delta t_{\mbox{\footnotesize HL}}=-6.35$~ms. The data x1 and x2 are of Hanford and Livingston data, respectively. }
&
{\small (a2) Output of ICA for GW190521\_074359.} \\[3em]
\includegraphics[width=.45\textwidth]
{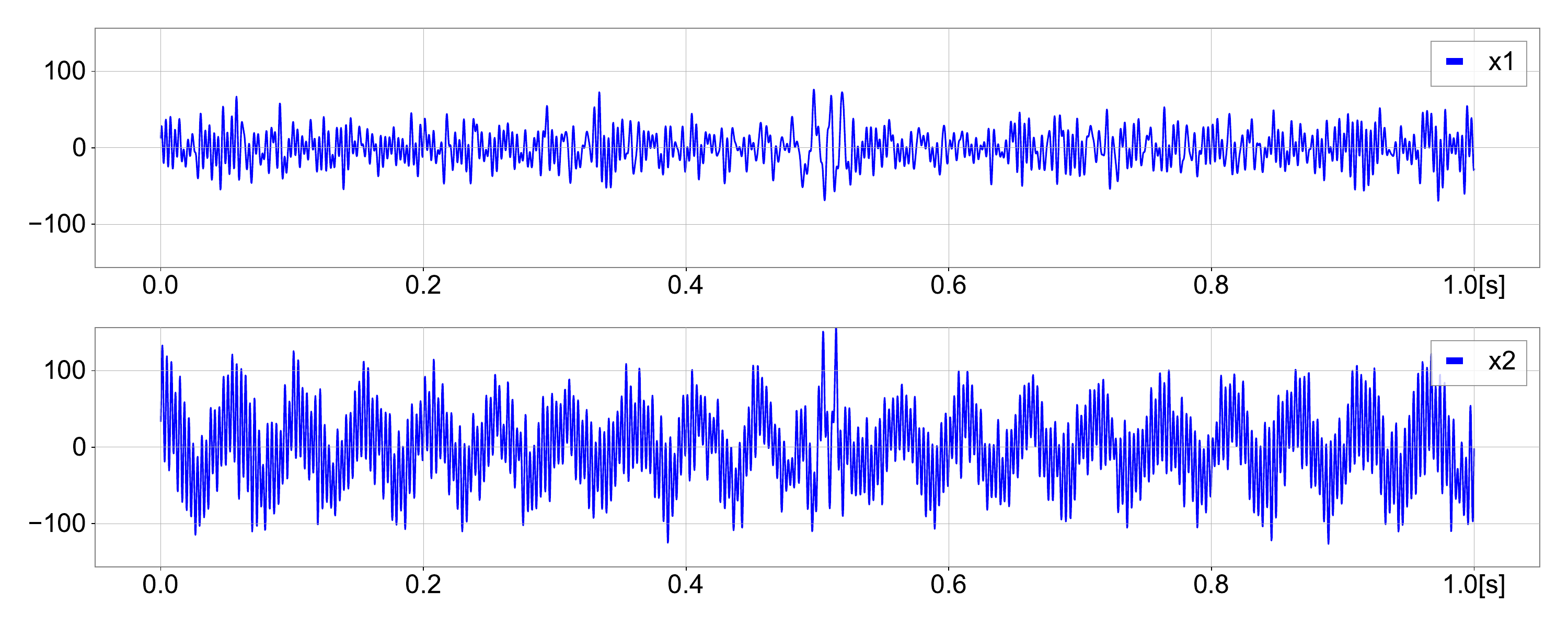}
&
\includegraphics[width=.45\textwidth]
{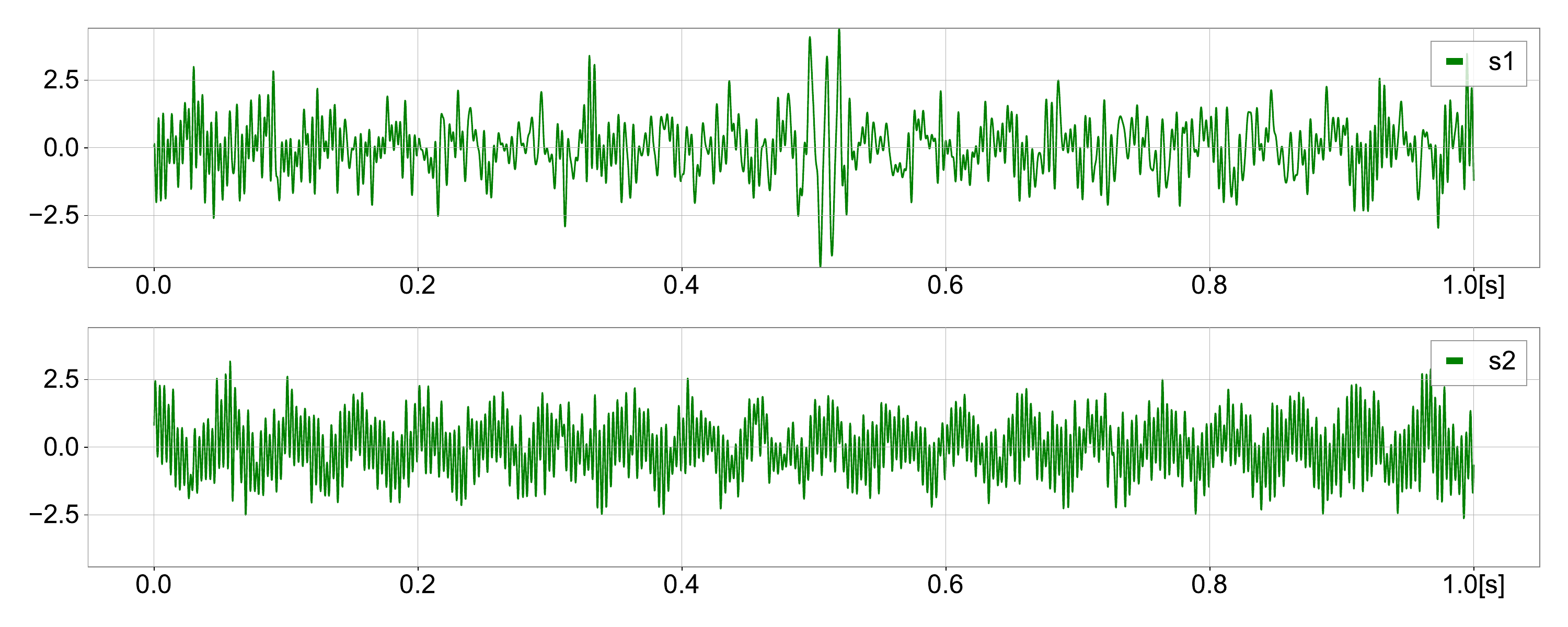}
\\
{\small (b1) Input signals of GW191109\_010717 with $\Delta t_{\mbox{\footnotesize HL}}=+3.17$~ms. The data x1 and x2 are of Hanford and Livingston data, respectively.}
&
{\small (b2) Output of ICA for GW191109\_010717.}
\\[3em]
\includegraphics[width=.45\textwidth]
{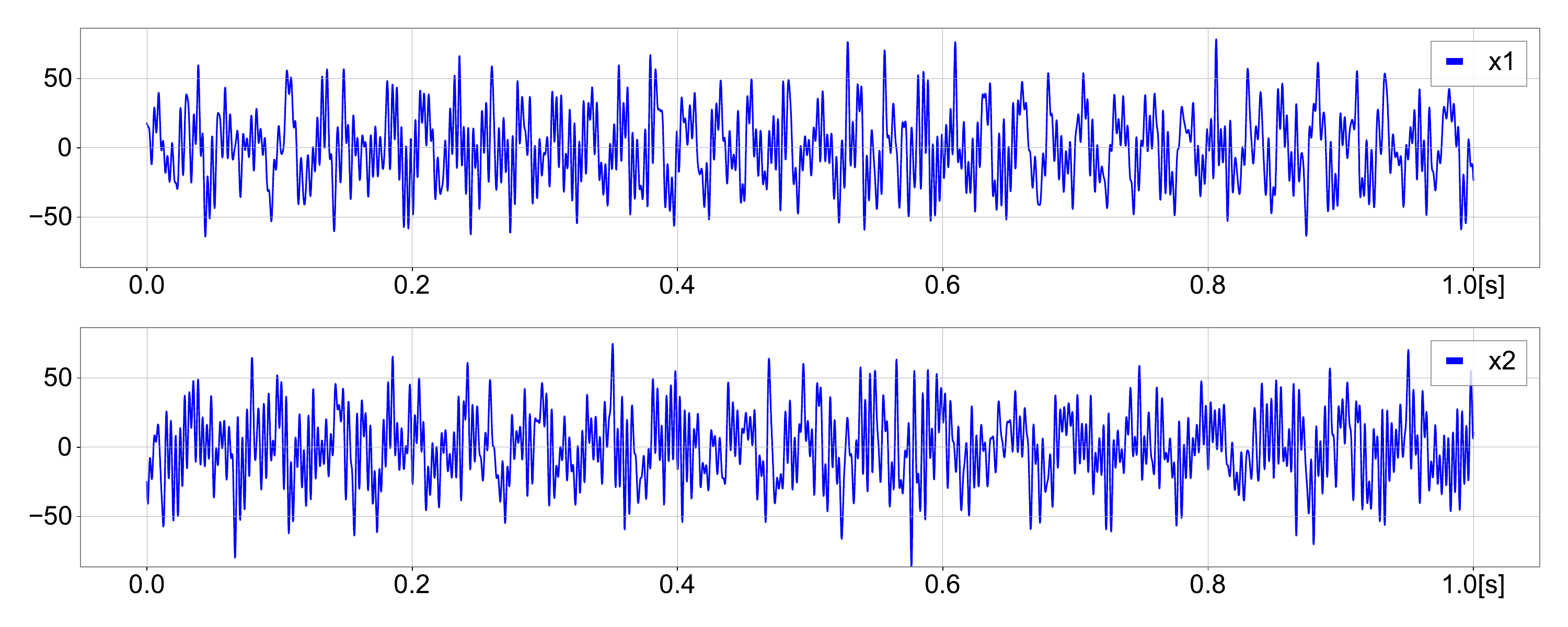}
&
\includegraphics[width=.45\textwidth]
{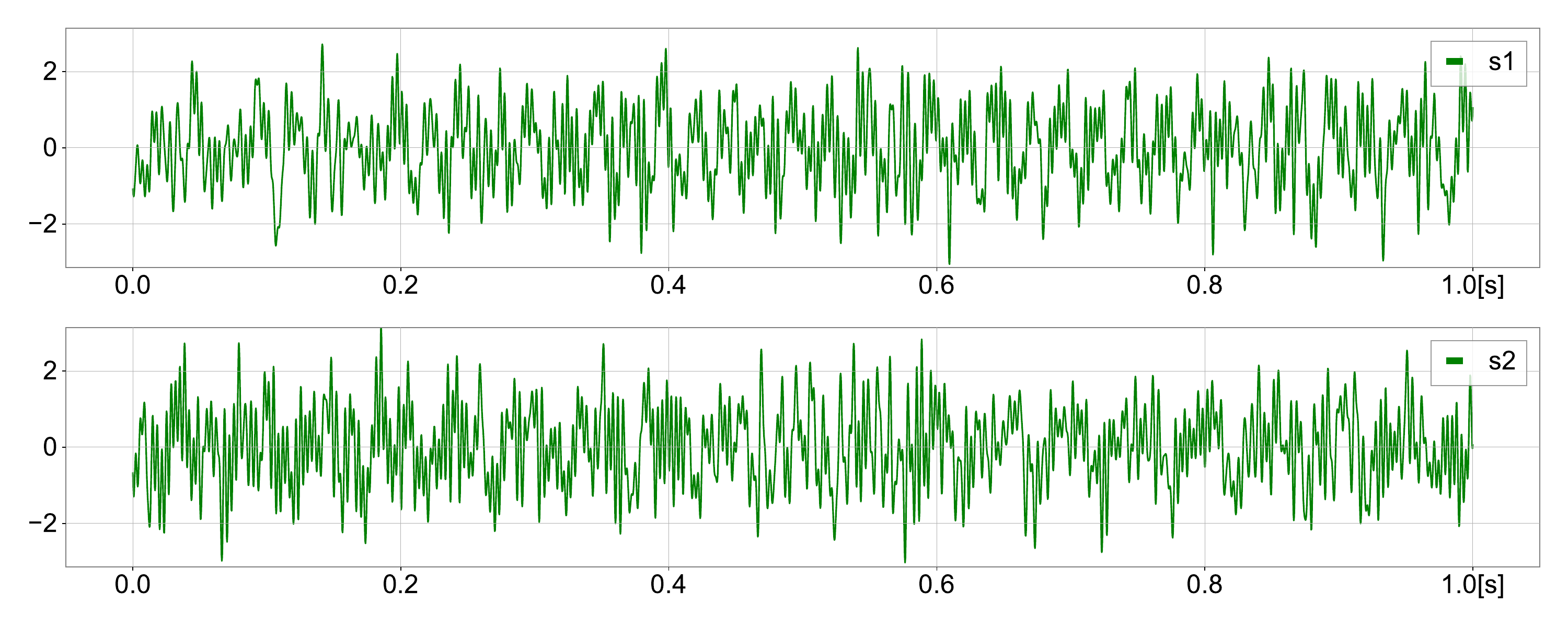}
\\
{\small (c1) Input signals of GW191204\_171526 with $\Delta t_{\mbox{\footnotesize HL}}=-2.44$~ms. The data x1 and x2 are of Hanford and Livingston data, respectively.}
&
{\small (c2) Output of ICA for GW191204\_171526.} \\[3em]
\includegraphics[width=.45\textwidth]
{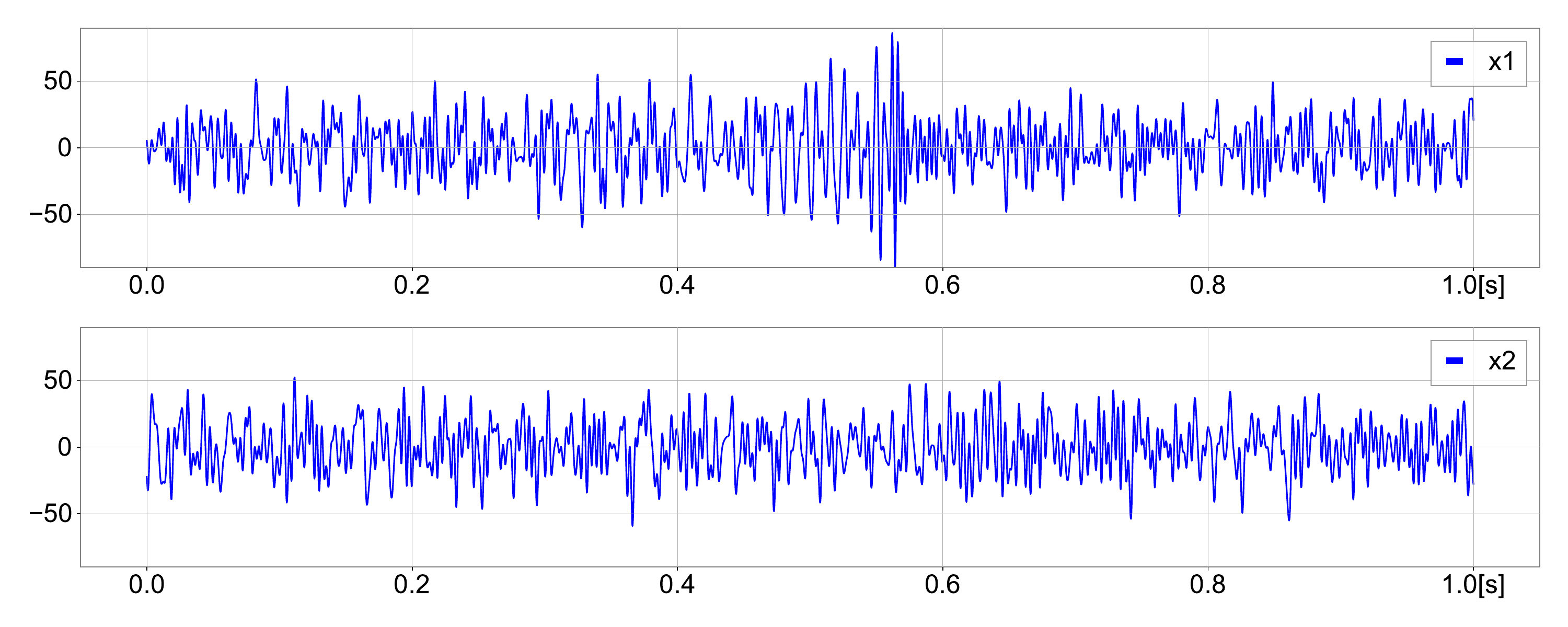}
&
\includegraphics[width=.45\textwidth]
{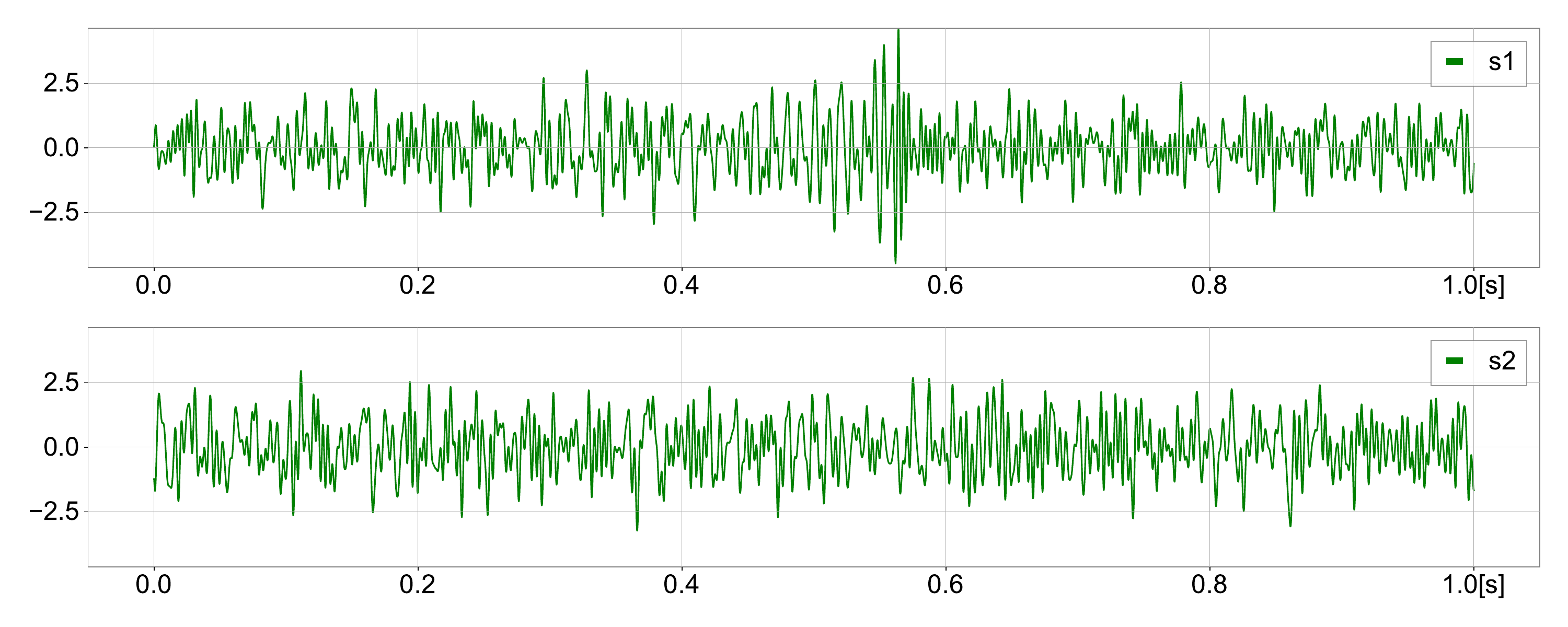}
\\
{\small (d1) Input signals of GW191216\_213338 with $\Delta t_{\mbox{\footnotesize HV}}=-11.0$~ms. The data x1 and x2 are of Hanford and Virgo data, respectively.}
&
{\small (d2) Output of ICA for GW191216\_213338. }\\[3em]
\end{tabular}
\caption{
Input and Output data of ICA analysis. 
\label{fig:otherevents}}
\end{figure}

We continue a similar analysis for other GW events from binary BHs with higher SNR events in O1-O3. 
Figure \ref{fig:otherevents} shows the input data (x1, x2, x3) and the output data (s1, s2, s3) that show the largest ${\cal A}$.  
For three-detector events, we made 2-dimensional search for shifting data, which requires 20100 combinations at maximum, and the computation time is approximately 20 h. 
The ICA results are summarized in Tables \ref{table:modelGWreal} and \ref{table:modelGWreal2}. 

\begin{table}
\begin{center}
\caption{Results of the wave extractions by ICA for large SNR events in O1-O3. The column obs shows which detector (Hanford/Livingston/Virgo) observed.  SNR is the network signal-to-noise ratio  (centered value) which is announced in GWOSC (https://gwosc.org).  $\Delta t_{\rm HL}$ is the time shift between Hanford data and Livingston data, $t_{\rm L}-t_{\rm H}$, in ms when ICA shows the best separation of the signal. ${\cal A}$ is the signal strength defined by eq. (\ref{SES}). ${\cal R}$ is the residuals of the extracted waveform and estimated inspiral waveform, (\ref{eq:residual}),  between [$t_c-0.15$~ms, $t_c$].  See table \ref{table:modelGWreal2} for comparisons of chirp-mass and red-shift.  \label{table:modelGWreal} }
\footnotesize
\begin{tabular}{ccc|ccc|cc|c}
\br
event & obs & SNR & $\Delta t_{\rm HL}$ (ms) & $\Delta t_{\rm HV}$ (ms) &$\Delta t_{\rm LV}$ (ms) &${\cal A}$&${\cal R}/10^{-12}$ 
& ref.
\\
\mr
GW150914& HL &26.0&  $-7.32 \pm ^{1.5} _{1.5}$& -- & -- & 4.19 & ~5.88  &Fig.\ref{fig:GW150914} \\
GW190521\_074359& HL &  32.8 & $-6.35 \pm ^{0.98} _{0.49}$ & -- & -- & 1.83 & 10.3 & Fig.\ref{fig:otherevents}(a)\\
GW191109\_010717& HL &  47.5 & $3.17 \pm ^{0.98} _{0.73}$ & -- & -- & 3.40 & 18.4 & Fig.\ref{fig:otherevents}(b) \\
GW191204\_171526& HL &  8.55 & $-2.44 \pm ^{0.49} _{0.73}$& -- & --& 2.07 & ~3.27 & Fig.\ref{fig:otherevents}(c)\\
GW191216\_213338& HV &  8.33 & -- & $-11.0 \pm ^{1.5} _{0.73}$ & -- & 3.08 & ~2.09 & Fig.\ref{fig:otherevents}(d)\\
GW200112\_155838& LV &  27.4 & -- & -- & $-23.2 \pm ^{0.49} _{0.24}$& 2.43 & 10.5 & Fig.\ref{fig:otherevents}(e)\\
\mr
GW170814& HLV &  24.1 & $-8.06 \pm ^{0.49} _{0.98}$& $0.98 \pm ^{2.4} _{0.24}$ & --& 3.54 & ~5.07 & Fig.\ref{fig:otherevents}(f)\\
GW190412& HLV &  13.3 & $-3.91 \pm ^{0.24} _{0.24}$& $-13.92 \pm ^{0.98} _{0.49}$ & --& 2.21 & ~4.40 & Fig.\ref{fig:otherevents}(g)\\
GW190521& HLV &  63.3 & $2.93 \pm ^{0.49} _{1.2}$& $-25.15 \pm ^{0.49} _{1.7}$ & --& 2.85 & 31.8  &Fig.\ref{fig:otherevents}(h)\\
GW190814& HLV &  6.11 & $2.20 \pm ^{0.49} _{0.24}$& $21.24 \pm ^{0.73} _{0.24}$ & --& 2.00 & ~1.65 &Fig.\ref{fig:otherevents}(i)\\
GW200129\_065458& HLV &  27.2 & $3.42 \pm ^{0.98} _{0.24}$& $-18.31 \pm ^{0.24} _{0.24}$ & --& 3.96 & 11.3 &Fig.\ref{fig:otherevents}(j)\\
GW200224\_222234& HLV &  31.1 & $-3.66 \pm ^{2.7} _{0.24}$& $-9.28 \pm ^{0.24} _{0.98}$ & --& 3.28 & 13.4 & Fig.\ref{fig:otherevents}(k)\\
GW200311\_115853& HLV &  26.6 & $-3.66 \pm ^{0.73} _{1.2}$& $-27.10 \pm ^{2.2} _{2.2}$ & --& 3.17 & ~4.34 & Fig.\ref{fig:otherevents}(l)\\
\br
\end{tabular}
\end{center}
\end{table}

\begin{table}
\begin{center}
\caption{\label{table:modelGWreal2} 
Comparisons of chirp mass, $M^{\rm source}_c$, shown in GWOSC and the one obtained by ICA,$M^{\rm obs}_c$ from the best fit inspiral-wave model.  The difference can be regard as redshift factor $(1+z_{\rm ICA})$. 
The redshift factor in GWOSC, $z$, is also shown. 
}
\footnotesize
\begin{tabular}{ccc|cc|cc|c}
\br
 &&&GWOSC&& ICA &&\\
\mr
event & obs & SNR & $M^{\rm source}_c/M_\odot$ & $z$ &
$M^{\rm obs}_c/M_\odot$& $z_{\rm ICA}$ & ref.
\\
\mr
GW150914& HL &26.0 & $28.6^{+1.7}_{-1.5}$ & $0.09^{+0.03}_{-0.03}$  
& 30.8 & $0.077^{+0.06}_{-0.06}$ &Fig.\ref{fig:GW150914} \\
GW190521\_074359& HL & 25.9 & $32.8^{+3.2}_{-2.8}$ & $0.21^{+0.10}_{-0.10}$& 36.4 &  $0.11^{+0.10}_{-0.10}$ &Fig.\ref{fig:otherevents}(a)\\
GW191109\_010717& HL & 17.3 & $47.5^{+9.6}_{-7.5}$ &$0.25^{+0.18}_{-0.12}$ & 53.7 &  $0.13^{+0.22}_{-0.19}$ &Fig.\ref{fig:otherevents}(b) \\
GW191204\_171526& HL & 17.5 & $8.56^{+0.41}_{-0.28}$ &$0.34^{+0.25}_{-0.18}$& 11.1 &$0.29^{+0.04}_{-0.06}$ &Fig.\ref{fig:otherevents}(c)\\
GW191216\_213338& HV & 18.6 & $8.33^{+0.22}_{-0.19}$ &$0.07^{+0.02}_{-0.03}$ & 9.00 &$0.08^{+0.03}_{-0.03}$  &Fig.\ref{fig:otherevents}(d)\\
GW200112\_155838& LV & 19.8 & $27.4^{+2.6}_{-2.1}$ &$0.24^{+0.07}_{-0.08}$& 32.7 & $0.19^{+0.10}_{-0.10}$ &Fig.\ref{fig:otherevents}(e)\\
\mr
GW170814& HLV & 17.7 & $24.1^{+1.4}_{-1.1}$ &$0.12^{+0.03}_{-0.04}$& 26.0 &  $0.08^{+0.05}_{-0.06}$ &Fig.\ref{fig:otherevents}(f)\\
GW190412& HLV & 19.8 & $13.3^{+0.5}_{-0.5}$ &$0.15^{+0.04}_{-0.04}$&  14.8 & $0.11^{+0.04}_{-0.04}$ &Fig.\ref{fig:otherevents}(g)\\
GW190521& HLV & 14.3 & $63.3^{+19.6}_{-14.6}$ &$0.56^{+0.36}_{-0.27}$&  81.7 & $0.29^{+0.39}_{-0.30}$ &Fig.\ref{fig:otherevents}(h)\\
GW190814& HLV & 25.3 & $6.11^{+0.06}_{-0.05}$ &$0.05^{+0.01}_{-0.01}$&  6.35 & $0.04^{+0.01}_{-0.01}$ &Fig.\ref{fig:otherevents}(i)\\
GW200129\_065458& HLV & 26.8 & $27.2^{+2.1}_{-2.3}$ &$0.18^{+0.05}_{-0.07}$&  30.6 &$0.13^{+0.10}_{-0.08}$  &Fig.\ref{fig:otherevents}(j)\\
GW200224\_222234& HLV & 20.0 & $31.1^{+3.3}_{-2.7}$ &$0.32^{+0.08}_{-0.11}$&  37.6 &$0.21^{+0.11}_{-0.12}$  &Fig.\ref{fig:otherevents}(k)\\
GW200311\_115853& HLV & 17.8 & $26.6^{+2.4}_{-2.0}$ &$0.23^{+0.05}_{-0.07}$& 31.0 &$0.17^{+0.09}_{-0.10}$ &Fig.\ref{fig:otherevents}(l)\\
\br
\end{tabular}
\end{center}
\end{table}

Table \ref{table:modelGWreal} can be used for understanding how ICA works. 
Roughly summarizing, the parameter ${\cal A}$, which is a measure of the quality of extraction, is almost the same regardless of the SNR. 
The residual ${\cal R}$, which is a measure of how the extracted wave can be matched with inspiral waveform, has negative correlation with SNR. 

Table \ref{table:modelGWreal2} can be used for understanding how the results of ICA are realistic. By making the matches with inspiral wave, we can specify  $M^{\rm obs}_c$ for each event.  With $M^{\rm source}_c$ in GWOSC table, we compare the redshift parameters in GWOSC table and $z_{\rm ICA}$.  As we see in the table \ref{table:modelGWreal2}, all $z$ parameters are consistent. 

\addtocounter{figure}{-1}
\begin{figure}
\begin{tabular}{p{0.5\textwidth}p{0.5\textwidth}}
\includegraphics[width=.45\textwidth]
{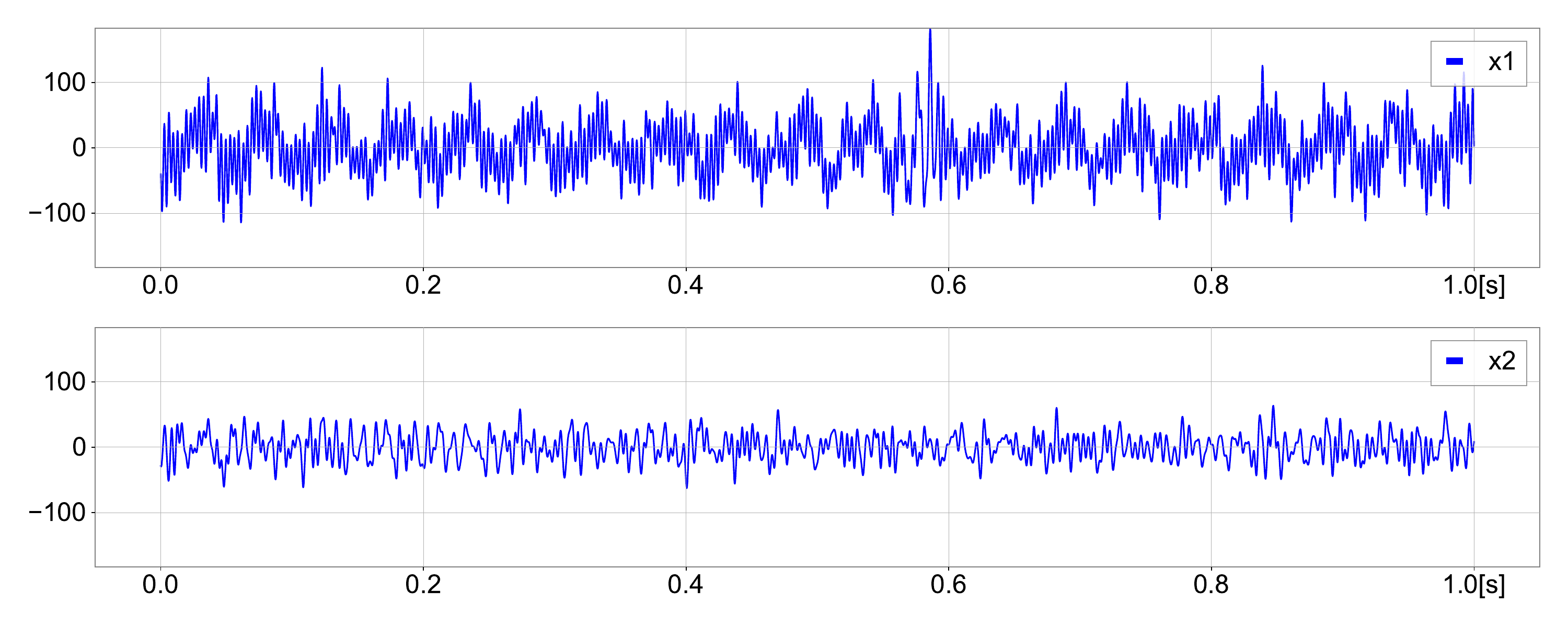}
&
\includegraphics[width=.45\textwidth]
{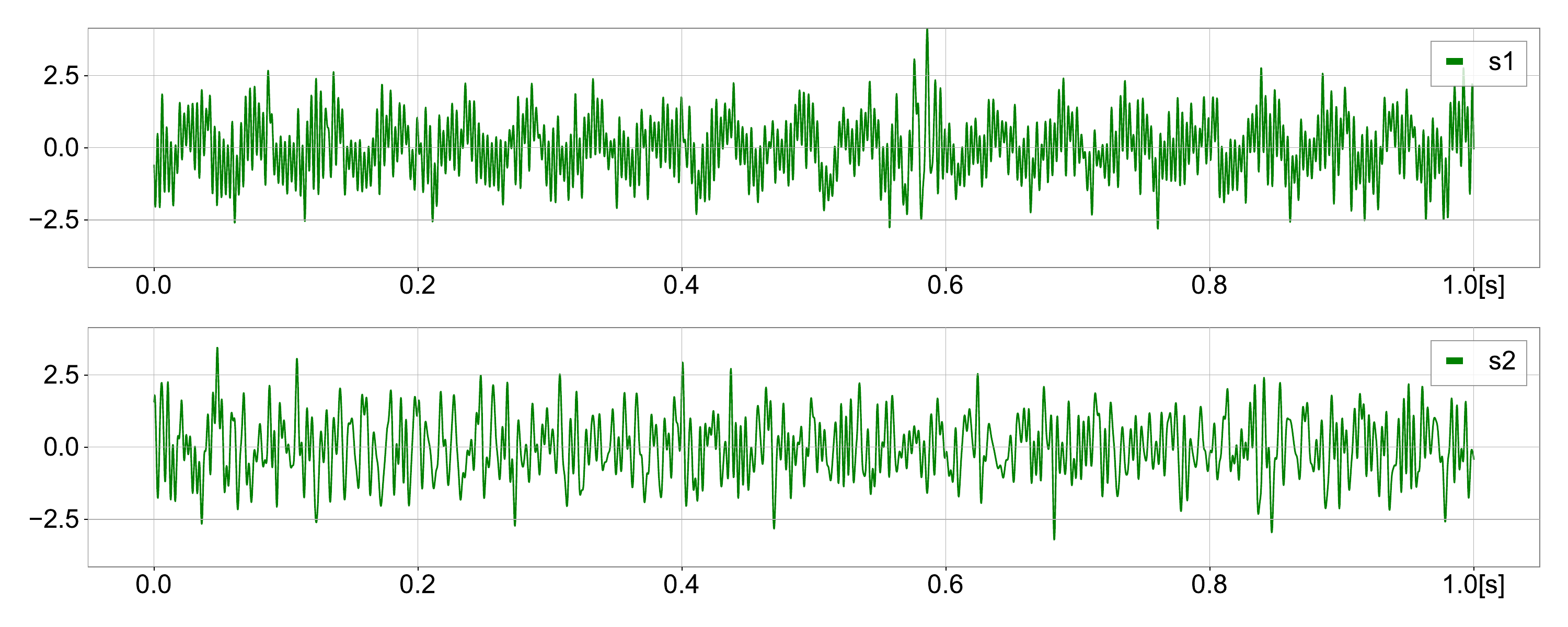}
\\
{\small (e1) Input signals of GW200112\_155838 with $\Delta t_{\mbox{\footnotesize LV}}=-23.2$~ms.  The data x1 and x2 are of Livingston and Virgo data, respectively.}
&
{\small (e2) Output of ICA for GW200112\_155838. }\\

\includegraphics[width=.45\textwidth]
{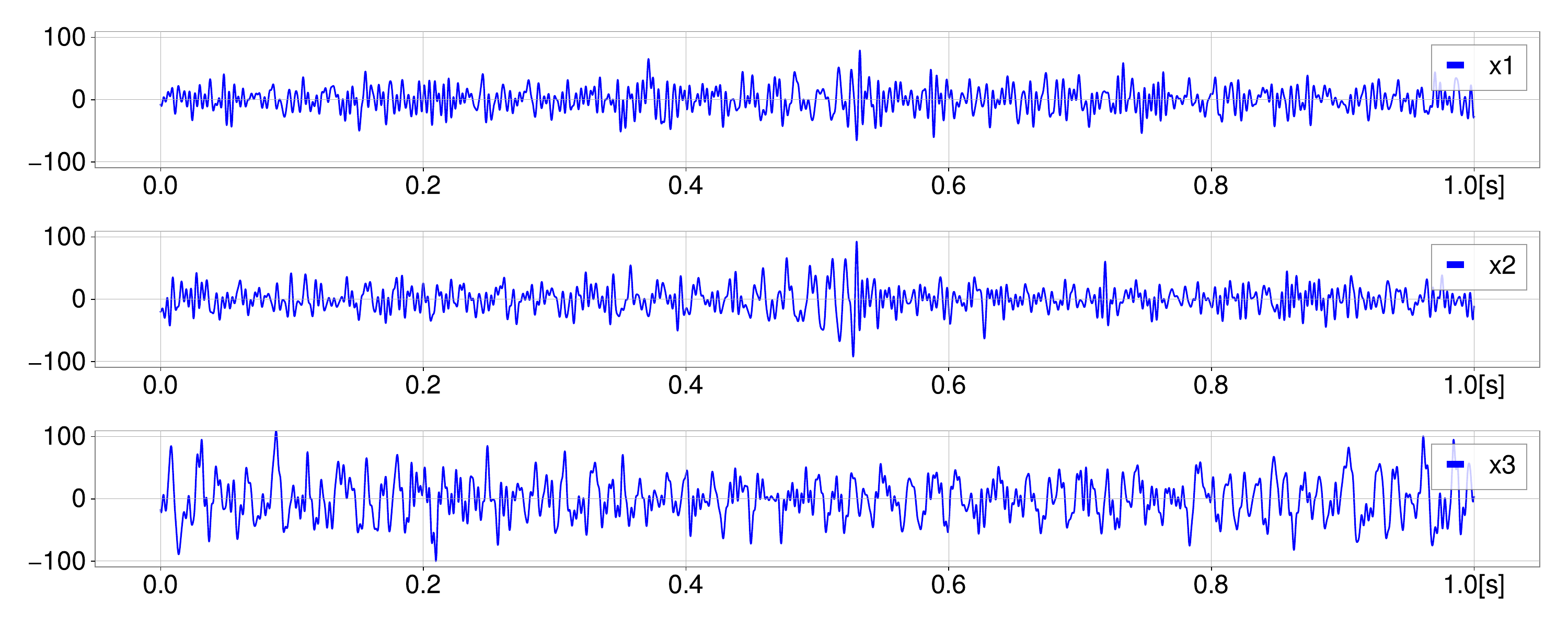}
&
\includegraphics[width=.45\textwidth]
{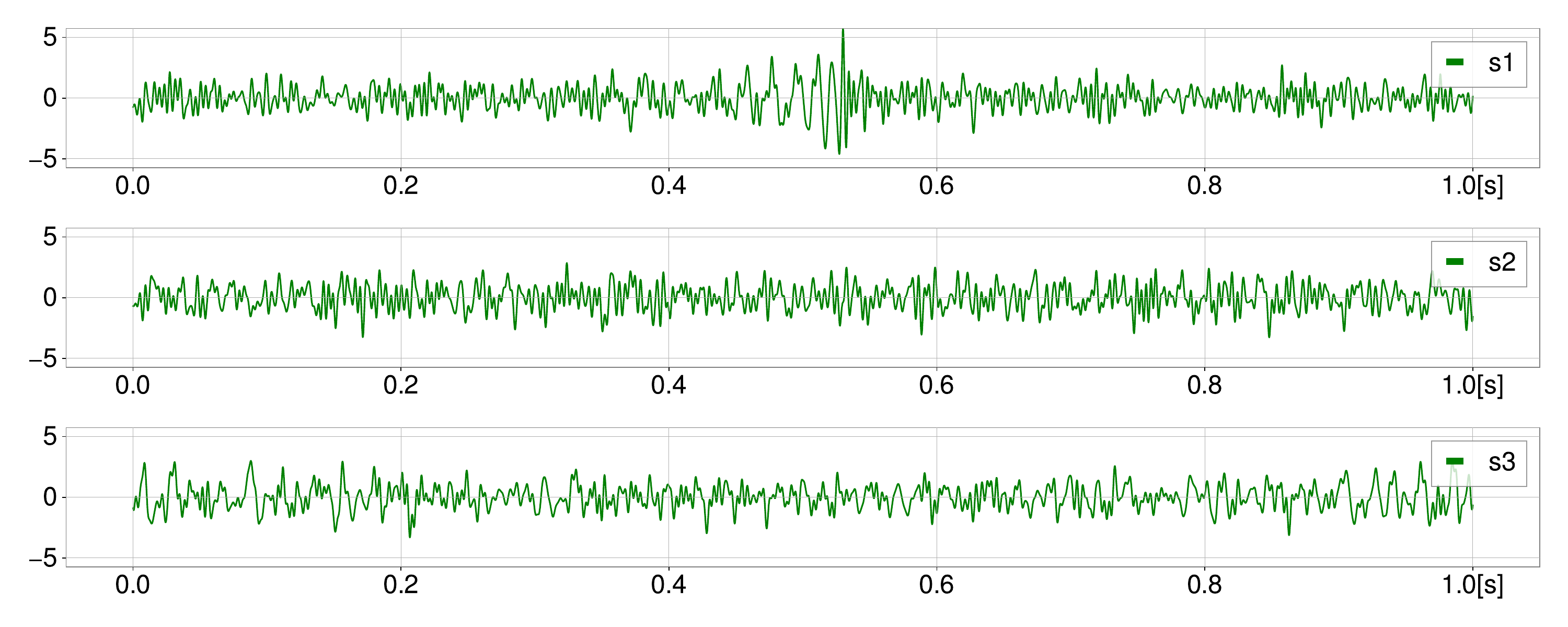}
\\
{\small (f1) Input signals of GW170814 with $\Delta t_{\mbox{\footnotesize HL}}=-8.06$~ms and $\Delta t_{\mbox{\footnotesize HV}}=+0.98$~ms. The data x1, x2, and x3 are of Hanford, Livingston, and Virgo, respectively. }
&
{\small (f2) Output of ICA for GW170814.} \\[3em]
\includegraphics[width=.45\textwidth]
{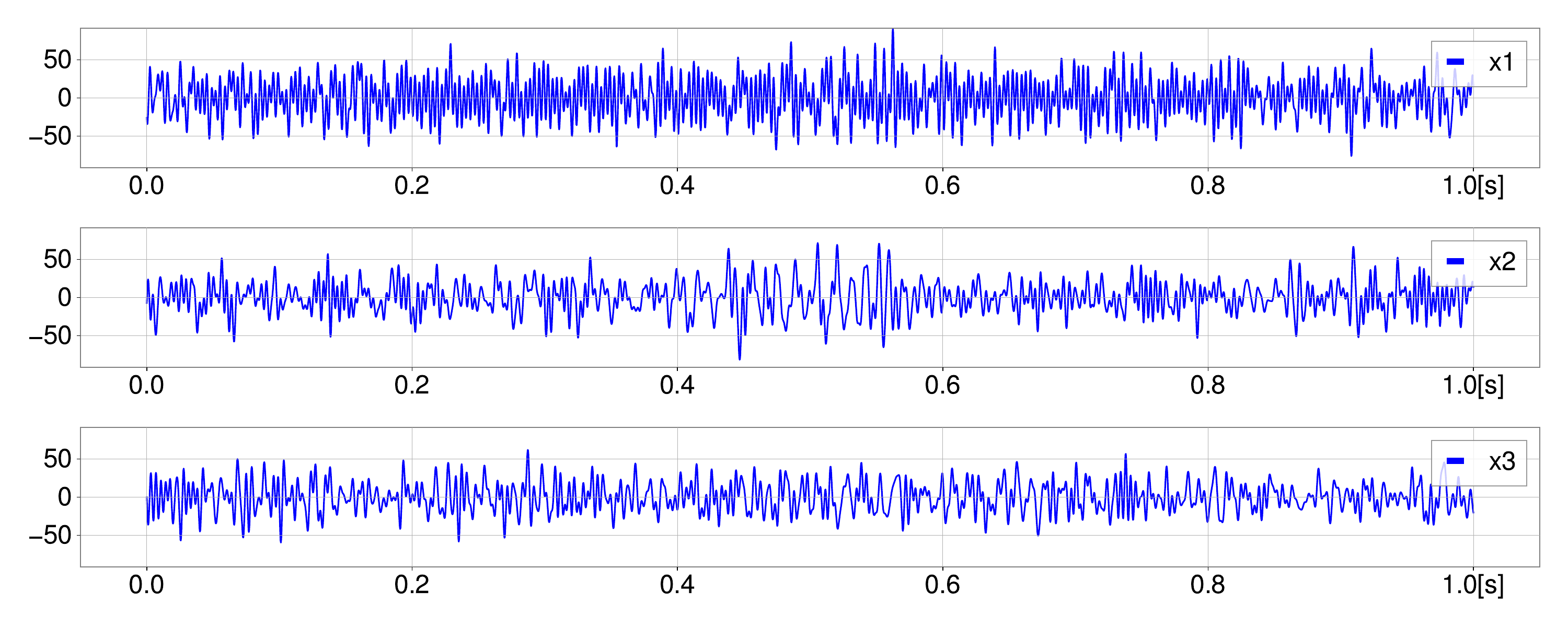}
&
\includegraphics[width=.45\textwidth]
{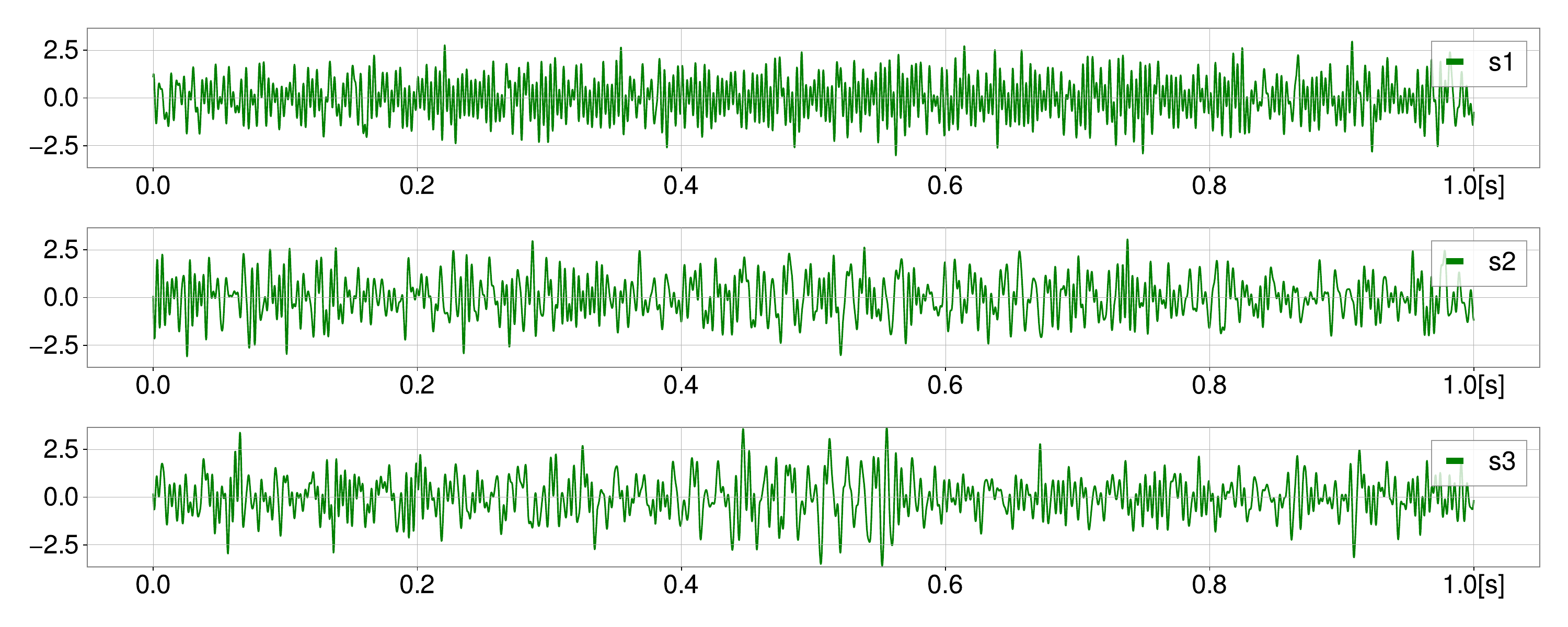}
\\
{\small (g1) Input signals of GW190412 with $\Delta t_{\mbox{\footnotesize HL}}=-3.91$~ms and $\Delta t_{\mbox{\footnotesize HV}}=-13.92$~ms. The data x1, x2, and x3 are of Hanford, Livingston, and Virgo, respectively. }
&
{\small (g2) Output of ICA for GW190412. }\\[3em]
\includegraphics[width=.45\textwidth]
{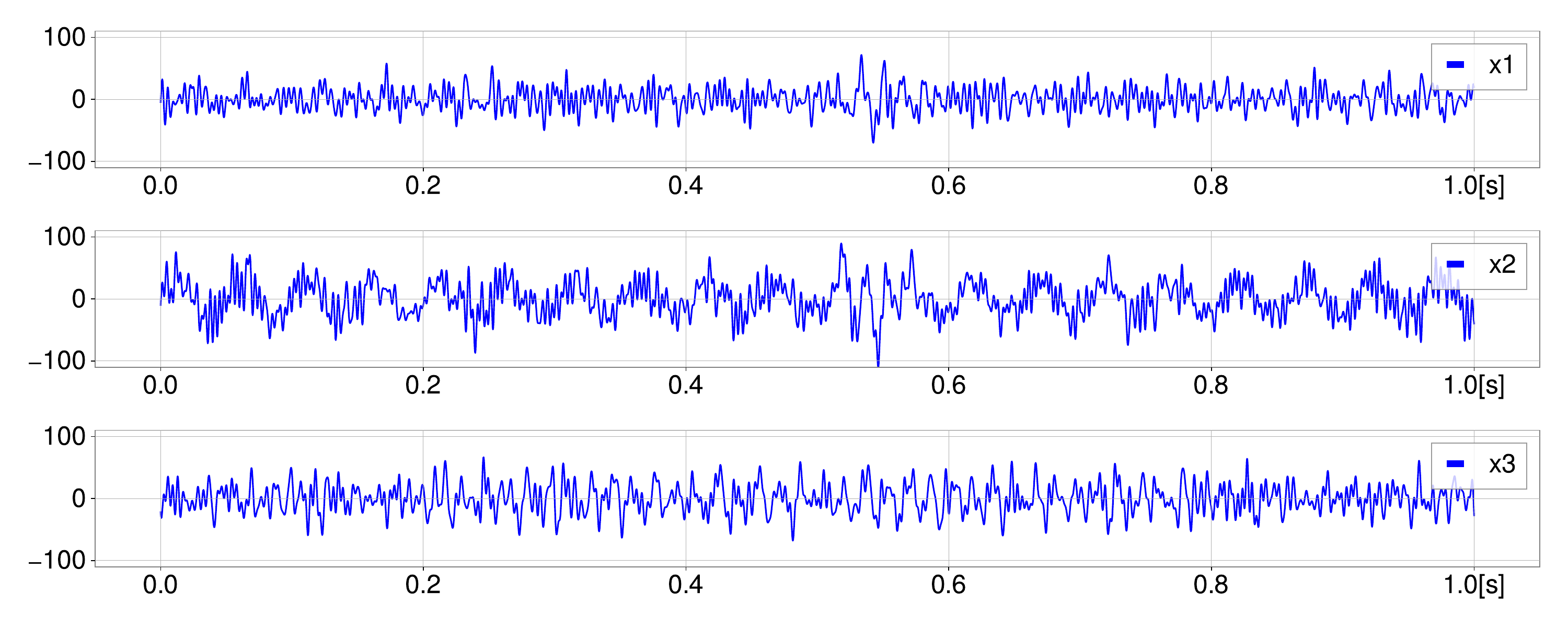}
&
\includegraphics[width=.45\textwidth]
{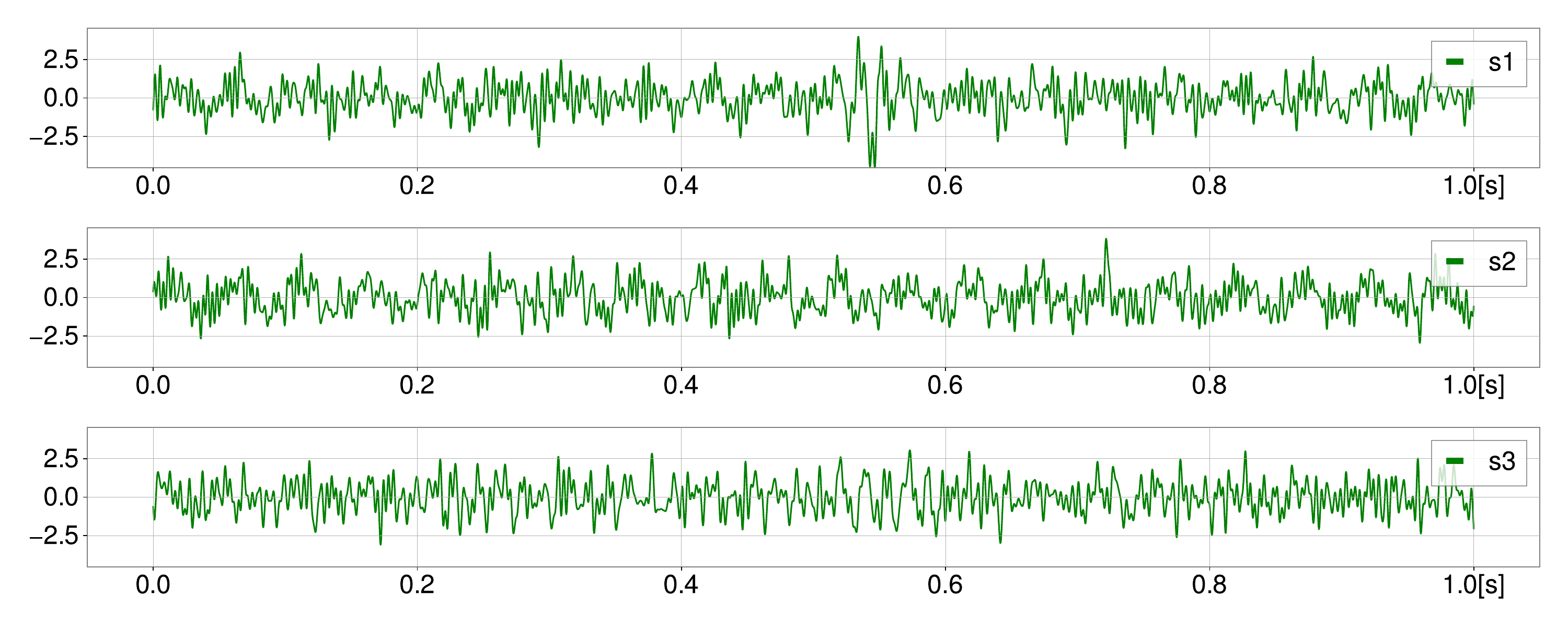}
\\
{\small (h1) Input signals of GW190521 with $\Delta t_{\mbox{\footnotesize HL}}=+2.93$~ms and $\Delta t_{\mbox{\footnotesize HV}}=-25.15$~ms. The data x1, x2, and x3 are of Hanford, Livingston, and Virgo, respectively. }
&
{\small (h2) Output of ICA for GW190521. }\\[3em]
\end{tabular}
\caption{
Input and Output data of ICA analysis (cont.)
}
\end{figure}

\addtocounter{figure}{-1}
\begin{figure}
\begin{tabular}{p{0.5\textwidth}p{0.5\textwidth}}
\includegraphics[width=.45\textwidth]
{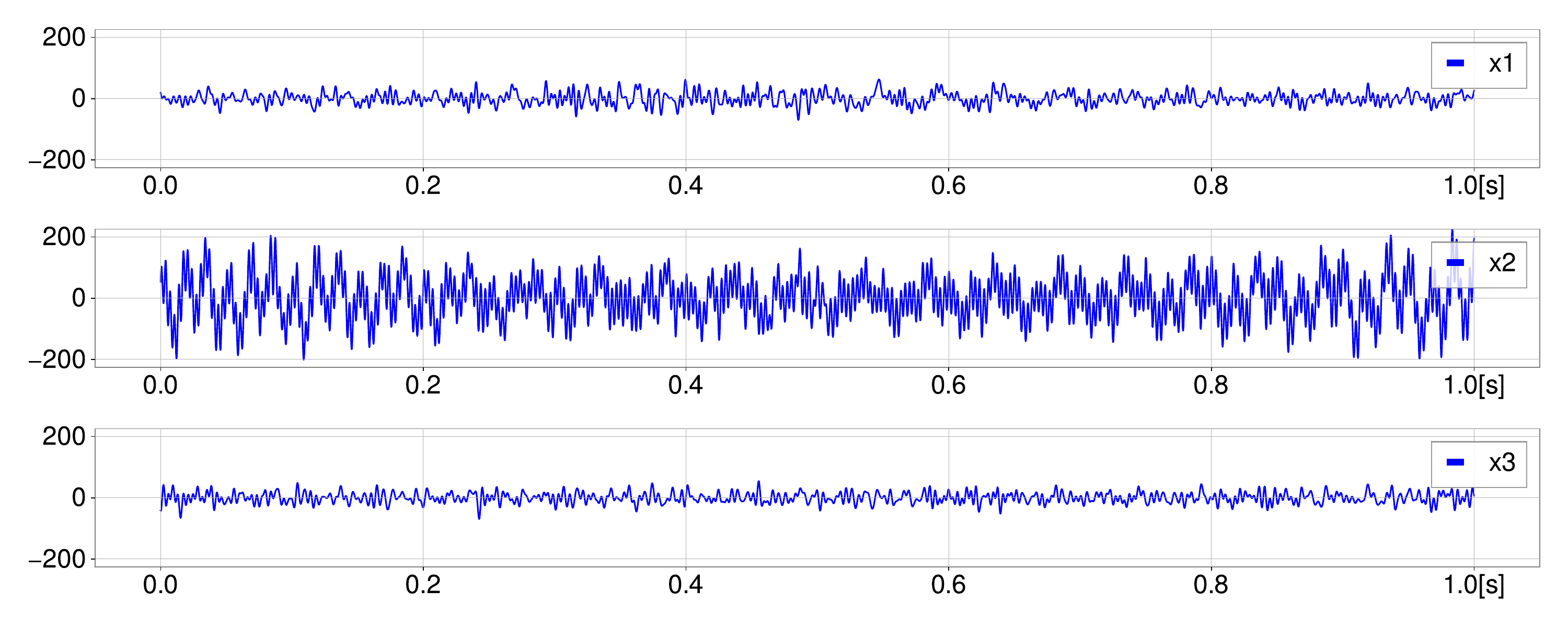}
&
\includegraphics[width=.45\textwidth]
{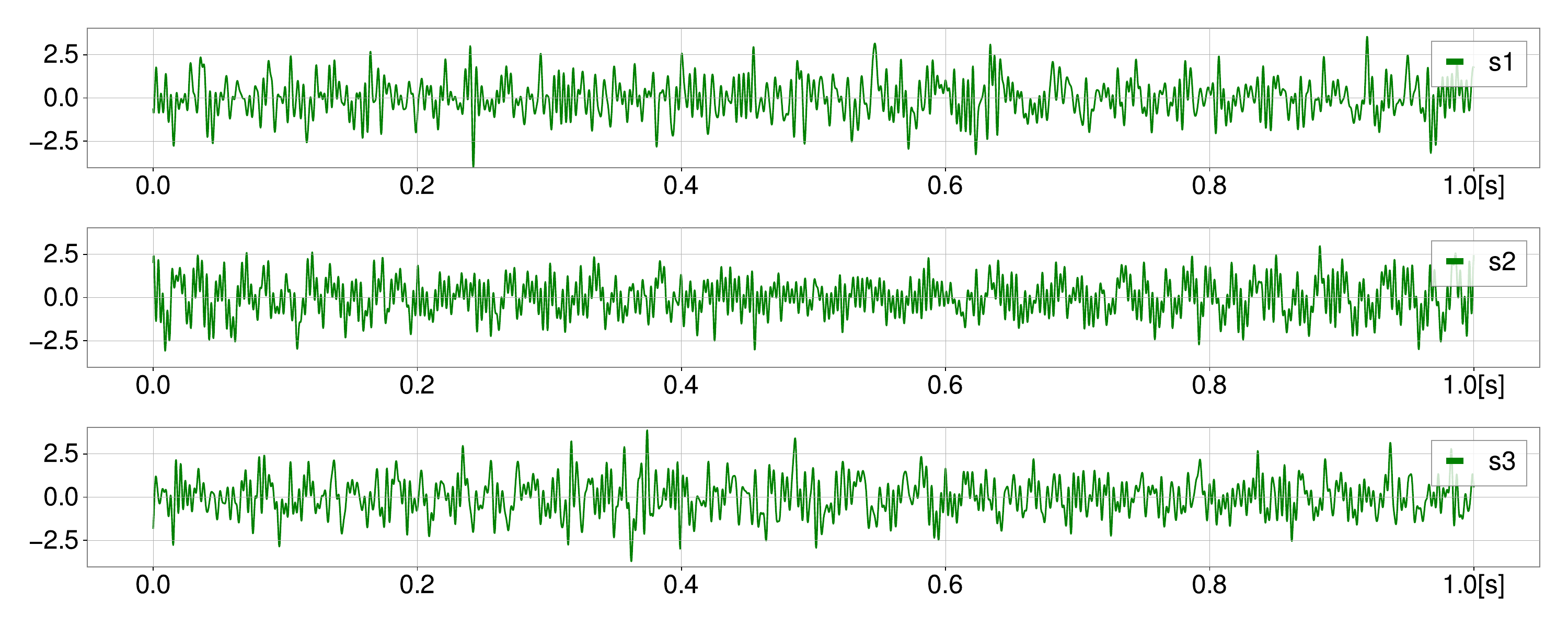}
\\
{\small (i1) Input signals of GW190814 with $\Delta t_{\mbox{\footnotesize HL}}=+2.20$~ms and $\Delta t_{\mbox{\footnotesize HV}}=+21.24$~ms. The data x1, x2, and x3 are of Hanford, Livingston, and Virgo, respectively. }
&
{\small (i2) Output of ICA for GW190814.} \\[3em]
\includegraphics[width=.45\textwidth]
{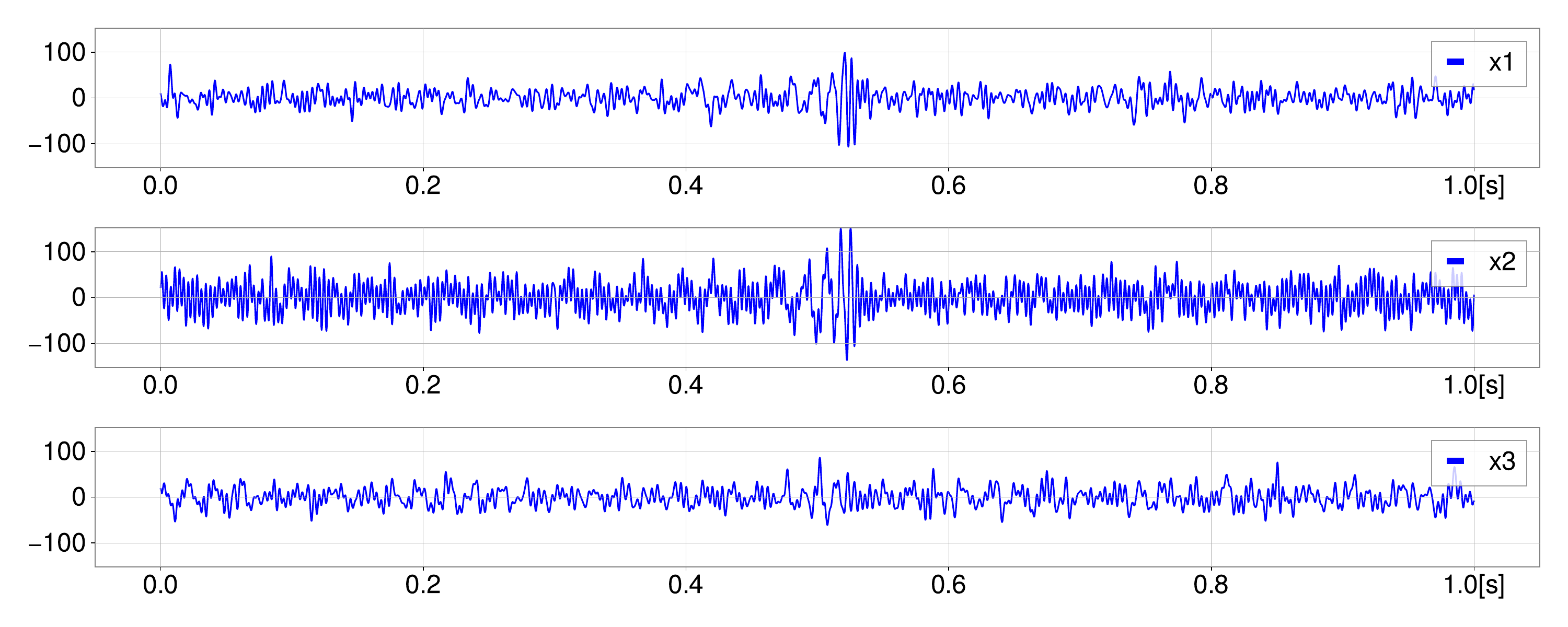}
&
\includegraphics[width=.45\textwidth]
{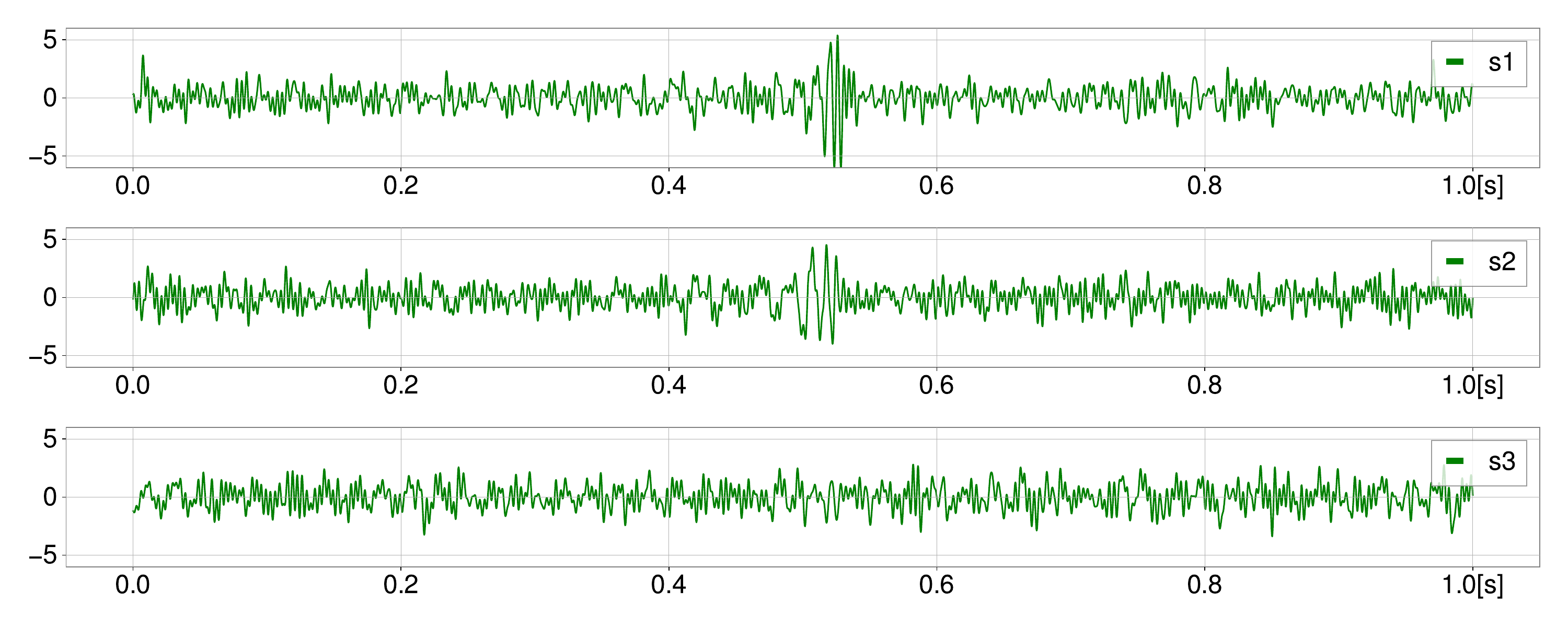}
\\
{\small (j1) Input signals of GW200129\_065458 with $\Delta t_{\mbox{\footnotesize HL}}=+3.42$~ms and $\Delta t_{\mbox{\footnotesize HV}}=-18.31$~ms. The data x1, x2, and x3 are of Hanford, Livingston, and Virgo, respectively. }
&
{\small (j2) Output of ICA for GW200129\_065458. }\\[3em]
\includegraphics[width=.45\textwidth]
{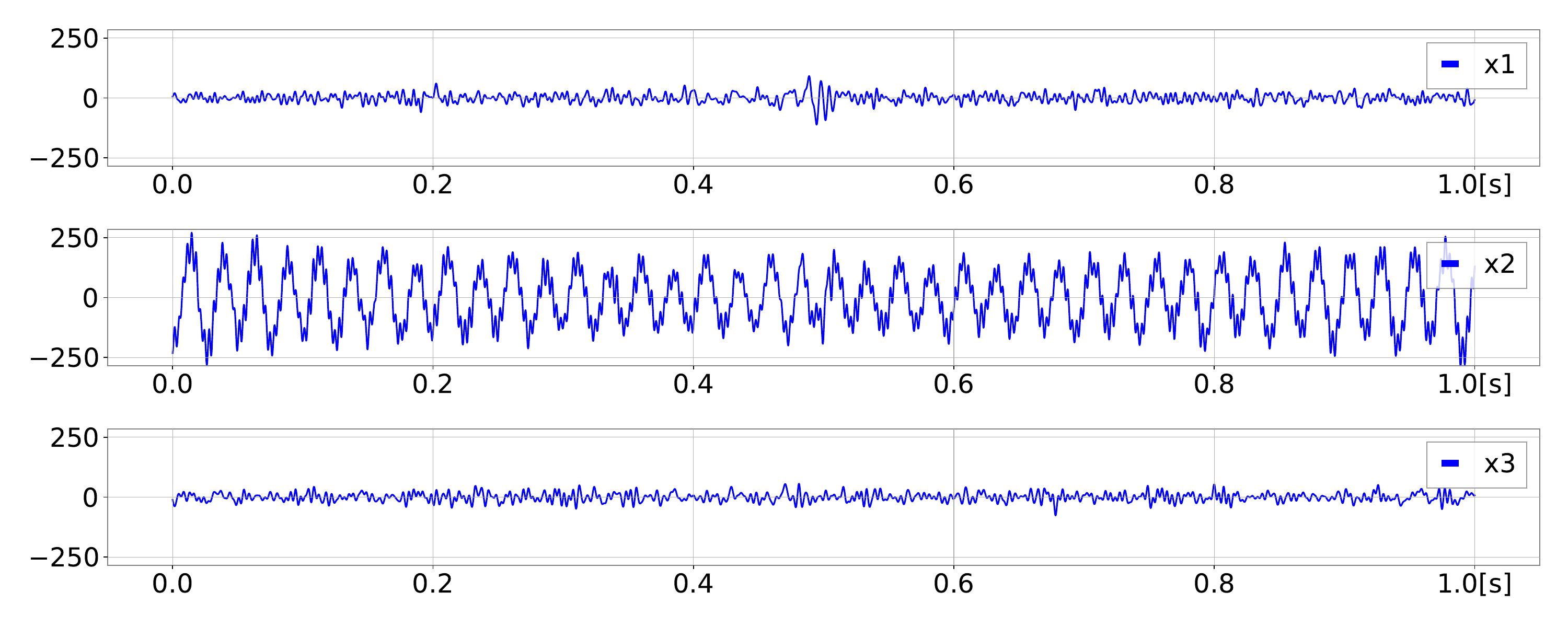}
&
\includegraphics[width=.45\textwidth]
{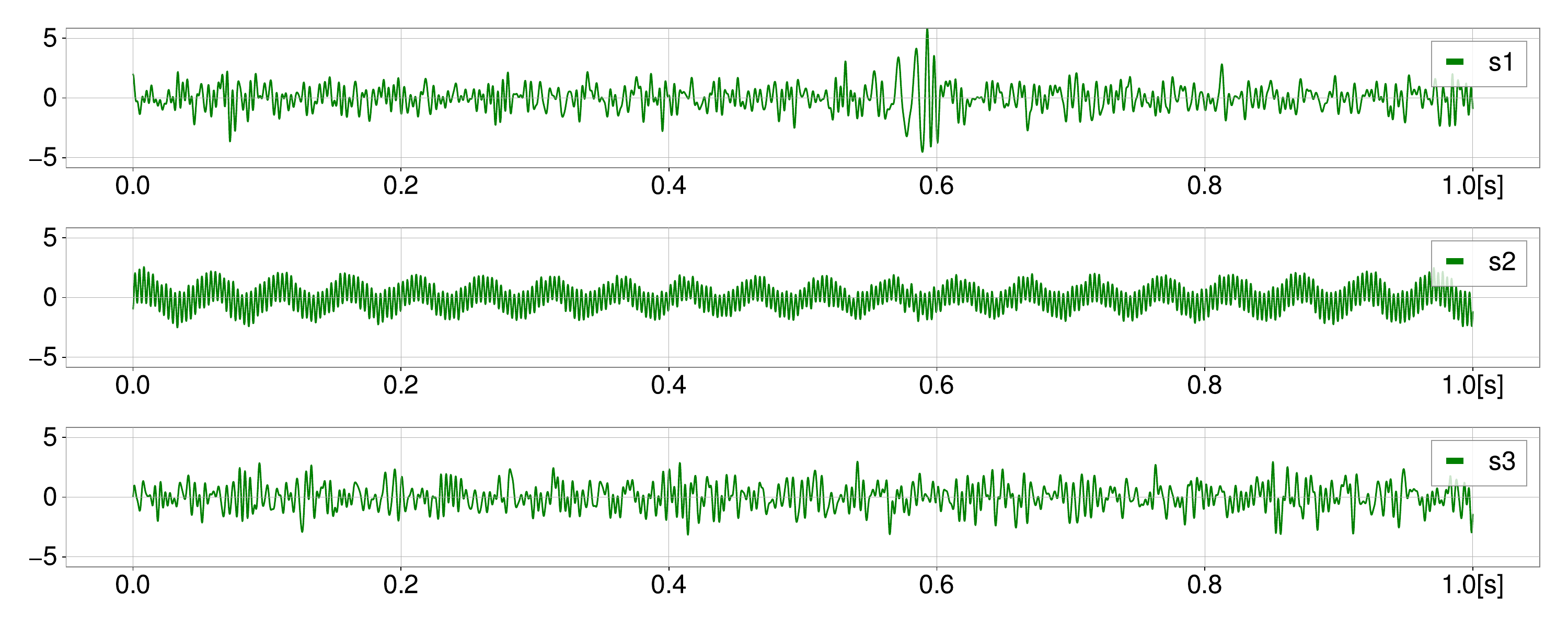}
\\
{\small (k1) Input signals of GW200224\_222234 with $\Delta t_{\mbox{\footnotesize HL}}=-3.66$~ms and $\Delta t_{\mbox{\footnotesize HV}}=-9.28$~ms. The data x1, x2, and x3 are of Hanford, Livingston, and Virgo, respectively. }
&
{\small (k2) Output of ICA for GW200224\_222234. }\\[3em]
\includegraphics[width=.45\textwidth]
{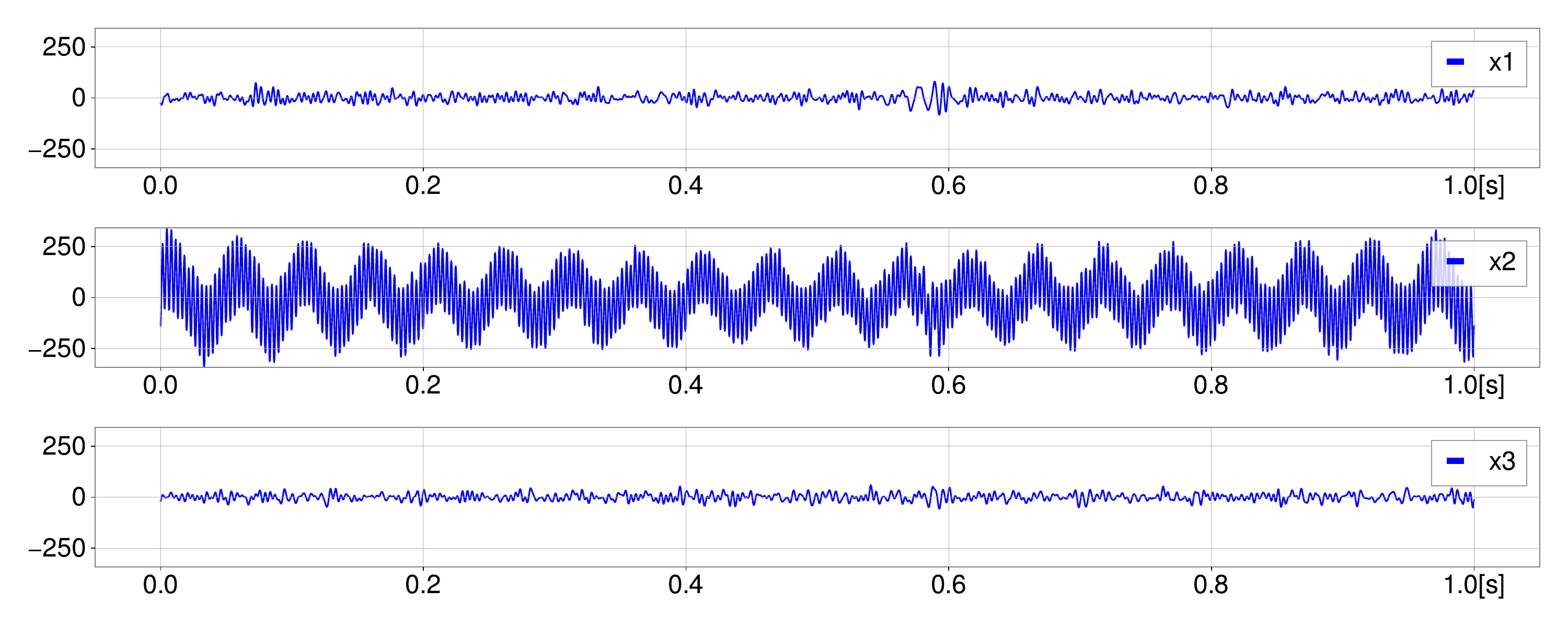}
&
\includegraphics[width=.45\textwidth]
{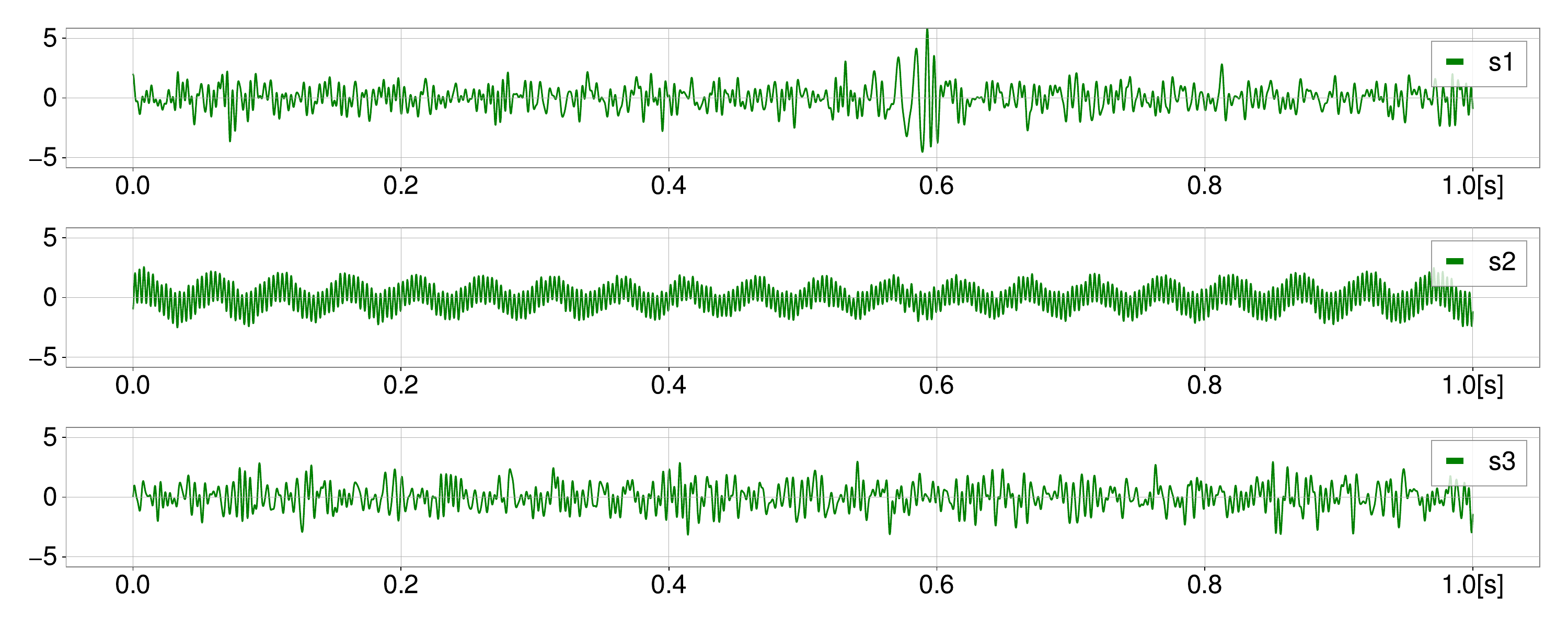}
\\
{\small (l1) Input signals of GW200311\_115853 with $\Delta t_{\mbox{\footnotesize HL}}=-3.66$~ms and $\Delta t_{\mbox{\footnotesize HV}}=-27.10$~ms. The data x1, x2, and x3 are of Hanford, Livingston, and Virgo, respectively. }
&
{\small (l2) Output of ICA for GW200311\_115853. }\\
\end{tabular}
\caption{
Input and Output data of ICA analysis (cont.)
}
\end{figure}

\clearpage
\section{Summary}
Independent component analysis (ICA) was applied to extract gravitational-wave signals. From the injection tests, we see that the extractions are available even from real interferometer data if the signal strength (SNR) is 15 or higher.  We then demonstrate this method for  inspiral-wave extraction for binary black-hole events, particularly 13 high-SNR events up to O3 (GWTC-3 catalog).  As shown in Table \ref{table:modelGWreal} the extractions were  performed for all events, and as shown in Table  \ref{table:modelGWreal2} the fitted waveforms show parameters $M_c$ and $z$ consistent with those reported by LIGO-Virgo-KAGRA collaboration papers.

ICA is based on the idea of how the signal is mathematically independent from others.  Only the non-Gaussian waves were extracted.  Additionally, the current FastICA method starts by normalizing and whitening the input data, which makes the output signals without amplitude information, and the phase can be reversed.  If we know the waveform similar to our applications to GWTC-3, we can determine the phase by evaluating the residual, while the amplitude itself remains undetermined. 

Despite these limitations, the method proposed here is attractive because it does not use any templates in advance.  If we could visualize the waveforms first, it would undoubtedly help in the theoretical understanding.
We believe that this ICA approach will contribute to testing theories of gravity and finding unknown GW signals. 

~

We thank Hirotaka Yuzurihara for his suggestions on the technical procedure. 
This work was supported by JSPS KAKENHI Grant Nos. 24K07029 and 18K03630.

\appendix
\section{FastICA using Kurtosis}\label{appA}
After whitening the input data $ {\bm x}$ to $ {\bm z}$ [Eq. (\ref{eq:whiten})], 
one possible measure of the non-Gaussianity of the output data is to use the kurtosis of $w_i^T {\bm z}$, 
\begin{equation}
 \mbox{kurt} ({\bm w}_i^T  {\bm z}) = E[({\bm w}_i^T  {\bm z}) ^4] - 3 \{ E[({\bm w}_i^T  {\bm z}) ^2] \}^2. \label{eq.kurt}
\end{equation}
A well-known algorithm, FastICA, detemines  ${\bm w}_i$ that maximizes  $\mbox{kurt} ({\bm w}_i^T  {\bm z})$.  
Using $E[({\bm w}_i^T  {\bm z}) ^2] = ||{\bm w}_i||^2$, the derivative of  Eq. (\ref{eq.kurt}) becomes 
\begin{equation}
\frac{\partial}{\partial {\bm w}_i} | \mbox{kurt} ({\bm w}_i^T  {\bm z}) | = \left( \begin{array}{c} 
E[4({\bm w}_i^T  {\bm z}) ^3 {\bm z}_1] \\
E[4({\bm w}_i^T  {\bm z}) ^3 {\bm z}_2] \\
\vdots
\end{array}\right) 
- 12 || {\bm w}_i ||^2 
{\bm w}_i.  \label{derivKurt}
\end{equation}
We can, then, find ${\bm w}_i$ as in Eq. (\ref{derivKurt}) can be satisfied by an iterative method, especially 
easily under the requirement that the norm of ${\bm w}_i$ is unity, $||{\bm w}_i||^2=1$. 
By repeating the process of finding another component $ {\bm w}_i$ as each $ {\bm w}_i$ satisfies its orthogonality, we can identify all possible source signals $\tilde{\bm s}(t)$.

However, the kurtosis method is sometimes sensitive to outliers.  An alternative method is to use a mimic function to the kurtosis which we described as the ``$g$-function" method in Eq. (\ref{eq.gfunc}). 




\end{document}